\documentclass[a4paper,usenames,dvipsnames,11pt]{article}
\pdfoutput=1

\usepackage{jheppub}
\usepackage{slashed}
\usepackage{mathrsfs,booktabs,multirow,tabularx}
\usepackage{stmaryrd}
\usepackage{xspace}
\usepackage{fancyvrb}
\usepackage[makeroom]{cancel}
\usepackage{amsmath}    
\usepackage{amssymb}    %
\usepackage{graphicx}   
\usepackage{verbatim}   
\usepackage{lscape}
\usepackage{subfig}
\usepackage{listings}

\def\beq{\begin{equation}}
\def\beqn{\begin{eqnarray}}
\def\eeq{\end{equation}}
\def\eeqn{\end{eqnarray}}

\def\lppdistr#1#2{\left(\frac{\log\left(#1\right)}{#1}\right)_{\!#2}}

\def\PDF#1#2{\Gamma_{\!#1/#2}}
\def\ePDF#1{\Gamma_{\!#1}}

\def\binomial#1#2{
\left(\!\!
\begin{array}{c}
#1\\
#2
\end{array}
\!\!\right)
}

\newcommand\sss{\scriptscriptstyle}

\newcommand\half{\frac{1}{2}}

\newcommand\aem{\alpha}
\newcommand\aemotpi{\frac{\aem}{2\pi}}
\newcommand\aemz{\alpha(\mu_0)}
\newcommand\aemzotpi{\frac{\aemz}{2\pi}}
\newcommand\aemmu{\alpha(\mu)}

\newcommand\gE{\gamma_{\sss\rm E}}

\newcommand{\bN}{\bar{N}}

\newcommand{\epem}{e^+e^-}
\newcommand{\lp}{e^+}
\newcommand{\lm}{e^-}
\newcommand{\lpm}{e^{\pm}}
\newcommand{\lmp}{e^{\mp}}
\newcommand{\bl}{\bar{e}}

\newcommand{\zp}{z_+}
\newcommand{\zm}{z_-}

\newcommand{\plp}{p_{\lp}}
\newcommand{\plm}{p_{\lm}}

\newcommand{\ord}{{\cal O}}

\newcommand\NF{n_{\sss F}}

\newcommand\hsig{\hat{\sigma}}
\newcommand\bsig{\bar{\sigma}}

\newcommand\APmat{{\mathbb P}}
\newcommand\Eop{{\mathbb E}}
\newcommand\Mmat{{\mathbb M}}

\newcommand\MSb{\overline{\rm MS}}

\newcommand\Fpdf{{\cal F}}
\newcommand\iLL{I^{\sss\rm LL}}
\newcommand\iNLL{I^{\sss\rm NLL}}
\newcommand\jLL{J^{\sss\rm LL}}
\newcommand\jNLL{J^{\sss\rm NLL}}
\newcommand\jbLL{\bar{J}^{\sss\rm LL}}
\newcommand\jbNLL{\bar{J}^{\sss\rm NLL}}
\newcommand\jhLL{\hat{J}^{\sss\rm LL}}
\newcommand\jhNLL{\hat{J}^{\sss\rm NLL}}
\newcommand\kLL{K^{\sss\rm LL}}
\newcommand\kNLL{K^{\sss\rm NLL}}
\newcommand\ILL{{\cal I}^{\sss\rm LL}}
\newcommand\INLL{{\cal I}^{\sss\rm NLL}}
\newcommand\JLL{{\cal J}^{\sss\rm LL}}
\newcommand\JNLL{{\cal J}^{\sss\rm NLL}}
\newcommand\KLL{{\cal K}^{\sss\rm LL}}
\newcommand\KNLL{{\cal K}^{\sss\rm NLL}}
\newcommand\ootimes{\,\overline{\otimes}\,}
\newcommand\ePDFs{\ePDF{\rm\sss S}}
\newcommand\ePDFns{\ePDF{\rm\sss NS}}
\newcommand\PV{P^{\rm\sss V}}
\newcommand\PS{P^{\rm\sss S}}
\newcommand\muz{\mu_0}
\newcommand\lbase{\ell}
\newcommand\qbase{q}
\newcommand\Lbase{{\cal L}}
\newcommand\cpar{C_5}
\newcommand\dencpar{M_1}
\newcommand\dendpar{M_2}
\newcommand\doc{C_4}
\newcommand\Cthree{C_1}
\newcommand\Cfour{C_2}
\newcommand\Cfive{C_3}
\newcommand\Coneone{C_{1,1}}
\newcommand\Ctwoone{C_{2,1}}
\newcommand\Cthreeone{C_{3,1}}
\newcommand\Conetwo{C_{1,2}}
\newcommand\Ctwotwo{C_{2,2}}
\newcommand\Cthreetwo{C_{3,2}}
\newcommand\Conethree{C_{1,3}}
\newcommand\Ctwothree{C_{2,3}}
\newcommand\Cthreethree{C_{3,3}}
\newcommand\Conefour{C_{1,4}}
\newcommand\Ctwofour{C_{2,4}}
\newcommand\Cthreefour{C_{3,4}}
\newcommand\Conej{C_{1,j}}
\newcommand\Ctwoj{C_{2,j}}
\newcommand\Cthreej{C_{3,j}}
\newcommand\DoneCtwo{D_{1,1}}
\newcommand\DoneCthree{D_{2,1}}
\newcommand\DonebCtwo{D_{1,2}}
\newcommand\DonebCthree{D_{2,2}}
\newcommand\DtwoCtwo{D_{1,3}}
\newcommand\DtwoCthree{D_{2,3}}
\newcommand\DtwobCtwo{D_{1,4}}
\newcommand\DtwobCthree{D_{2,4}}
\newcommand\DjCtwo{D_{1,j}}
\newcommand\DjCthree{D_{2,j}}
\newcommand\Fzero{F_0}
\newcommand\Fone{F_1}
\newcommand\Ftwo{F_2}
\newcommand\Fis{F_i}
\newcommand\rrdo{R_1}
\newcommand\rrcz{R_2}
\newcommand\rrno{R_3}
\newcommand\rrnt{R_4}
\newcommand\rrnth{R_5}
\newcommand{\hz}{\hat{z}}

\def\Li#1{\text{Li}_{#1}}
\def\jnum#1{J^{\sss\rm num}_{#1}}

\newcommand\Lzero{L_0}
\newcommand\GammaLL{\Gamma^{\sss\rm LL}}
\newcommand\GammaNLL{\Gamma^{\sss\rm NLL}}
\newcommand\invm{\mathfrak{M}}

\title{The partonic structure of the electron at the next-to-leading 
logarithmic accuracy in QED}

\author[a]{V. Bertone,}
\affiliation[a]{Dipartimento di Fisica, Universit\`a di Pavia and INFN, 
Sezione di Pavia,\\ Via Bassi 6, I-27100 Pavia, Italy}

\author[b,c]{M. Cacciari,}
\affiliation[b]{Sorbonne Universit\'e, CNRS, Laboratoire de Physique 
Th\'orique et Hautes \'Energies,\\ LPTHE, F-75005 Paris, France}
\affiliation[c]{Universit\'e de Paris, LPTHE, F-75005 Paris, France}

\author[d]{S. Frixione,}
\affiliation[d]{INFN, Sezione di Genova, Via Dodecaneso 33, I-16146, 
Genoa, Italy}

\author[b,c,e]{G. Stagnitto}
\affiliation[e]{Tif Lab, Dipartimento di Fisica, Universit\`a di Milano 
and INFN, Sezione di Milano,\\ Via Celoria 16, I-20133 Milano, Italy}

\emailAdd{valerio.bertone@cern.ch}
\emailAdd{cacciari@lpthe.jussieu.fr}
\emailAdd{Stefano.Frixione@cern.ch}
\emailAdd{gstagnit@lpthe.jussieu.fr}

\abstract{
By working in QED, we obtain the electron, positron, and photon Parton
Distribution Functions (PDFs) of the unpolarised electron at the 
next-to-leading logarithmic accuracy. The PDFs account for all of the
universal effects of initial-state collinear origin, and are key 
ingredients in the calculations of cross sections in the so-called
structure-function approach. We present both numerical and analytical 
results, and show that they agree extremely well with each other.
The analytical predictions are defined by means of an additive formula
that matches a large-$z$ solution that includes all orders in the QED
coupling constant $\aem$, with a small- and intermediate-$z$ solution
that includes terms up to ${\cal O}(\aem^3)$.
}


\preprint{
\begin{flushright}
TIF-UNIMI-2020-11\\
\today
\end{flushright}
}

\begin{document}
\maketitle
\flushbottom

\section{Introduction\label{sec:intro}}
The typical cross section relevant to $\epem$ collisions is in
principle entirely computable as a perturbative series in the QED
coupling constant $\aem$. In practice, however, this is hardly useful,
since the coefficients of such a series are very large, thus compensating
the suppression due to $\aem$ -- in other words, all terms of the series
might be of the same order numerically, which leads to a complete loss
of predictive power. The problem stems from the fact that the incoming
$\lpm$ particles tend to copiously radiate photons (which in turn may
convert into $\epem$ pairs, and so forth) at small angles w.r.t.~the
beamline. In perturbation theory, any zero-angle emission would induce
a divergent cross section, were it not for the screening effect provided
by the mass of the emitter and/or the emitted particle. Thus, when integrating 
over all possible emissions, the cross section will contain logarithms of
the ratio $m^2/E^2$, where $E$ is a scale of the order of the hardness
of the process, and $m$ is the screening mass (i.e.~that of the electron
in the case we are interested in). It is these logarithms that, by growing
large when $m^2/E^2\ll 1$, give the dominant contributions to the
perturbative coefficients, and ultimately prevent the series from 
being well behaved.

Fortunately, such $\log(m^2/E^2)$ terms are universal, and because of this 
they can be taken into account to all orders in $\aem$ by a process-independent 
resummation procedure. In the so-called structure-function approach, the
physical cross section is then written by means of a factorisation
formula such as the following one\footnote{Throughout this paper we
sum over lepton and photon polarisations. However, the techniques we 
shall employ can be extended to deal with polarised particles.}:
\beqn
d\bsig_{\epem}(\plp,\plm,m^2)&=&\sum_{ij=\lpm,\gamma}\int d\zp d\zm\,
\PDF{i}{\lp}(\zp,\mu^2,m^2)\,\PDF{j}{\lm}(\zm,\mu^2,m^2)
\nonumber\\*&&\phantom{\sum_{ij}\int}\times
d\hsig_{ij}(\zp\plp,\zm\plm,\mu^2)\,.
\label{master0}
\eeqn
The quantities $\PDF{i}{\lpm}$ are called the Parton Distribution Functions 
(PDFs) of the electron or the positron, a name that originates from 
the analogy of eq.~(\ref{master0}) with its QCD counterpart. PDFs collect 
and resum all of the $\log(m^2/E^2)$ terms; conversely, the short-distance 
cross sections \mbox{$d\hsig_{ij}$} do not contain any such logarithms, 
and are expected to be well-behaved order by order in perturbation theory.
Neither $\PDF{i}{\lpm}$ nor \mbox{$d\hsig_{ij}$} are physical quantities; 
their definitions always involve some degree of arbitrariness, which is
parametrised by the mass scale $\mu$, that is only constrained by the 
requirement $\mu\sim E$, and by the chosen factorisation scheme.

Fuller details on the usage of the factorisation formula~(\ref{master0})
in calculations relevant to $\epem$ colliders and on its physical
meaning can be found e.g.~in ref.~\cite{Frixione:2019lga}. In particular,
we shall adopt the notation of ref.~\cite{Frixione:2019lga}, whereby
the incoming $e^\pm$ are called {\em particles} (with $d\bsig_{\epem}$
thus being a particle-level cross section, defined so as to retain only
terms that do not vanish in the $m/E\to 0$ limit), and the objects $i$ and 
$j$ are called {\em partons} (so that $d\hsig_{ij}$ is a parton-level cross 
section). This allows one to distinguish easily between an electron that
stems from one of the collider beams, and an electron that initiates
the hard collisions, and that stems from the PDF $\PDF{\lm}{\lm}$.

We point out that it is somehow customary in QED to call $\PDF{\lm}{\lm}$
($\PDF{\lp}{\lp}$) the electron (positron) structure function. This is 
motivated by the fact that, by ignoring the contributions of partons whose 
species is {\em not} the same as that of the incoming particle, and by working 
at the first order in perturbation theory, a structure function (which is an 
observable) can be made to coincide with the PDF relevant to the case where 
particle and parton have the same identity, by means of a suitable definition
of such a PDF. This position is not tenable at higher perturbative orders, 
and when more parton species are allowed in any given particle. Therefore, 
``structure functions'' will not be used in this paper, and we shall only 
refer to PDFs.

The crucial point that we need to stress here is that in QED $\lpm$ PDFs, 
at variance with hadronic PDFs, are entirely calculable with perturbative 
techniques. 
Presently they are known in close analytical forms~\cite{Skrzypek:1990qs,
Skrzypek:1992vk,Cacciari:1992pz} which are leading-logarithmic (LL) accurate,
and that include all-order in $\aem$ contributions in the region 
$z_\pm\simeq 1$ (which is responsible for the bulk of the cross section), 
matched with up to $\ord(\aem^3)$ terms for any values of $z_\pm$; both of 
these forms exploit leading-order (LO) initial conditions. The goal of the 
present work 
is to improve on the results of refs.~\cite{Skrzypek:1990qs,Skrzypek:1992vk,
Cacciari:1992pz} by extending them to the next-to-leading logarithm accuracy
(NLL) starting from the next-to-leading order (NLO) initial conditions
computed in ref.~\cite{Frixione:2019lga}. In keeping with what was done
in the literature, we shall present predictions both for all-order PDFs 
in the $z_\pm\simeq 1$ region, and for up to $\ord(\aem^3)$ NLL terms 
valid for any $z_\pm$. By working at the NLL+NLO accuracy, the mixing
between the electron/positron and the photon PDFs is taken into proper 
account, as are running-$\aem$ effects. Our results are obtained with 
both analytical and numerical methods, which are compared and used to 
validate each other.

This paper is organised as follows. For those readers who are not interested 
in the technical procedures which underpin this work but only in their final 
outcomes, we summarise our final results in sect.~\ref{sec:syn}. The 
details of the derivations of such results are then given in the 
remainder of the paper. In sect.~\ref{sec:evol} we introduce
the evolution equations for the PDFs that we are going to solve, and
report the associated initial conditions. Sect.~\ref{sec:eop} briefly
describes the evolution-operator formalism. Analytical solutions are
computed in sect.~\ref{sec:sols}, for any $z_\pm$ values in
sect.~\ref{sec:rec} (the resulting lengthy expressions are partly
collected in appendix~\ref{sec:recres}), and for $z_\pm\simeq 1$ in
sect.~\ref{sec:asy} (with additional details reported in 
appendix~\ref{sec:asyph}), while a description of the codes employed to
obtain numerical results is given in sect.~\ref{sec:num}. The solutions
of sects.~\ref{sec:rec} and~\ref{sec:asy} are combined (``matched'')
in sect.~\ref{sec:match}. Our analytical and numerical predictions are 
extensively compared in sect.~\ref{sec:res}. Finally, we conclude and 
give a short outlook in sect.~\ref{sec:concl}. Additional material
is collected in the appendices.

\section{Synopsis of results\label{sec:syn}}
The $\lpm$ PDFs that we shall compute in this paper result from solving the 
evolution equations of eqs.~(\ref{APeqsga})--(\ref{APeqns}), with the initial
conditions given in eqs.~(\ref{G0sol})--(\ref{Gpossol2}). The all-order,
large-$z$ solutions and one of the numerical codes we shall employ 
require the use of an evolution operator, whose RGE is presented in 
eq.~(\ref{matAPmell4}). The latter can be solved in a closed form in 
the case of a one-dimensional flavour space; such a closed form is reported 
either in eq.~(\ref{Esol1}) or in eq.~(\ref{Esol1ex}), the two differing by 
terms of $\ord(\aem^3)$ and with the latter form suitable to take the 
fixed-$\aem$ limit. The two-dimensional flavour space case is discussed
in appendix~\ref{sec:asyph}. The PDF solutions valid for any $z$ and including
up to $\ord(\aem^3)$ contributions are represented in terms of sets of basis
functions, which are obtained by solving recursive equations; these are
given in eqs.~(\ref{JLLsol}) and~(\ref{JNLLsol}), and their explicit
solutions partly in appendix~\ref{sec:recres}, and partly in an 
ancillary file (see below). Conversely, the all-order, large-$z$ solutions 
for the PDFs are reported in eq.~(\ref{LLsol3}) (LL accurate for singlet
and non-singlet), eq.~(\ref{gaLLsol4run}) (LL accurate for photon),
eq.~(\ref{NLLsol3run}) (NLL accurate with running $\aem$ for singlet
and non-singlet), eq.~(\ref{gaNLLsol4run}) (NLL accurate with running 
$\aem$ for photon), and eq.~(\ref{NLLsol3}) (NLL accurate with fixed 
$\aem$ for singlet and non-singlet; the corresponding result for the
photon is obtained from eq.~(\ref{gaNLLsol4run}), but is not reported
explicitly). The all-$z$ and large-$z$ solutions are then matched in
an additive manner, as is shown in eq.~(\ref{matchsol}).
The {\tt arXiv} submission of the present work will be accompanied by
two ancillary files, that will contain the main results for the PDFs
as {\tt Mathematica} formulae, and some analytical results too long
to fit in this paper. Furthermore, a numerical code that returns
the PDFs will be made public, to be downloaded at: 

$\phantom{aaaaaaaaaaaaaaaaa}${\tt https://github.com/gstagnit/ePDF}.

\section{Evolution equations and initial conditions\label{sec:evol}}
By working in QED the cases of the electron and of the positron PDFs
are identical. Thus, in order to be definite in this paper we shall only 
consider the PDFs of the electron, which allows us to simplify the notation 
of ref.~\cite{Frixione:2019lga} in the following way:
\beq
\ePDF{i}(z,\mu^2)\big[{\rm this~paper}\big]\;\equiv\;
\PDF{i}{\lm}(z,\mu^2)\big[{\rm ref}.~\protect
\text{\cite{Frixione:2019lga}}\big]\,.
\eeq
The evolution equations are therefore~\cite{Gribov:1972ri,Lipatov:1974qm,
Altarelli:1977zs,Dokshitzer:1977sg}:
\beq
\frac{\partial\ePDF{i}(z,\mu^2)}{\partial\log\mu^2}=
\frac{\aem(\mu)}{2\pi}\left[P_{ij}\otimes\ePDF{j}\right](z,\mu^2)\,.
\label{APeq2}
\eeq
Henceforth, we shall omit to write the explicit $z$ and/or $\mu$ dependences
when no confusion can possibly arise. By working with a single fermion family, 
eq.~(\ref{APeq2}) becomes:
\beqn
\frac{\partial\ePDF{\lm}}{\partial\log\mu^2}&=&
\frac{\aem}{2\pi}\Big(
P_{\lm\lm}\otimes\ePDF{\lm}+
P_{\lm\lp}\otimes\ePDF{\lp}+
P_{\lm\gamma}\otimes\ePDF{\gamma}\Big)\,,
\label{lmAPeq1}
\\
\frac{\partial\ePDF{\lp}}{\partial\log\mu^2}&=&
\frac{\aem}{2\pi}\Big(
P_{\lp\lm}\otimes\ePDF{\lm}+
P_{\lp\lp}\otimes\ePDF{\lp}+
P_{\lp\gamma}\otimes\ePDF{\gamma}\Big)\,,
\label{lmAPeq2}
\\
\frac{\partial\ePDF{\gamma}}{\partial\log\mu^2}&=&
\frac{\aem}{2\pi}\Big(
P_{\gamma\lm}\otimes\ePDF{\lm}+
P_{\gamma\lp}\otimes\ePDF{\lp}+
P_{\gamma\gamma}\otimes\ePDF{\gamma}\Big)\,.
\label{lmAPeq3}
\eeqn
This system of equations can be simplified by introducing the singlet
and non-singlet combinations:
\beq
\ePDFs=\ePDF{\lm}+\ePDF{\lp}\,,\;\;\;\;\;\;\;\;
\ePDFns=\ePDF{\lm}-\ePDF{\lp}\,.
\label{snsdef}
\eeq
Equations~(\ref{lmAPeq1})--(\ref{lmAPeq3}) are then re-written as follows:
\beqn
\frac{\partial}{\partial\log\mu^2}\binomial{\ePDFs}{\ePDF{\gamma}}&=&
\frac{\aem}{2\pi}\APmat_{\rm\sss S}\otimes\binomial{\ePDFs}{\ePDF{\gamma}}\,,
\label{APeqsga}
\\
\frac{\partial\ePDFns}{\partial\log\mu^2}&=&
\frac{\aem}{2\pi}P_{\rm\sss NS}\otimes\ePDFns\,,
\label{APeqns}
\eeqn
which show, as is customary, that the non-singlet component decouples
(i.e.~evolves independently) from the singlet-photon system. The evolution
kernels in eqs.~(\ref{APeqsga}) and~(\ref{APeqns}) are defined by starting
from the elementary Altarelli-Parisi kernels. One uses the following
decomposition:
\beq
P_{\lpm\lpm}=\PV_{ee}+\PS_{ee}\,,\;\;\;\;\;\;\;\;
P_{\lpm\lmp}=\PV_{e\bl}+\PS_{ee}\,.
\eeq
Furthermore, in QED:
\beq
P_{\lpm\gamma}=P_{e\gamma}\,,\;\;\;\;\;\;\;\;
\phantom{+\PS_{ee}}\;\;\;
P_{\gamma\lpm}=P_{\gamma e}\,.
\eeq
Thus, by introducing the quantities\footnote{As is indicated in
eq.~(\ref{PSS}), we work here with a single fermion family. However,
we find it useful to keep a parametrical dependence on $\NF$ in view
of future work involving more than one flavour family.}:
\beqn
P_{\Sigma\Sigma}&=&P_{\lpm\lpm}+P_{\lpm\lmp}\equiv
\PV_{ee}+\PV_{e\bl}+2\NF\PS_{ee}\,,
\;\;\;\;\;\;\;\;
\NF=1\,,
\label{PSS}
\\
P_{\Sigma\gamma}&=&2\NF P_{e\gamma}\,,
\label{PSga}
\\
P_{\gamma\Sigma}&=&P_{\gamma e}\,,
\label{Pgaga}
\eeqn
one finally defines the evolution kernels:
\beqn
\APmat_{\rm\sss S}&=&\left(
\begin{array}{cc}
P_{\Sigma\Sigma} & P_{\Sigma\gamma} \\
P_{\gamma\Sigma} & P_{\gamma\gamma} \\
\end{array}
\right)\,,
\label{kersga}
\\
P_{\rm\sss NS}&=&P_{\lpm\lpm}-P_{\lpm\lmp}\equiv
\PV_{ee}-\PV_{e\bl}\,.
\label{kerns}
\eeqn
After solving the evolution equations for the singlet and non-singlet
components, one recovers the solutions for the electron and the positron
by inverting eq.~(\ref{snsdef}):
\beq
\ePDF{\lm}=\half\left(\ePDFs+\ePDFns\right)\,,\;\;\;\;\;\;\;\;
\ePDF{\lp}=\half\left(\ePDFs-\ePDFns\right)\,.
\label{snsinv}
\eeq
The electron PDFs can be expanded perturbatively. We denote the first
two coefficients of such an expansion in the same way as in
ref.~\cite{Frixione:2019lga}, namely:
\beq
\ePDF{i}=\ePDF{i}^{[0]}+\aemotpi\,\ePDF{i}^{[1]}+\ord(\aem^2)\,.
\label{Gexp}
\eeq
The evolution equations are supplemented by the initial conditions
computed up to $\ord(\aem)$ in ref.~\cite{Frixione:2019lga}. These
read as follows:
\beqn
\ePDF{i}^{[0]}(z,\muz^2)&=&\delta_{i\lm}\delta(1-z)\,,
\label{G0sol}
\\
\ePDF{\lm}^{[1]}(z,\muz^2)&=&\left[\frac{1+z^2}{1-z}\left(
\log\frac{\muz^2}{m^2}-2\log(1-z)-1\right)\right]_+ +K_{ee}(z)\,,
\label{G1sol2}
\\
\ePDF{\gamma}^{[1]}(z,\muz^2)&=&\frac{1+(1-z)^2}{z}\left(
\log\frac{\muz^2}{m^2}-2\log z-1\right) +K_{\gamma e}(z)\,,
\label{Ggesol2}
\\
\ePDF{\lp}^{[1]}(z,\muz^2)&=&0\,,
\label{Gpossol2}
\eeqn
where $\muz\simeq m$, and $m$ is the electron mass. 
The rightmost terms on the r.h.s.~of eqs.~(\ref{G1sol2}) and~(\ref{Ggesol2}) 
are associated with, and fully determined by, the scheme used to subtract 
the initial-state collinear singularities. In this paper, we work in the 
$\MSb$ scheme, which implies:
\beq
K_{ee}(z)=K_{\gamma e}(z)=0\;\;\;\;\Longleftrightarrow\;\;\;\;\MSb\,.
\label{KkerMS}
\eeq
We conclude this section with a general remark on evolution. Throughout
this paper, by ``evolution'' we understand the one governed by RGE's.
This implies, in particular, that the contributions of $\epem$ low-energy
data to the running of $\aem$ are {\em locally} neglected (i.e.~for scales 
of the order of the masses of light hadronic resonances). However, nothing
prevents one from taking into account such contributions in an inclusive
way. Namely, by starting from a precise determination of $\aem=\aem_H$ 
that does include low-energy contributions, and that can be associated 
with a scale $\mu_H$ (just) larger than the mass of the heaviest hadronic
resonance, one can backward-evolve $\aem_H=\aem(\mu_H)$ from $\mu_H$ down to 
$\muz$, thus determining the value of $\aem(\muz)$ that is employed in this 
work. By doing so, the possible local effects of the resonances on 
the evolution of the PDFs are still neglected, but this is not important: 
in the factorisation formulae such as eq.~(\ref{master0}) where the PDFs 
are used, the scales are meant to be hard and therefore never assume values 
comparable to the masses of the light hadronic resonances.

\section{Evolution operator\label{sec:eop}}
As far as the evolution in $\mu$ is concerned, eqs.~(\ref{APeqsga}) 
and~(\ref{APeqns}) are identical. We shall thus start dealing with the 
former one, which has a more involved flavour structure; the results will
then be applied to the non-singlet case as well, by simply considering
a one-dimensional flavour space. We re-write eq.~(\ref{APeqsga}) by 
means of a simpler notation, where all of the irrelevant indices are dropped:
\beq
\frac{\partial\Gamma(z,\mu^2)}{\partial\log\mu^2}=
\frac{\aem(\mu)}{2\pi}\big[\APmat\otimes\Gamma\big](z,\mu^2)\,,
\label{APeq5}
\eeq
and $\Gamma$ is a column vector. We define the Mellin transform of
any function $f(z)$ whose domain is $[0,1]$ as follows:
\beq
M[f]\equiv f_N=\int_0^1 dz\,z^{N-1} f(z)\,.
\eeq
If $f(z)$ is the convolution of two functions $g(z)$ and $h(z)$:
\beq
f(z)=g\otimes_z h=\int_0^1 dx\,dy\,\delta(z-xy)g(x)h(y)\,,
\label{mell1}
\eeq
then:
\beq
M[g\otimes h]=M[g]\,M[h]
\;\;\;\;\Longleftrightarrow\;\;\;\;
[g\otimes h]_N=g_N\,h_N\,.
\label{Mellprod}
\eeq
The evolution kernels are expanded in a perturbative series whose
coefficients follow the same conventions as those in eq.~(\ref{Gexp}),
namely:
\beq
\APmat(x,\mu)=\sum_{k=0}^\infty\left(\frac{\aem(\mu)}{2\pi}\right)^k
\APmat^{[k]}(x)\,.
\label{APmatex}
\eeq
Equation~(\ref{APeq5}) is, in Mellin space:
\beq
\frac{\partial \Gamma_N(\mu^2)}{\partial\log\mu^2}=
\frac{\aem(\mu)}{2\pi}\,\APmat_N(\mu)\,\Gamma_N(\mu^2)=
\sum_{k=0}^\infty\left(\frac{\aem(\mu)}{2\pi}\right)^{k+1}
\APmat_N^{[k]}\,\Gamma_N(\mu^2)\,.
\label{matAPmell}
\eeq
By denoting by $\Gamma_{0,N}\equiv \Gamma_N(\muz^2)$ the PDF initial conditions
at the reference scale $\mu_0$, and by introducing the evolution
operator $\Eop_N(\mu^2,\muz^2)$ such that:
\beq
\Gamma_N(\mu^2)=\Eop_N(\mu^2,\muz^2)\,\Gamma_{0,N}\,,\;\;\;\;\;\;
\Eop_N(\muz^2,\muz^2)=I\,,
\label{FvsE}
\eeq
eq.~(\ref{matAPmell}) becomes:
\beq
\frac{\partial \Eop_N(\mu^2,\muz^2)}{\partial\log\mu^2}\,\Gamma_{0,N}=
\sum_{k=0}^\infty\left(\frac{\aem(\mu)}{2\pi}\right)^{k+1}
\APmat_N^{[k]}\,\Eop_N(\mu^2,\muz^2)\,\Gamma_{0,N}\,.
\label{matAPmell2}
\eeq
Since eq.~(\ref{matAPmell2}) must be true regardless of the specific
choice for $\Gamma_{0,N}$, it is equivalent to:
\beqn
\frac{\partial \Eop_N(\mu^2,\muz^2)}{\partial\log\mu^2}&=&
\sum_{k=0}^\infty\left(\frac{\aem(\mu)}{2\pi}\right)^{k+1}
\APmat_N^{[k]}\,\Eop_N(\mu^2,\muz^2)
\nonumber\\*
&=&
\frac{\aem(\mu)}{2\pi}\left[\APmat_N^{[0]}+
\frac{\aem(\mu)}{2\pi}\,\APmat_N^{[1]}\right]\Eop_N(\mu^2,\muz^2)
+\ord(\aem^2)\,.
\label{matAPmell3}
\eeqn
Following ref.~\cite{Furmanski:1981cw}, it is appropriate to 
introduce the variable\footnote{\label{ft:t}This differs by 
a minus sign w.r.t.~that of QCD, since it is convenient to 
still have $t>0$ for $\mu>\mu_0$.}:
\beq
t=\frac{1}{2\pi b_0}\log\frac{\aem(\mu)}{\aem(\mu_0)}\,.
\label{tdef}
\eeq
We use the following definition of the QED $\beta$ function:
\beq
\frac{\partial\aem(\mu)}{\partial\log\mu^2}=\beta(\aem)=
b_0\aem^2+b_1\aem^3+\ldots\,,
\label{betaQED}
\eeq
with
\beq
b_0=\frac{\NF}{3\pi}\,,\;\;\;\;\;\;\;\;
b_1=\frac{\NF}{4\pi^2}\,,
\label{b0b1}
\eeq
and $\NF$ the number of active charged fermion families. Equation~(\ref{tdef}) 
implies that:
\beq
\frac{\partial}{\partial\log\mu^2}=\frac{1}{2\pi b_0}\,
\frac{\beta(\aem(\mu))}{\aem(\mu)}\,\frac{\partial}{\partial t}\,,
\label{dmudt}
\eeq
and thus:
\beq
\frac{\partial\aem(\mu)}{\partial t}=2\pi b_0\aem(\mu)
\;\;\;\;\Longrightarrow\;\;\;\;
\aem(\mu)=\aem(\mu_0)e^{2\pi b_0 t}\,.
\label{aemvst}
\eeq
With eq.~(\ref{dmudt}), eq.~(\ref{matAPmell3}) becomes\footnote{As the
argument of the evolution operator, we shall use $t$ interchangeably
with the pair $(\mu,\muz)$: the physical meaning is identical,
and one is quickly reminded of the variable in which the actual
evolution is carried out.}:
\beqn
\frac{\partial \Eop_N(t)}{\partial t}&=&
\frac{b_0\aem^2(\mu)}{\beta(\aem(\mu))}
\sum_{k=0}^\infty\left(\frac{\aem(\mu)}{2\pi}\right)^k
\APmat_N^{[k]}\,\Eop_N(t)
\nonumber\\*
&=&
\left[\APmat_N^{[0]}+\frac{\aem(\mu)}{2\pi}\left(
\APmat_N^{[1]}-\frac{2\pi b_1}{b_0}\,\APmat_N^{[0]}\right)
\right]\Eop_N(t)+\ord(\aem^2)\,.
\label{matAPmell4}
\eeqn
Note that, from eq.~(\ref{FvsE}), $\Eop_N(t=0)=I$.

If the flavour space is one-dimensional (as for the non-singlet evolution), 
eq.~(\ref{matAPmell4}) can be solved analytically. Notation-wise, we deal 
with this case by performing the formal replacements:
\beq
\Eop_N\;\longrightarrow\;E_N\,,
\;\;\;\;\;\;\;\;
\APmat_N^{[k]}\;\longrightarrow\;P_N^{[k]}\,.
\label{NSrepl}
\eeq
By exploiting eq.~(\ref{aemvst}), one readily obtains:
\beq
\log E_N=P_N^{[0]}\,t+\frac{1}{4\pi^2 b_0}
\big(\aem(\mu)-\aem(\mu_0)\big)
\left(P_N^{[1]}-\frac{2\pi b_1}{b_0}\,P_N^{[0]}\right)+\ord(\aem^3)\,.
\label{Esol1}
\eeq
By construction, the $\ord(\aem^3)$ terms neglected in eq.~(\ref{Esol1})
stem from the truncation of the series that gives the evolution kernels
in eq.~(\ref{APmatex}); conversely, the relationship between $\aem(\mu)$
and $\aem(\muz)$ is treated exactly thanks to the usage of the variable
$t$. If one wants to expose explicitly the large logarithms that originate
from having $\mu\gg\muz$, one can use the following series expansions:
\beqn
\aem(\mu_0)&=&\aem(\mu)-
\aem^2(\mu)b_0L+
\aem^3(\mu)\left(b_0^2L^2-b_1L\right)
+\ord(\aem^4)\,,
\label{a0ina}
\\
t&=&\frac{\aem(\mu)}{2\pi}\,L-
\frac{\aem^2(\mu)}{4\pi}\left(b_0L^2-\frac{2b_1}{b_0}L\right)+
\ord(\aem^3)\,,
\label{tina}
\eeqn
having defined:
\beq
L=\log\frac{\mu^2}{\mu_0^2}\,.
\label{Ldef}
\eeq
By employing these results, eq.~(\ref{Esol1}) becomes:
\beq
\log E_N=\frac{\aem(\mu)}{2\pi}P_N^{[0]}L
+\left(\frac{\aem(\mu)}{2\pi}\right)^2 \left(
P_N^{[1]}L-\pi b_0 P_N^{[0]}L^2\right)
+\ord(\aem^3)\,.
\label{Esol1ex}
\eeq
This result is useful because, at variance with that of eq.~(\ref{Esol1}),
it allows one to consider the case of a non-running $\aem$, which can simply
be obtained from eq.~(\ref{Esol1ex}) in the limit $b_0\to 0$. As a 
consistency check, it is immediate to verify that, by taking such a limit,
one arrives at a form for $\log E_N$ which could have been directly obtained
from eq.~(\ref{matAPmell3}), by working in a one-dimensional flavour space
and by freezing $\aem(\mu)$ there.

\section{Analytical solutions for the PDFs\label{sec:sols}}
In this section we obtain the NLL-accurate PDFs of the electron in closed 
analytical forms in two different ways: by solving the evolution equations
order by order in perturbation theory (sect.~\ref{sec:rec}), and by using 
the properties of the evolution operator to obtain the asymptotic behaviour 
in the $z\to 1$ region to all orders in $\aem$ (sect.~\ref{sec:asy}). These 
two results can then be combined in order to obtain predictions which are 
numerically well-behaved in the whole of the $z$ range (sect.~\ref{sec:match}).

\subsection{Recursive solutions\label{sec:rec}}
Following ref.~\cite{Cacciari:1992pz}, perturbative solutions
for the evolution equations can conveniently be obtained by re-writing
eq.~(\ref{APeq5}) in an integral form:
\beq
\frac{\partial\Fpdf(z,\mu^2)}{\partial\log\mu^2}=
\frac{\aem(\mu)}{2\pi}\big[\APmat\ootimes\Fpdf\big](z,\mu^2)\,,
\label{APeq5int}
\eeq
with\footnote{The use of a $\Theta$ function in eq.~(\ref{Fdef}) 
guarantees its validity also when $\Gamma$ is a distribution, and 
thus allows one to take into account its possible endpoint contributions. 
Conversely, while $\Fpdf$ should also be treated as a distribution,
we shall regard it as an ordinary function, because in the large-$z$
region we shall in any case employ the asymptotic solutions whose
results, given in sect.~\ref{sec:asy}, are more accurate there.}:
\beq
\Fpdf(z,\mu^2)=\int_0^1 dy\,\Theta(y-z)\,\Gamma(y,\mu^2)
\;\;\;\;\Longrightarrow\;\;\;\;
\Gamma(z,\mu^2)=-\,\frac{\partial}{\partial z}\Fpdf(z,\mu^2)\,,
\label{Fdef}
\eeq
and the modified convolution operator defined as follows:
\beq
g\ootimes_z h=\int_0^1 dx\,\Theta(x-z)\,g(x)\,h(z/x)=
\bar{g}\otimes_z h\,,
\;\;\;\;\;\;\;\;
\bar{g}(x)=xg(x)\,,
\label{modconv}
\eeq
which is a valid definition regardless of whether $g(x)$ is a 
distribution or an ordinary function. 
Note that $\Fpdf$ is a column vector, and that eq.~(\ref{APeq5int})
has a matrix structure, in the flavour space. As was the case for
eq.~(\ref{APeq5}), this implies that all of the results to be obtained
in the following can be applied to the limiting situation of a
one-dimensional flavour space as well.

The procedure of ref.~\cite{Cacciari:1992pz} is LL-accurate.
In order to generalise it to the NLL accuracy we are interested in
in this work, it is best to first consider the case of non-running $\aem$.
With this assumption, the solution of eq.~(\ref{APeq5int})
can formally be written as follows:
\beq
\Fpdf(z,\mu^2)=\Fpdf(z,\mu_0^2)+
\frac{\aem}{2\pi}\int_{\log\mu_0^2}^{\log\mu^2}d\log\mu^{\prime^2}\,
\left[\APmat\ootimes\Fpdf\right](z,\mu^{\prime^2})\,.
\label{Fsol}
\eeq
From this equation, $\Fpdf$ can be obtained by representing it
by means of a power series:
\beq
\Fpdf(z,\mu^2)=\sum_{k=0}^\infty \frac{\eta_0^k}{2^kk!}
\left(\ILL_k(z)+\frac{\aem}{2\pi}\,\INLL_k(z)\right),
\label{Fexp}
\eeq
where:
\beq
\eta_0=\frac{\aem}{\pi}L\,,
\;\;\;\;\;\;\;\;\;\;
L=\log\frac{\mu^2}{\mu_0^2}\,,
\label{eta0def}
\eeq
and with $\ILL_k$ and $\INLL_k$ two sets of unknown functions\footnote{More
precisely, eq.~(\ref{Fexp}) implies that for any $k$, $\ILL_k$ and $\INLL_k$ 
are two-dimensional column vectors in the singlet-photon flavour space, whose 
elements are functions of $z$, and c-number functions in the non-singlet
flavour space. An extra flavour index will be included in the notation
when distinguishing the flavour components will be important (see
appendix~\ref{sec:recres}).}.
By replacing eq.~(\ref{Fexp}) into eq.~(\ref{Fsol}), the two sides
of the latter equation become two series in $\eta_0$: one then equates
the coefficients relevant to the same power of $\eta_0$ on the two
sides, thereby obtaining equations that can be solved for $\ILL_k$ 
and $\INLL_k$ (recursively in $k$). The r.h.s.~of eq.~(\ref{Fexp}) is simply 
an expansion in terms of \mbox{$\aem L$}, and thus $\eta_0$ is a convenient 
expansion parameter, irrespective of the logarithmic accuracy one is working 
at. Indeed, eq.~(\ref{Fexp}) can be extended by adding further contributions
to its r.h.s., that are suppressed by higher powers of $\aem$. Conversely,
by keeping only the $\ILL_k$ contributions, one recovers what was done
in ref.~\cite{Cacciari:1992pz}.
The recursive solutions for $\ILL_k$ and $\INLL_k$ stemming from
eq.~(\ref{Fexp}) read as follows:
\beqn
\ILL_k&=&\APmat^{[0]}\ootimes\ILL_{k-1}\,,
\label{ILLsol}
\\
\INLL_k&=&\APmat^{[0]}\ootimes\INLL_{k-1}+\APmat^{[1]}\ootimes\ILL_{k-1}\,,
\label{INLLsol}
\eeqn
with:
\beq
\ILL_0=\Fpdf^{[0]}(z,\mu_0^2)\,,\;\;\;\;\;\;\;\;
\INLL_0=\Fpdf^{[1]}(z,\mu_0^2)\,.
\label{Iini}
\eeq
The quantities in eq.~(\ref{Iini}) must be obtained by direct computation 
by using the definition in eq.~(\ref{Fdef}), with the perturbative expansion 
of eq.~(\ref{Gexp}) and the initial conditions of 
eqs.~(\ref{G0sol})--(\ref{Gpossol2}). By doing so, we obtain:
\beqn
\ILL_{{\rm\sss S},\,0}=\ILL_{{\rm\sss NS},\,0}&=&1\,,
\label{ILLSNSini}
\\
\ILL_{\gamma,\,0}&=&0\,,
\label{ILLgamini}
\\
\INLL_{{\rm\sss S},\,0}=\INLL_{{\rm\sss NS},\,0}&=&
2z+(1-2z-z^2)\log(1-z)-2\log^2(1-z)
\nonumber\\*&+&
\Big[z+z^2/2+2\log(1-z)\Big]\log\frac{\muz^2}{m^2}\,,
\label{INLLSNSini}
\\
\INLL_{\gamma,\,0}&=&
-2(1-z)+(2-4z+z^2)\log z+2\log^2 z
\nonumber\\*&-&
\Big[\half\,(3-4z+z^2)+2\log z\Big]\log\frac{\muz^2}{m^2}\,.
\label{INLLgamini}
\eeqn

The key to the simplicity of the solutions in eqs.~(\ref{ILLsol}) 
and~(\ref{INLLsol}) is the fact that the
dependence on $\mu$ on the r.h.s.~of eq.~(\ref{Fexp}) is entirely
parametrised by $L$, which in turn allows one to compute the integral
on the r.h.s.~of eq.~(\ref{Fsol}) in a trivial manner:
\beq
\int_{\log\mu_0^2}^{\log\mu^2}d\log\mu^{\prime^2}
\left.\eta_0^k\right|_{\mu\to\mu^\prime}=
\left(\frac{\aem}{\pi}\right)^{-1}\frac{\eta_0^{k+1}}{k+1}\,.
\label{eta0int}
\eeq
Unfortunately, things are not so simple when $\aem$ is running. In this
case, as was already done in sect.~\ref{sec:eop}, it is convenient
to use the variable $t$ introduced in eq.~(\ref{tdef}). 
Owing to eq.~(\ref{dmudt}), the analogue of eq.~(\ref{Fsol}) reads
as follows:
\beq
\Fpdf(z,t)=\Fpdf(z,0)+
\int_0^t du\,\frac{b_0\aem^2(u)}{\beta(\aem(u))}
\left[\APmat\ootimes\Fpdf\right](z,u)\,.
\label{Fsolt}
\eeq
As a consequence of this, we shall use the representation:
\beq
\Fpdf(z,t)=\sum_{k=0}^\infty \frac{t^k}{k!}
\left(\JLL_k(z)+\frac{\aem(t)}{2\pi}\,\JNLL_k(z)\right),
\label{Fexpt}
\eeq
rather than that of eq.~(\ref{Fexp}). Thus:
\beqn
\frac{b_0\aem^2(t)}{\beta(\aem(t))}\,\APmat\ootimes\Fpdf&=&
\sum_{k=0}^\infty \frac{t^k}{k!}\,\Bigg\{
\APmat^{[0]}\ootimes\JLL_k
+\frac{\aem(t)}{2\pi}\Bigg[\APmat^{[0]}\ootimes\JNLL_k+
\APmat^{[1]}\ootimes\JLL_k
\nonumber\\*&&\phantom{aaaaP^{[0]}\ootimes\JLL_k}
-\frac{2\pi b_1}{b_0}\,\APmat^{[0]}\ootimes\JLL_k\Bigg]
+\ord(\aem^2)\Bigg\}\,.
\eeqn
The r.h.s.~of eq.~(\ref{Fsolt}) therefore features two independent
classes of integrals, namely:
\beqn
a_k&=&\int_0^t du\,u^k=\frac{t^{k+1}}{k+1}\,,
\label{akint}
\\
b_k&=&\int_0^t du\,u^k\aem(u)\,.
\label{bkint}
\eeqn
In order to evaluate eq.~(\ref{bkint}), we make repeated use of 
eq.~(\ref{aemvst}). Then:
\beq
b_k=\aem(0)\int_0^t du\,u^k\,e^{2\pi b_0 u}=
\aem(t)e^{-2\pi b_0 t}\sum_{j=0}^\infty\frac{(2\pi b_0)^j}{j!}
\int_0^t du\,u^k\,u^j\,.
\eeq
By direct computation:
\beq
e^{-2\pi b_0 t}\sum_{j=0}^\infty\frac{(2\pi b_0)^j}{(k+1+j)j!}\,t^{k+1+j}=
\frac{t^{k+1}}{k+1}\sum_{p=0}^\infty d_{k,p}\,t^p\,,
\eeq
with:
\beq
d_{k,p}=(-)^p(2\pi b_0)^p\,\frac{\Gamma(k+2)}{\Gamma(k+2+p)}\,.
\label{dkp}
\eeq
We have thus:
\beq
\Fpdf(z,t)-\Fpdf(z,0)=
\sum_{k=0}^\infty\frac{1}{k!}\left(g_k\,\frac{t^{k+1}}{k+1}+
\frac{\aem(t)}{2\pi}\,h_k\,\sum_{p=0}^\infty\frac{t^{k+1+p}}{k+1}\,d_{k,p}
\right),
\label{Frhs}
\eeq
where:
\beqn
g_k&=&\APmat^{[0]}\ootimes\JLL_k\,,
\\
h_k&=&\APmat^{[0]}\ootimes\JNLL_k+\APmat^{[1]}\ootimes\JLL_k
-\frac{2\pi b_1}{b_0}\,\APmat^{[0]}\ootimes\JLL_k\,.
\eeqn
The r.h.s.~of eq.~(\ref{Frhs}) can be simplified by means of algebraic
manipulations of the summation indices:
\beq
\sum_{k=0}^\infty\frac{1}{k!}\,g_k\,\frac{t^{k+1}}{k+1}=
\sum_{k=1}^\infty\frac{t^k}{k!}\,g_{k-1}\,,
\label{aksimp}
\eeq
and:
\beq
\sum_{k=0}^\infty\frac{1}{k!}\,h_k\,\sum_{p=0}^\infty
\frac{t^{k+1+p}}{k+1}\,d_{k,p}=
\sum_{k=1}^\infty\,\frac{t^k}{k!}\,
\sum_{p=0}^{k-1}(-)^p(2\pi b_0)^p\,h_{k-1-p}\,,
\label{bksimp}
\eeq
since from eq.~(\ref{dkp}):
\beq
\frac{d_{k-1-p,p}}{(k-p)!}=\frac{(-)^p(2\pi b_0)^p}{k!}\,.
\eeq
The initial conditions must then be written as follows:
\beqn
\Fpdf(z,0)&=&\Fpdf^{[0]}(z,\mu_0^2)+
\frac{\aem(t)e^{-2\pi b_0 t}}{2\pi}\,\Fpdf^{[1]}(z,\mu_0^2)
\nonumber\\*&=&
\Fpdf^{[0]}(z,\mu_0^2)+\frac{\aem(t)}{2\pi}\,\Fpdf^{[1]}(z,\mu_0^2)
\sum_{k=0}^\infty \frac{(-)^k(2\pi b_0)^k}{k!}\,t^k\,.
\label{Fpdf0ex}
\eeqn
By replacing the results of eqs.~(\ref{aksimp}), (\ref{bksimp}),
and~(\ref{Fpdf0ex}) into eq.~(\ref{Frhs}), and by using the representation
of eq.~(\ref{Fexpt}) for $\Fpdf(z,t)$, we obtain the sought recursion
relations:
\beqn
\JLL_k&=&\APmat^{[0]}\ootimes\JLL_{k-1}\,,
\label{JLLsol}
\\
\JNLL_k&=&(-)^k(2\pi b_0)^k\Fpdf^{[1]}(\mu_0^2)
\label{JNLLsol}
\\*&&
+\sum_{p=0}^{k-1}(-)^p(2\pi b_0)^p
\Bigg(\APmat^{[0]}\ootimes\JNLL_{k-1-p}+\APmat^{[1]}\ootimes\JLL_{k-1-p}
\nonumber\\*&&\phantom{\sum_{p=0}^{k-1}(-)^p(2\pi b_0)^p\Bigg(}
-\frac{2\pi b_1}{b_0}\,\APmat^{[0]}\ootimes\JLL_{k-1-p}\Bigg)\,,
\nonumber
\eeqn
with:
\beq
\JLL_0=\Fpdf^{[0]}(z,\mu_0^2)\,,\;\;\;\;\;\;\;\;
\JNLL_0=\Fpdf^{[1]}(z,\mu_0^2)\,.
\label{Jini}
\eeq
These results generalise those obtained in the case of non-running
$\aem$, which can be obtained from them. Indeed, in the limit of
fixed $\aem$, which at the NLL can be achieved by letting $b_0\to 0$
and $b_1\to 0$ (with $b_1/b_0\to 0$), we have $t\to\eta_0/2$, whence 
eq.~(\ref{Fexpt}) coincides with eq.~(\ref{Fexp}), if one identifies $\JLL$ 
with $\ILL$ and $\JNLL$ with $\INLL$. This is justified, since 
eqs.~(\ref{ILLsol}) and~(\ref{JLLsol}) are identical, and the recursive 
relation of eq.~(\ref{JNLLsol}) coincides with that of eq.~(\ref{INLLsol}) 
when $\aem$ is not running.

After solving eqs.~(\ref{ILLsol}) and~(\ref{INLLsol}) for $\ILL_k$
and $\INLL_k$, with the definition in eq.~(\ref{Fdef}) one arrives 
at the following representation of the PDF in the case of fixed $\aem$:
\beq
\Gamma(z,\mu^2)=\sum_{k=0}^\infty \frac{\eta_0^k}{2^kk!}
\left(\iLL_k(z)+\frac{\aem}{2\pi}\,\iNLL_k(z)\right),
\label{PDFexp}
\eeq
where
\beq
\iLL_k(z)=-\,\frac{d}{dz}\,\ILL_k(z)\,,
\;\;\;\;\;\;\;\;
\iNLL_k(z)=-\,\frac{d}{dz}\,\INLL_k(z)\,.
\eeq
Analogously, in the case of running $\aem$:
\beq
\Gamma(z,\mu^2)=\sum_{k=0}^\infty \frac{t^k}{k!}
\left(\jLL_k(z)+\frac{\aem(t)}{2\pi}\,\jNLL_k(z)\right),
\label{PDFexpt}
\eeq
with
\beq
\jLL_k(z)=-\,\frac{d}{dz}\,\JLL_k(z)\,,
\;\;\;\;\;\;\;\;
\jNLL_k(z)=-\,\frac{d}{dz}\,\JNLL_k(z)\,.
\label{jfromJ}
\eeq
We point out that with, for example, \mbox{$\aem(\mu)=1/128$} and 
\mbox{$\aem(\mu_0)=1/137$} we have \mbox{$t\simeq 0.1/\NF$}. 
Furthermore, since:
\beq
2\pi b_0=\frac{2}{3}\,\NF\,,\;\;\;\;\;\;\;\;
\frac{2\pi b_1}{b_0}=\frac{3}{2}\,,
\eeq
the numerical coefficients in front of the convolution products
and of the initial conditions in eq.~(\ref{JNLLsol}) are of order one.
Therefore, the series of eq.~(\ref{Fexpt}) is expected to be poorly
convergent only for $z\to 1$ and $z\to 0$, owing to the possible presence 
of $\log^p(1-z)$ and $\log^pz$ terms in the $\JLL$ and $\JNLL$ functions.

Part of the explicit results for the functions $\jLL_k$ 
(with \mbox{$0\le k\le 3$}) and $\jNLL_k$ (with \mbox{$0\le k\le 2$}) 
are reported in appendix~\ref{sec:recres}, and part in one of the
ancillary files.

\subsection{Asymptotic large-$z$ solutions\label{sec:asy}}
The electron PDF is equal to $\delta(1-z)$ at the LO 
(see eq.~(\ref{G0sol})); while the LL evolution of such an
initial condition does smooth its behaviour, resulting in a tail
that extends down to $z=0$~\cite{Skrzypek:1990qs,Skrzypek:1992vk,
Cacciari:1992pz}, the PDF remains very peaked towards $z=1$, 
where it has an integrable singularity. This implies that the
perturbative expansion of the LL-accurate solution features
\mbox{$\log(1-z)$} terms at each order: if one truncates such
a perturbative series, one exposes a non-integrable divergence at $z=1$, 
regardless of the order at which the truncation occurs. The same is true 
when NLO initial conditions and NLL-accurate evolution are considered,
as is explicitly shown by the results in appendix~\ref{sec:lzexp}.

In order to address this issue, the $\log(1-z)$ terms must be resummed. 
This can conveniently be done by exploiting the evolution-operator formalism 
presented in sect.~\ref{sec:eop}, whose usage is simplified by the
observation that the large-$z$ region corresponds to the large-$N$
region in Mellin space:
\beq
z\;\rightarrow\;1
\;\;\;\;\;\;\longleftrightarrow\;\;\;\;\;\;
N\;\rightarrow\;\infty\,.
\label{largezN}
\eeq
Thus in this section, when dealing with Mellin transforms and their
inverse, we shall often implicitly assume eq.~(\ref{largezN}).

A second crucial observation is that the $z\to 1$ asymptotic behaviours
of the singlet and non-singlet components are actually identical.
This implies that also in the former case one can effectively work 
as if the evolution operator were a c-number and not a matrix,
thereby allowing one to exploit the closed-form solutions of
eqs.~(\ref{Esol1}) and~(\ref{Esol1ex}). Therefore, we shall start
by understanding the non-singlet notation in sects.~\ref{sec:asyLL}
and~\ref{sec:asyNLL} in order to derive the main results relevant to
the asymptotic $z\to 1$ region; we shall return to and comment on the 
singlet-photon case in sect.~\ref{sec:asyNLLs} and in 
appendix~\ref{sec:asyph}.

\subsubsection{LL solution for non-singlet\label{sec:asyLL}}
Given that the LL-accurate result has been available for a long
while~\cite{Gribov:1972ri}, this case is presented here only to show how 
the evolution-operator formalism helps find the asymptotic solution in a 
straightforward manner. At the LL we are entitled to neglect the 
running\footnote{Whenever the coupling constant is not running, we simply 
denote its fixed value by $\aem$, i.e.~we remove its argument $\mu$ from 
the notation.} of $\aem$. Thus, the appropriate form for the evolution 
operator is obtained by keeping only the $\ord(\aem)$ term of 
eq.~(\ref{Esol1ex}), with $\aem(\mu)\to\aem$ there, supplemented by 
the LO initial condition:
\beq
\Gamma_{0,N}^{[0]}=1\,.
\label{G0N0}
\eeq
From eqs.~(\ref{FvsE}) and~(\ref{G0N0}) we obtain:
\beq
\Gamma(z,\mu^2)=M^{-1}\big[\exp\big(\log E_N\big)\big]\,.
\label{LLsol}
\eeq
A direct calculation in the large-$N$ region leads to:
\beq
P_N^{[0]}\;\stackrel{N\to\infty}{\longrightarrow}\;-2\log\bN+2\lambda_0\,,
\label{P0asy2}
\eeq
where all terms suppressed by at least one inverse power of $N$ have been
neglected, and we have defined:
\beq
\bN=N\,e^{\gE}\,,\;\;\;\;\;\;\;\;
\lambda_0=\frac{3}{4}\,.
\label{bNdef}
\eeq
We point out that $\bN$ is a quantity that routinely appears in the 
computation of Mellin transforms, and which helps retain some universal 
subleading terms. Therefore:
\beq
\log E_N=\frac{\aem}{2\pi}\,P_N^{[0]}\,L
\;\stackrel{N\to\infty}{\longrightarrow}\;
-\eta_0\left(\log\bN-\lambda_0\right)\,,
\label{logENLL}
\eeq
with $\eta_0$ defined in eq.~(\ref{eta0def}). Equation~(\ref{logENLL}),
when substituted into eq.~(\ref{LLsol}), implies:
\beq
M\left[\Gamma(z,\mu^2)\right]=
N^{-\eta_0}e^{-\gE\eta_0}e^{\lambda_0\eta_0}\,.
\label{LLsol2}
\eeq
The inverse Mellin transform can now be evaluted by using the following
result, valid for any $\kappa>0$:
\beq
M\left[(1-z)^{-1+\kappa}\right]=
\frac{\Gamma(\kappa)\Gamma(N)}{\Gamma(\kappa+N)}
\;\stackrel{N\to\infty}{\longrightarrow}\;
\Gamma(\kappa)N^{-\kappa}\,.
\label{omzM}
\eeq
The comparison of eq.~(\ref{omzM}) with eq.~(\ref{LLsol2}) allows
one to arrive at the final result~\cite{Gribov:1972ri}:
\beq
\Gamma(z,\mu^2)=
\frac{e^{-\gE\eta_0}e^{\lambda_0\eta_0}}{\Gamma(1+\eta_0)}\,
\eta_0(1-z)^{-1+\eta_0}\,.
\label{LLsol3}
\eeq
This is identical to what is nowadays a rather standard form, except for
an exponentiated term of pure-soft origin (stemming from the use of 
$\beta_{exp}=\beta$, rather than of $\beta_{exp}=\eta$, as defined 
e.g.~in eq.~(67) of ref.~\cite{Beenakker:1996kt}). Such a term clearly 
cannot be obtained by means of the collinear resummation carried out here.

\subsubsection{$\MSb$ NLL solution for non-singlet\label{sec:asyNLL}}
At the NLL, the PDF initial conditions must be set as given in 
eqs.~(\ref{G0sol}) and~(\ref{G1sol2}), with $K_{ee}=0$ in the latter 
equation (see eq.~(\ref{KkerMS})). By exploiting the property of the 
Mellin transform of eq.~(\ref{Mellprod}), we have:
\beqn
\Gamma(z,\mu^2)&=&
\left(\delta(1-x)+\frac{\aem(\mu_0^2)}{2\pi}
\left[\frac{1+x^2}{1-x}\left(
\log\frac{\mu_0^2}{m^2}-2\log(1-x)-1\right)\right]_+\right)
\nonumber\\*&&\otimes_z\;
M^{-1}\big[\exp\big(\log E_N\big)\big]\,,
\label{NLLsol}
\eeqn
with $\log E_N$ given in eq.~(\ref{Esol1}) (where running-$\aem$ effects
are also included). With eq.~(\ref{P0asy2}) and its NLO analogue:
\beq
P_N^{[1]}\;\stackrel{N\to\infty}{\longrightarrow}\;
\frac{20}{9}\NF\log\bN+\lambda_1\,,
\label{P1asy2}
\eeq
where:
\beq
\lambda_1=\frac{3}{8}-\frac{\pi^2}{2}+6\zeta_3-\frac{\NF}{18}(3+4\pi^2)\,,
\label{lambda1}
\eeq
we can cast the logarithm of the evolution operator in the same form
as in eq.~(\ref{logENLL}), namely:
\beq
\log E_N\;\stackrel{N\to\infty}{\longrightarrow}\;
-\xi_1\log\bN+\hat{\xi}_1\,,
\label{ENxi1}
\eeq
having defined:
\beqn
\xi_1&=&
2t-\frac{\aem(\mu)}{4\pi^2 b_0}\Big(1-e^{-2\pi b_0 t}\Big)
\left(\frac{20}{9}\NF+\frac{4\pi b_1}{b_0}\right)
\label{xiR1def}
\\*&=&
2\left[1-\frac{\aem(\mu)}{\pi}\left(\frac{5}{9}\NF+\frac{\pi b_1}{b_0}
\right)\right]t
\nonumber\\*&&\phantom{aaa}
+\frac{\aem(\mu)}{\pi}\left(\frac{10}{9}\pi b_0\NF+2b_1\pi^2\right)t^2
+\ord(t^3)\,,
\\
\hat{\xi}_1&=&\frac{3}{2}\,t+\frac{\aem(\mu)}{4\pi^2 b_0}
\Big(1-e^{-2\pi b_0 t}\Big)
\left(\lambda_1-\frac{3\pi b_1}{b_0}\right)
\label{chi1Rdef}
\\&=&
\frac{3}{2}\left[1+\frac{\aem(\mu)}{\pi}\left(\frac{\lambda_1}{3}\,
-\frac{\pi b_1}{b_0}\right)\right]t
\nonumber\\*&&\phantom{aaa}
-\frac{\aem(\mu)}{\pi}\left(\frac{\pi b_0}{2}\,\lambda_1-
\frac{3}{2}\,\pi^2 b_1\right)t^2+\ord(t^3)\,.
\eeqn
Equation~(\ref{ENxi1}) implies that we can follow the same steps
that have led us to eq.~(\ref{LLsol3}), and therefore:
\beq
M^{-1}\big[\exp\big(\log E_N\big)\big]=
\frac{e^{-\gE\xi_1}e^{\hat{\xi}_1}}{\Gamma(1+\xi_1)}\,
\xi_1(1-y)^{-1+\xi_1}\,.
\label{NLLsol2}
\eeq
We must now replace this result into eq.~(\ref{NLLsol}). In this
way, two independent convolution integrals emerge:
\beqn
I_+(z)&=&\half\left[\frac{1+x^2}{1-x}\right]_+\otimes_z
(1-y)^{-1+\kappa}\,,
\label{Ipdef}
\\
I_{\rm L}(z)&=&\half\left[\frac{1+x^2}{1-x}\log(1-x)\right]_+\otimes_z
(1-y)^{-1+\kappa}\,.
\label{ILdef}
\eeqn
A tedious but otherwise relatively straightforward procedure
leads to the following results:
\beqn
I_+(z)&=&(1-z)^{-1+\kappa}\Big[A(\kappa)+\log(1-z)+\frac{3}{4}\Big]\,,
\label{Ipi3}
\\
I_{\rm L}(z)&=&(1-z)^{-1+\kappa}\Big[B(\kappa)+A(\kappa)\log(1-z)
+\half\log^2(1-z)-\frac{7}{8}\Big]\,,
\label{ILi3}
\eeqn
where, inside the square brackets, we have neglected terms that vanish
at $z\to 1$. We have introduced the two functions:
\beqn
A(\kappa)&=&\sum_{k=1}^\infty\,\frac{1}{k\,k!}\,
\frac{\Gamma(1-\kappa+k)}{\Gamma(1-\kappa)}=
-\gE-\psi_0(\kappa)\,,
\label{Ares}
\\
B(\kappa)&=&-\sum_{k=1}^\infty\,\frac{1}{k^2\,k!}
\frac{\Gamma(1-\kappa+k)}{\Gamma(1-\kappa)}
\nonumber
\\*&=&
\half\,\gE^2+\frac{\pi^2}{12}+\gE\,\psi_0(\kappa)+\half\,\psi_0(\kappa)^2
-\half\,\psi_1(\kappa)\,,
\label{Bres}
\eeqn
where:
\beq
\psi_j(z)=\frac{d^{j+1}\log\Gamma(z)}{dz^{j+1}}\,.
\eeq
By putting everything back together, we arrive at the final result:
\beqn
&&\!\!\!\!\!\Gamma(z,\mu^2)=
\frac{e^{-\gE\xi_1}e^{\hat{\xi}_1}}{\Gamma(1+\xi_1)}\,
\xi_1(1-z)^{-1+\xi_1}
\label{NLLsol3run}
\\*&&\phantom{aaa}\times
\Bigg\{1+\frac{\aem(\mu_0)}{\pi}\Bigg[\left(\log\frac{\mu_0^2}{m^2}-1\right)
\left(A(\xi_1)+\frac{3}{4}\right)-2B(\xi_1)+\frac{7}{4}
\nonumber\\*&&\phantom{\times aaa1+\frac{\aem}{\pi}\Bigg[}\;
+\left(\log\frac{\mu_0^2}{m^2}-1-2A(\xi_1)\right)\log(1-z)
-\log^2(1-z)\Bigg]\Bigg\}\,,
\nonumber
\eeqn
which is therefore the NLL-accurate counterpart of eq.~(\ref{LLsol3}).

A couple of observations about eq.~(\ref{NLLsol3run}) are in order.
Firstly, owing to eqs.~(\ref{xiR1def}) and~(\ref{tina}), we have
\mbox{$\xi_1\simeq\eta_0$}. With $\muz$ and $\mu$ of the order of
the electron mass and of a few hundred GeV's, respectively, one
obtains $\eta_0\sim 0.05$. Therefore, both the LL and the NLL solutions
are still very peaked towards $z=1$, where they diverge with an integrable
singularity. Furthermore, the $z\to 1$ behavior of eq.~(\ref{NLLsol3run})
is worse than that of eq.~(\ref{LLsol3}) because of the presence of the
explicit \mbox{$\log^p(1-z)$} terms in the former equation\footnote{One 
can show that such terms are in part an artifact of the $\MSb$ scheme;
we shall return to this point in a forthcoming paper~\cite{BCCFS2}.}. 
Secondly, the small numerical value of $\xi_1$ just mentioned implies 
that the following expansions:
\beqn
A(\kappa)&=&\frac{1}{\kappa}+\ord(\kappa)\,,
\label{Aexp}
\\
B(\kappa)&=&-\frac{\pi^2}{6}+2\zeta_3\kappa+\ord(\kappa^2)\,,
\label{Bexp}
\eeqn
are rather accurate approximations of the complete results of
eqs.~(\ref{Ares}) and~(\ref{Bres}). Equation~(\ref{Aexp}), in particular,
implies that numerically the $\log(1-z)$ term is much larger 
than the (formally dominant) $\log^2(1-z)$ one, even for $z$ values
that are {\em extremely} close to one. This fact might be significant
when performing the integral of the convolution between electron
PDFs and short-distance cross sections.

From eq.~(\ref{NLLsol3run}) one can also readily obtain
a LL-accurate solution, where at variance with that of
eq.~(\ref{LLsol3}) the effects due to the running of $\aem$
are included. Explicitly:
\beq
\Gamma(z,\mu^2)=
\frac{e^{-\gE\xi_0}e^{\hat{\xi}_0}}{\Gamma(1+\xi_0)}\,
\xi_0(1-z)^{-1+\xi_0}\,,
\label{LLsol3run2}
\eeq
where
\beq
\xi_0=2t\,,\;\;\;\;\;\;\;\;
\hat{\xi}_0=\frac{3}{2}\,t\,;
\label{xiR0chi0Rdef}
\eeq
this is again consistent with the findings of ref.~\cite{Gribov:1972ri}.
Finally, the running of $\aem$ can formally be switched off in the 
NLL-accurate solution. In order to do so, one must repeat the procedure
that leads to eq.~(\ref{NLLsol3run}); however, rather than using the
expression of the evolution operator given in eq.~(\ref{Esol1}), one
must use that of eq.~(\ref{Esol1ex}). By doing so, one arrives at:
\beqn
&&\!\!\!\!\!\Gamma(z,\mu^2)=
\frac{e^{-\gE\eta_1}e^{\hat{\eta}_1}}{\Gamma(1+\eta_1)}\,
\eta_1(1-z)^{-1+\eta_1}
\label{NLLsol3}
\\*&&\phantom{aaa}\times
\Bigg\{1+\frac{\aem}{\pi}\Bigg[\left(\log\frac{\mu_0^2}{m^2}-1\right)
\left(A(\eta_1)+\frac{3}{4}\right)-2B(\eta_1)+\frac{7}{4}
\nonumber\\*&&\phantom{\times aaa1+\frac{\aem}{\pi}\Bigg[}\;
+\left(\log\frac{\mu_0^2}{m^2}-1-2A(\eta_1)\right)\log(1-z)
-\log^2(1-z)\Bigg]\Bigg\}.\phantom{aa}
\nonumber
\eeqn
where
\beqn
\eta_1&=&\eta_0\left(1-\frac{5\aem}{9\pi}\,\NF\right)\,,
\label{eta1def}
\\
\hat{\eta}_1&=&\eta_0\left(\lambda_0+\frac{\aem}{4\pi}\,\lambda_1\right)\,.
\label{heta1def}
\eeqn

\subsubsection{The singlet and photon cases\label{sec:asyNLLs}}
The key result relevant to the evolution of the singlet and photon
sector in the $z\to 1$ region is the following:
\beq
\APmat_{{\rm\sss S},N}\;\stackrel{N\to\infty}{\longrightarrow}\;
\left(
\begin{array}{cc}
-2\log\bN+2\lambda_0 & 0 \\
0 & -\frac{2}{3}\,\NF \\
\end{array}
\right)
+\frac{\aem}{2\pi}
\left(
\begin{array}{cc}
\frac{20}{9}\NF\log\bN+\lambda_1 & 0 \\
0 & -\NF \\
\end{array}
\right)+\ord(\aem^2)\,,
\label{APmatasy2}
\eeq
that is obtained by means of a direct computation starting from the
definitions given in sect.~\ref{sec:evol}. Equation~(\ref{APmatasy2})
implies that the singlet and the photon evolve independently from
each other in this limit. Since the kernel evolution is a diagonal matrix, 
so is the evolution operator, and therefore the solutions for its elements on 
the diagonal are given by either eq.~(\ref{Esol1}) or eq.~(\ref{Esol1ex}). 

Let us start by considering the singlet. The singlet-singlet elements of 
the $\ord(\aem^0)$ and $\ord(\aem)$ matrices in eq.~(\ref{APmatasy2})
are identical to eqs.~(\ref{P0asy2}) and~(\ref{P1asy2}) respectively.
Thus, the solutions of eqs.~(\ref{NLLsol3run}), (\ref{LLsol3run2}),
and~(\ref{NLLsol3}) are also valid for the singlet.

As far as the photon is concerned, eq.~(\ref{b0b1}) and the photon-photon 
elements in eq.~(\ref{APmatasy2}) imply that the second term on the 
r.h.s.~of eq.~(\ref{Esol1}) is equal to zero. Therefore:
\beq
M^{-1}\big[E_{\gamma\gamma,N}\big]=\frac{\aem(\muz)}{\aem(\mu)}\,
\delta(1-z)\,,
\label{Eggsol}
\eeq
having used eq.~(\ref{tdef}).
The convolution with the initial conditions of eqs.~(\ref{G0sol})
and~(\ref{Ggesol2}) is thus trivial, and the final result reads as follows:
\beq
\ePDF{\gamma}(z,\mu^2)=\frac{1}{2\pi}\,
\frac{\aem(\muz)^2}{\aem(\mu)}\,\frac{1+(1-z)^2}{z}\left(
\log\frac{\muz^2}{m^2}-2\log z-1\right)\,.
\label{gaNLLsol3run}
\eeq
We can observe the presence of an $\aem(\mu)$ term in the denominator of 
eq.~(\ref{gaNLLsol3run}) which, typically, will cancel an analogous 
factor in the short-distance cross sections (given that these, for
consistency with the present results, will have to be computed in
the $\MSb$ scheme). Thus, one sees the natural emergence of quantities 
that are employed in the so-called $\aem(0)$ scheme (see 
e.g.~ref.~\cite{Denner:1991kt}) -- this is the same mechanism 
that has been anticipated in ref.~\cite{Frederix:2016ost} in 
the case of photon fragmentation functions. We stress that this properties
stems from the computation of both the PDFs and the cross sections
in the same collinear subtraction scheme, {\em and} from the solution
of the evolution equations for the PDFs.

Unfortunately, eq.~(\ref{gaNLLsol3run}) does not give a good description
of the true large-$z$ behaviour of the photon PDF. This is because the
off-diagonal terms of the evolution kernel imply that such a PDF
receives a contribution that primarily stems from the initial conditions
of the electron PDF. As we have seen previously, these are much more
peaked towards $z=1$ than their photon counterparts, so much so that 
this behaviour compensates the fact that the off-diagonal elements
of the evolution kernel are suppressed w.r.t.~the diagonal ones, which
are the only ones that have been taken into account in eq.~(\ref{APmatasy2}).
It then follows that, in order to improve on the solution in
eq.~(\ref{gaNLLsol3run}), one needs to solve the evolution equations
of the singlet-photon system by including those off-diagonal elements.
In turn, this entails a significant increase in complexity, and for
this reason we refrain from discussing the relevant procedure here --
all of the results are reported in appendix~\ref{sec:asyph}.

\section{Numerical solutions for the PDFs\label{sec:num}}
The numerical evolution for the PDFs is achieved by first solving the
evolution equation for the evolution operator in Mellin space. More
specifically, we solve the equation given in the first line of
eq.~(\ref{matAPmell4}) without expanding the $\beta$ function
in the denominator. As is discussed in sect.~\ref{sec:evol}, the
introduction of the singlet and non-singlet combinations,
eq.~(\ref{snsdef}), allows for a decoupling of the evolution equations
that is well-suited for a numerical implementation. As has been done thus
far, in the following we shall implicitly refer to the two-dimensional 
singlet-photon case, keeping in mind that the non-singlet case is obtained 
by considering a one-dimensional flavour space.

The numerical solution of eq.~(\ref{matAPmell4}) for the evolution
operator $\Eop_N$ is obtained by means of a discretised path-ordered
product~\cite{Bonvini:2012sh}. The evolution range $[0,t]$ is
partitioned into $n$ intervals $[t_i,t_{i+1}]$, with $t_0=0$ and $t_n=t$,
and the evolution operator is written as follows\footnote{The product on 
the r.h.s.~of eq.~(\ref{eq:pathorderEop}) must be understood as a product 
among matrices. Thus, by reading the product from left to right one finds
decreasing values of the index $i$.}:
\beq
\Eop_N(t) = \prod_{i=n-1}^{0} \Eop_N(t_{i+1},t_i)\,.
\label{eq:pathorderEop}
\eeq
If the interval $\Delta t_{i} = t_{i+1} - t_{i}$ is small enough, the
evolution operator $\Eop_N(t_{i+1},t_i)$ relevant to it can 
be evaluated by using the trapezoidal approximation:
\beq
\Eop_N(t_{i+1},t_i) \simeq I +\frac{\Delta t_{i}}2 
\sum_{l=i}^{i+1}\frac{b_0\aem^2(\mu_0)e^{4\pi b_0 t_l}}
{\beta(\aem(\mu_0) e^{2\pi b_0 t_l})}
\sum_{k=0}^\infty\left(\frac{\aem(\mu_0) e^{2\pi b_0 t_l}}{2\pi}\right)^k
\APmat_N^{[k]}\,,
\label{Eoptrap}
\eeq
where we have used eq.~(\ref{aemvst}). We have found that a number of
intervals $n=20$ is appropriate for giving stable results for 
$t$ values relevant to applications to TeV-range colliders.
At the perturbative order at which we are working, the sum over $k$
on the r.h.s.~of eq.~(\ref{Eoptrap}) is restricted to $k\le 1$.
After having obtained the evolution operator, the PDFs at the hard scale
$\mu$ are computed in the Mellin space according to eq.~(\ref{FvsE}). Finally, 
in order to invert the PDFs from the Mellin to the $z$ space, we employ 
a numerical algorithm based on the so-called Talbot path. Details on the 
implementation of this method can be found in ref.~\cite{DelDebbio:2007ee}.
The computer program that implements what has been described thus far 
was used to obtain all of the numerical results presented in
sect.~\ref{sec:res} and appendix~\ref{sec:recres}.

In the non-singlet case one can devise an alternative procedure. Namely,
one can exploit the analytical $N$-space solution for the evolution operator, 
given in eq.~(\ref{Esol1}), multiply it by the Mellin-transformed initial
conditions, and then invert the result thus obtained back to the $z$ space
by means of a numerical contour integration. We have implemented this
strategy in a computer program\footnote{This builds upon the code originally
written by the authors of ref.~\cite{Mele:1990cw}.} fully independent
from the one described above, and verified that the two are in
perfect agreement.

\section{Matching\label{sec:match}}
The best analytical prediction is obtained by matching the recursive 
solution of eq.~(\ref{PDFexpt}), that is valid for all $z$ values but in 
practice can be computed only up to a certain $\ord(\aem^n)$ (here, $n=3$), 
with the solutions of eqs.~(\ref{NLLsol3run}) (for singlet and non-singlet)
and~(\ref{gaNLLsol4run}) (for photon),
that retain all orders in $\aem$ but are sensible only when $z\simeq 1$. 
In order to distinguish these two classes of solutions, we now denote them 
as follows\footnote{We shall henceforth consider the case of NLL solutions 
with running $\aem$, which constitutes our most accurate scenario. However, 
the procedure is unchanged in the case of NLL solutions with fixed $\aem$, 
or in the case of LL solutions\label{ft:types}.}:
\beqn
\Gamma_{\rm rec}(z)&=&
\Gamma(z)\big[\protect{\rm eq.}~(\ref{PDFexpt})\big]\,,
\label{PDFrecdef}
\\
\Gamma_{\rm asy}(z)&=&
\Gamma(z)\big[\protect{\rm eq.}~(\ref{NLLsol3run})\big]
\;\;\;\;\;\;\;\;\;\;\;\;\;\,{\rm non\!-\!singlet}\,,
\label{PDFasydef1}
\\
\Gamma_{\rm asy}(z)&=&
\binomial{\;\Gamma(z)\big[\protect{\rm eq.}~(\ref{NLLsol3run})\big]}
{\ePDF{\gamma}(z)\big[\protect{\rm eq.~(\ref{gaNLLsol4run})}\big]}
\;\;\;\;\;\;\;\;{\rm singlet\!-\!photon}\,.
\label{PDFasydef2}
\eeqn
We remind the reader that eq.~(\ref{PDFexpt}) implicitly encompasses
the non-singlet, singlet, and photon cases by means of the $\jLL_k$ and 
$\jNLL_k$ functions (see sect.~\ref{sec:rec}). 

We define the matched PDFs with the additive formula\footnote{Additive
matching has been considered in refs.~\cite{Kuraev:1985hb,Nicrosini:1986sm,
Cacciari:1992pz}; refs.~\cite{Skrzypek:1990qs,Skrzypek:1992vk} 
adopt a multiplicative one.}:
\beq
\Gamma_{\rm mtc}(z)=\Gamma_{\rm rec}(z)+\Big(\Gamma_{\rm asy}(z)-
\Gamma_{\rm subt}(z)\Big)\,G(z)\,,
\label{matchsol}
\eeq
where $G(z)$ is a largely arbitrary function that must obey the
following condition
\beq
\lim_{z\to 1}G(z)=1\,,
\label{Gztoo}
\eeq
and that might optionally be used to power-suppress at small $z$ the
difference in round brackets in eq.~(\ref{matchsol}) -- we shall give more
details on this point later. The quantity $\Gamma_{\rm subt}(z)$ (that we 
call ``subtraction term'') is responsible for removing the double counting,
i.e.~the contributions which are present both in the recursive and in 
the asymptotic solutions. We shall eventually construct it explicitly, but 
we anticipate the obvious fact that it must feature the dominant $z\to 1$
contributions to the PDF (which, in turn, are present in both the
recursive and the asymptotic solutions, as is discussed in
appendix~\ref{sec:lzexp}).

Before proceeding we stress that, although general, the arguments that
follow from eq.~(\ref{matchsol}) are best understood if the PDFs are
strongly peaked at $z\to 1$, which is indeed what happens for the
singlet and non-singlet components, but not for the photon (at least
to a certain extent). Thus, we shall first understand the two former 
cases, and return to the latter one only towards the end of this section.

We want the matched PDF to coincide with the asymptotic or the recursive 
solution for those $z$ values appropriate for either of the latter two
quantities. This is equivalent to requiring:
\beqn
\Gamma_{\rm mtc}(z)&\sim&\Gamma_{\rm asy}(z)\;\;\;\;\;\;\;\;
z\simeq 1\,,
\label{Gmtcasy}
\\
\Gamma_{\rm mtc}(z)&\sim&\Gamma_{\rm rec}(z)\;\;\;\;\;\;\;\;
z~~{\rm elsewhere}\,.
\label{Gmtcelse}
\eeqn
Given eq.~(\ref{Gztoo}), eq.~(\ref{Gmtcasy}) is satisfied when:
\beq
\Big|\Gamma_{\rm rec}(z)-\Gamma_{\rm subt}(z)\Big|\ll
\Big|\Gamma_{\rm asy}(z)\Big|\,,
\;\;\;\;\;\;\;\;z\simeq 1\,.
\label{GGllGlarge}
\eeq
Conversely, there are two ways in which the behaviour in eq.~(\ref{Gmtcelse})
can be achieved.
\begin{itemize}
\item[{\em a)}]
$G(z)$ can be expanded in series around $z=0$, and is such that:
\beq
\lim_{z\to 0}G(z)=0\,,
\label{Gztoz}
\eeq
in addition to satisfying eq.~(\ref{Gztoo}).
\item[{\em b)}] One has:
\beq
\Big|\Gamma_{\rm asy}(z)-\Gamma_{\rm subt}(z)\Big|\ll
\Big|\Gamma_{\rm rec}(z)\Big|\,,
\label{GGllGsmall}
\eeq
for small and intermediate $z$ values. When eq.~(\ref{GGllGsmall}) holds,
one can set:
\beq
G(z)\equiv 1\,.
\label{Gzeqo}
\eeq
\end{itemize}
The option of item {\em a)} stems from the observation that
both $\Gamma_{\rm asy}(z)$ and $\Gamma_{\rm subt}(z)$ are only sensible
when the \mbox{$\log^p(1-z)$} terms they contain are large. When this
is not the case, i.e.~at small- and intermediate-$z$ values, one 
can suppress them (in fact, one must, if eq.~(\ref{Gmtcelse}) is
to be fulfilled) by means of power-suppressed terms, here parametrised
by $G(z)$, without affecting the formal accuracy of the
matched PDF. However, this has the potential drawback of introducing
in $\Gamma_{\rm mtc}(z)$ a dependence on the arbitrary quantity $G(z)$,
which must remain small in order not to lose predictive power.
This issue is avoided if the option of item {\em b)} is viable. This
has the drawback that it relies on the condition in eq.~(\ref{GGllGsmall}),
that might be problematic since it constrains $\Gamma_{\rm subt}(z)$ in a 
$z$ region which is not the natural domain of such a function.

\begin{figure}[thb]
  \begin{center}
  \includegraphics[width=0.47\textwidth]{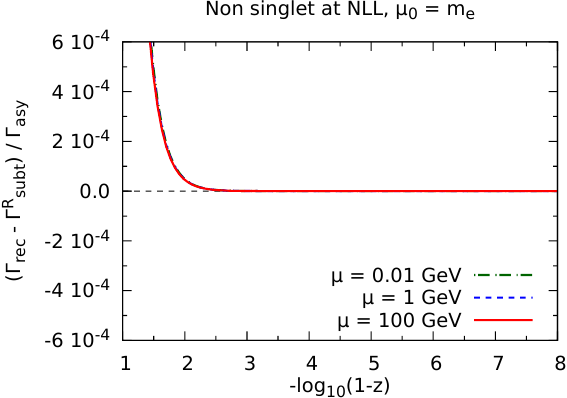}                                                                    
$\phantom{a}$
  \includegraphics[width=0.47\textwidth]{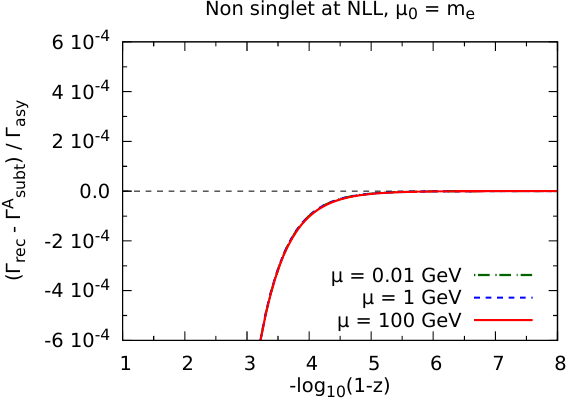}
\caption{\label{fig:RmSoA} 
Ratio of the l.h.s.~of eq.~(\ref{GGllGlarge}) over its r.h.s. (without the
absolute values), for the two choices of the subtraction term 
(eq.~(\ref{subtRz}), left panel; eq.~(\ref{subtAz}), right panel), and 
for three different hard-scale values: $\mu=0.01$~GeV (dot-dashed green), 
$\mu=1$~GeV (dashed blue), and $\mu=100$~GeV (solid red). As is indicated, 
the scale on the $y$ axis of these plots is in units of $10^{-4}$.
}
  \end{center}
\end{figure}

Although there is significant freedom in the construction of the
subtraction term, the recursive and asymptotic solutions provide us
with two obvious candidates. Namely, we can set either
\beq
\Gamma_{\rm subt}(z)\equiv\Gamma_{\rm subt}^{\rm R}(z)=\overline{\Gamma}(z)
\big[{\rm\protect{\rm eq.}~(\ref{PDFexptLb})}\big]
\label{subtRz}
\eeq
or 
\beq
\Gamma_{\rm subt}(z)\equiv\Gamma_{\rm subt}^{\rm A}(z)=\overline{\Gamma}(z)
\big[{\rm\protect{\rm eq.}~(\ref{lzPDFexptf})}\big]\,.
\label{subtAz}
\eeq
In other words: with eq.~(\ref{subtRz}) we use all of the contributions
to the recursive solution which are non-vanishing when $z\to 1$, while
with eq.~(\ref{subtAz}) we employ the $\ord(\aem^3)$ expansion of the
asymptotic solution. Therefore, as it follows from the discussion in 
appendix~\ref{sec:lzexp}, $\Gamma_{\rm subt}^{\rm A}(z)$ essentially
contains a subset of the terms present in $\Gamma_{\rm subt}^{\rm R}(z)$.
More precisely:
\beqn
\Gamma_{\rm subt}^{\rm R}(z)\;\;&\longleftrightarrow&\;\;
\Big\{\lbase_i(z),\qbase_i(z)\Big\}\equiv
\left\{\frac{\log^i(1-z)}{1-z}\,,\,\log^i(1-z)\right\}\,,
\label{GsR}
\\
\Gamma_{\rm subt}^{\rm A}(z)\;\;&\longleftrightarrow&\;\;
\Big\{\lbase_i(z)\Big\}\equiv
\left\{\frac{\log^i(1-z)}{1-z}\right\}\,.
\label{GsA}
\eeqn
By construction (see eq.~(\ref{jhdef})), we have
\beq
\lim_{z\to 1}\Big(\Gamma_{\rm rec}(z)-\Gamma_{\rm subt}^{\rm R}(z)\Big)=0\,,
\label{limRmR0}
\eeq
and therefore eq.~(\ref{GGllGlarge}) automatically holds when the subtraction
term is defined by means of the recursive solution. Conversely,
\beq
\Gamma_{\rm rec}(z)-\Gamma_{\rm subt}^{\rm A}(z)\;\simeq\;
\aem\,\qbase_2(z)\;\stackrel{z\to 1}{\longrightarrow}\;\infty\,.
\eeq
However, in spite of this, eq.~(\ref{GGllGlarge}) holds also in this
case, since:
\beq
\frac{\Gamma_{\rm rec}(z)-\Gamma_{\rm subt}^{\rm A}(z)}
{\Gamma_{\rm asy}(z)}\;\simeq\;
\frac{\qbase_2(z)+\ldots}{\lbase_2(z)+\ldots}
\;\stackrel{z\to 1}{\longrightarrow}\;0\,.
\label{ratRmA0oA}
\eeq
The conclusion is that eq.~(\ref{GGllGlarge}) is satisfied with both
choices of the subtraction term. The difference between adopting
$\Gamma_{\rm subt}^{\rm R}(z)$ or $\Gamma_{\rm subt}^{\rm A}(z)$
is that with the former function the matched PDF will converge towards 
the asymptotic solution at $z$ values relatively smaller than those
relevant to the latter function. This can be seen in fig.~\ref{fig:RmSoA},
where the ratio of the l.h.s.~over the r.h.s.~of eq.~(\ref{GGllGlarge})
(without the absolute values) is plotted as a function of 
\mbox{$-\log_{10}(1-z)$} for the two choices of the subtraction 
term considered here, and for three different
hard scales $\mu$. Note that the scale on the $y$ axis of the plots
in fig.~\ref{fig:RmSoA} is in units of $10^{-4}$.
The curves in fig.~\ref{fig:RmSoA} are relevant to the non-singlet
component. We point out that their analogues for the singlet component
are qualitatively and quantitatively very similar to those shown here.
Apart from being in keeping with the expectations emerging from
eqs.~(\ref{limRmR0})--(\ref{ratRmA0oA}), fig.~\ref{fig:RmSoA} shows
that, even in the case of eq.~(\ref{subtRz}), the matched PDF will attain 
its asymptotic form only for values of $z$ which are extremely close to one;
in other words, non-logarithmic contributions are almost always important.
This being the case, by choosing $\Gamma_{\rm subt}^{\rm A}(z)$ as a
subtraction term rather than $\Gamma_{\rm subt}^{\rm R}(z)$ (which, as
was anticipated, ``delays'' the onset of the asymptotic regime in the
matched PDF) one renders the transition between the asymptotic and 
recursive solutions less abrupt; this turns out to be beneficial
in order to reproduce the results of the numerical evolution.

\begin{figure}[thb]
  \begin{center}
  \includegraphics[width=0.47\textwidth]{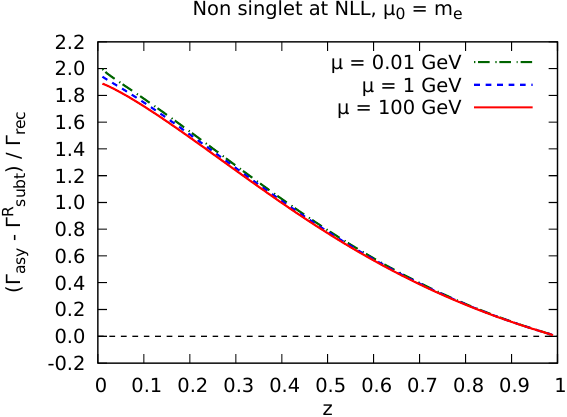}
$\phantom{aa}$
  \includegraphics[width=0.47\textwidth]{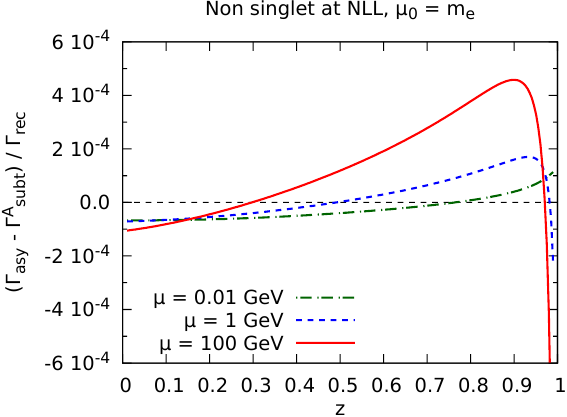}
\caption{\label{fig:AmSoR} 
Same as in fig.~\ref{fig:RmSoA}, for eq.~(\ref{GGllGsmall}). As is indicated, 
the scale on the $y$ axis of the plot on the right panel is in units of 
$10^{-4}$.
}
  \end{center}
\end{figure}

As far as the small- and intermediate-$z$ region is concerned, we
observe that:
\beqn
\Gamma_{\rm asy}(z)-\Gamma_{\rm subt}^{\rm R}(z)&=&\ord(\aem)\,,
\label{AmRz}
\\
\Gamma_{\rm asy}(z)-\Gamma_{\rm subt}^{\rm A}(z)&=&\ord(\aem^4)\,.
\label{AmAz}
\eeqn
Equation~(\ref{AmRz}) implies that it is unlikely that, if the subtraction 
term is defined by means of the recursive solution, one can avoid the use
of the $G(z)$ function. Conversely, eq.~(\ref{AmAz}) implies that
the definition by means of the asymptotic solution has a better
chance of satisfying eq.~(\ref{GGllGsmall}), thus bypassing the
need to introduce $G(z)$. Note that the difference in eq.~(\ref{AmAz}) is 
of $\ord(\aem^4)$ as a direct consequence of the fact that  we have computed 
$\Gamma_{\rm subt}^{\rm A}(z)$ to $\ord(\aem^3)$ (see eq.~(\ref{lzPDFexptf})).
In fig.~\ref{fig:AmSoR} we plot the ratio of the l.h.s.~over the r.h.s.~of 
eq.~(\ref{GGllGsmall}) (without the absolute values), by using the same 
layout as in fig.~\ref{fig:RmSoA}.
In order to be definite, we have considered again the non-singlet
component in fig.~\ref{fig:AmSoR}, and have verified that the singlet
one gives results which are extremely similar to those of the non-singlet.
Figure~\ref{fig:AmSoR} confirms our expectation based on eqs.~(\ref{AmRz}) 
and~(\ref{AmAz}).

We now turn to discussing the case of the photon PDF, which is
quite different from that of the singlet and the non-singlet. The starting
point is the same as for the latter PDFs, namely the definition of
the subtraction term with either eq.~(\ref{subtRz}) or~(\ref{subtAz}),
since those formulae are the general parametrisations of the perturbative
expansion of the recursive or the asymptotic solutions, respectively, 
whose actual values are determined by the parameters (given in 
appendices~\ref{sec:abJhLL}, \ref{sec:abJhNLL}, \ref{sec:SNSasyexp},
and~\ref{sec:gaasyexp}) specific to the particle which is being
considered. Indeed, in the case of the photon, the analogues of
eqs.~(\ref{GsR}) and~(\ref{GsA}) are:
\beqn
\Gamma_{\rm subt}^{\rm R}(z)\;\;&\longleftrightarrow&\;\;
\Big\{\qbase_i(z)\Big\}\,,
\label{gaGsR}
\\
\Gamma_{\rm subt}^{\rm A}(z)\;\;&\longleftrightarrow&\;\;
\Big\{\qbase_i(z)\Big\}\,.
\label{gaGsA}
\eeqn
Actually, because of eqs.~(\ref{cLLlim}) and~(\ref{cNLLlim}),
one can make a stronger statement, namely:
\beq
\Gamma_{\rm subt}^{\rm R}(z)=\Gamma_{\rm subt}^{\rm A}(z)\,.
\label{gaSReqA}
\eeq
We stress that eq.~(\ref{gaSReqA}) is not a property inherent to the
photon PDF, but a consequence of having been able to keep the relevant
subleading terms in the computation of its large-$z$ form as carried
out in appendix~\ref{sec:asyph}. In order to be definite, for consistency
with the case of the singlet/non-singlet we shall label the subtraction
term with ``A'' here.

\begin{figure}[thb]
  \begin{center}
  \includegraphics[width=0.47\textwidth]{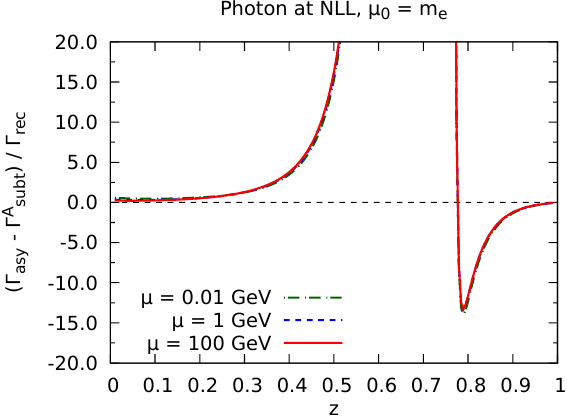}
$\phantom{aa}$
  \includegraphics[width=0.47\textwidth]{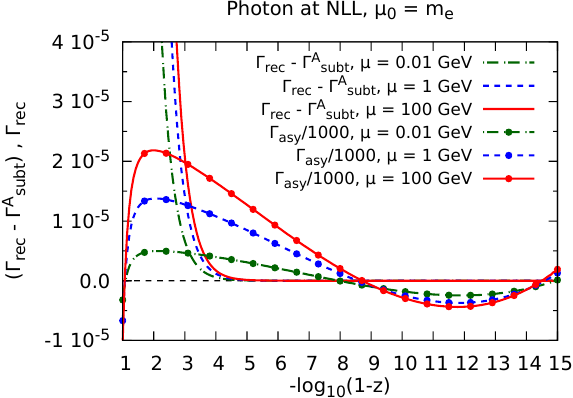}
\caption{\label{fig:techga} 
Plots assessing the validity of eq.~(\ref{GGllGsmall}) (left panel)
and eq.~(\ref{GGllGlarge}) (right panel), in the case of the photon PDF. 
See the text for details.}
  \end{center}
\end{figure}
The analogues of the right-hand side panels of fig.~\ref{fig:RmSoA} 
and of fig.~\ref{fig:AmSoR} are presented in the right and left
panels of fig.~\ref{fig:techga}, respectively. We start from the right-hand 
side one, in order to assess the validity of eq.~(\ref{GGllGlarge}).
Unfortunately, it turns out that at large $z$'s the NLL photon PDF
becomes negative in a certain range, and it crosses twice the zero.
For this reason, we cannot consider the ratio of the two sides of
eq.~(\ref{GGllGlarge}) as was done in fig.~\ref{fig:RmSoA}, but only plot 
separately \mbox{$\Gamma_{\rm rec}(z)-\Gamma_{\rm subt}(z)$}
and \mbox{$\Gamma_{\rm asy}(z)$}; these two quantities are displayed
in fig.~\ref{fig:techga} by adopting identical patterns (each associated 
with a different hard scale $\mu$), with the curves relevant to 
\mbox{$\Gamma_{\rm asy}(z)$}
overlaid by full circles. Furthermore, in order for the latter curves 
to fit into the layout, they have been multiplied by a constant factor
equal to $10^{-3}$. The plot clearly shows how eq.~(\ref{GGllGlarge}) 
is safely fulfilled\footnote{Strictly speaking, no such conclusion is 
possible in an extremely narrow neighbourhood of the points at which 
the PDF crosses zero, where it is however not relevant, since all 
quantities of interest are vanishingly small there.}.

We now consider the left panel of fig.~\ref{fig:techga}, in order
to assess the validity of eq.~(\ref{GGllGsmall}); given that for all of
the $z$ values employed in the plot the photon PDFs is positive, we
can compute the ratio of the two sides of eq.~(\ref{GGllGsmall}) 
(without the absolute values) as
was done previously in fig.~\ref{fig:AmSoR}. It is immediate to see 
that the conclusions are the opposite of those valid in the cases
of the singlet and non-singlet -- namely, in a very large range in $z$
the subtraction term and the asymptotic solution do {\em not} agree
with each other\footnote{See appendix~\ref{sec:asyph}, in particular
the comments after eq.~(\ref{Mmoinv5res}), for a discussion on the
origin of this behaviour.}. Thus, in the case of the photon the 
use of a damping function $G(z)$ is unavoidable. In order to define
it, it is useful to introduce the function:
\beq
\hz(z)=-\log_{10}(1-z)\,,
\label{hzdef}
\eeq
by means of which we set:
\beqn
G(z)=\left\{
\begin{array}{ll}
1               &\phantom{aaaa} \hz_1\le\hz(z)\,,\\
G_p\!\left(\frac{\hz(z)-\hz_0}{\hz_1-\hz_0}\right) 
   &\phantom{aaaa} \hz_0\le\hz(z)<\hz_1\,,\\
0               &\phantom{aaaa} \hz(z) < \hz_0\,,\\
\end{array}
\right.
\;\;\;\;\;\;\;\;\;\;
G_p(v)=\frac{v^p}{v^p+(1-v)^p}\,.
\label{Gdef}
\eeqn
This is a smooth function that obeys eqs.~(\ref{Gztoo}) and~(\ref{Gztoz}),
and where $\hz_0$, $\hz_1$, and $p$ are free parameters. The physical
meaning of the parameters $\hz_0$ and $\hz_1$ is that, for $z$ such that 
\mbox{$\hz(z)<\hz_0$} (\mbox{$\hz(z)>\hz_1$}), the matched PDF
coincides with the recursive (the asymptotic) solution. As a matter of 
fact, eqs.~(\ref{hzdef}) and~(\ref{Gdef}) stem from the observation that it 
is $\hz(z)$, and not $z$, the natural variable to carry out the matching, 
and this is because the large-$z$ behaviour of the PDFs is achieved when 
logarithmic terms grow much larger than non-logarithmic ones.

\begin{figure}[thb]
  \begin{center}
  \includegraphics[width=0.60\textwidth]{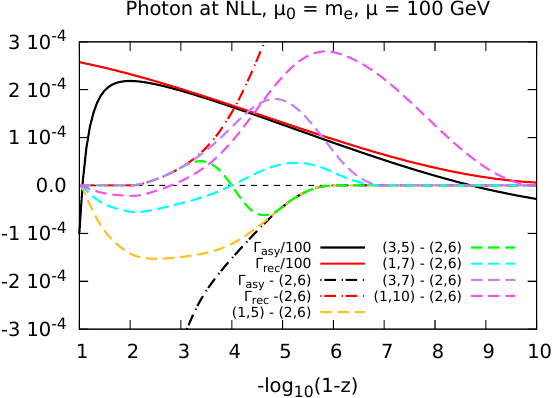}
\caption{\label{fig:gamatch} 
Study of the dependence of the matched photon PDF upon the parameters
of the matching function $G(z)$, defined in eq.~(\ref{Gdef}). We have 
set $\mu=100$~GeV.
}
  \end{center}
\end{figure}
In principle, the parameters $\hz_0$, $\hz_1$, and $p$ are unconstrained.
In order to choose them sensibly, we plot in fig.~\ref{fig:gamatch} 
the asymptotic and recursive solutions as solid black and red
curves, respectively (both are multiplied by a factor of $10^{-2}$, 
for reasons that will soon become clear). For the matching to work
reasonably well, the transition between the recursive and the asymptotic
solutions must occur in a region where these two predictions are as close
as possible to each other (which we interpret as the signal that both
give a reasonable description of the ``true'' photon PDF). From 
fig.~\ref{fig:gamatch}, we gather that such a region is
\mbox{$2\lesssim\hz\lesssim 6$}; this suggests to set $\hz_0=2$
and $\hz_1=6$. However, it is clear that there is (and there must be)
a certain flexibility in these choices. The agreement between the
asymptotic and recursive solutions quickly worsens in the range
\mbox{$z\in(1,2)$}, which implies that $\hz_0=1$ must be considered
as an extreme choice. On the other hand, for $\hz>6$ the asymptotic and 
recursive solutions do tend to stay relatively close to each other,
to the extent that even $\hz_1=10$ appears to be an acceptable choice.
As far as $p$ is concerned, the larger this parameter the more abrupt
is the transition between the two regimes. We have therefore opted to
employ $p=2$, which essentially corresponds to the slowest
transition compatible with the derivatives of $G(z)$ being continuous.
In order to assess the impact of the choices of $\hz_0$ and $\hz_1$
on the matched PDF, we have computed the latter for several values of
these parameters. In fig.~\ref{fig:gamatch} we display as dashed curves 
the differences between any of the matched predictions (relevant to
\mbox{$(\hz_0,\hz_1)=(1,5)$}, $(3,5)$, $(1,7)$, $(3,7)$, and $(1,10)$)
minus the one obtained with \mbox{$(\hz_0,\hz_1)=(2,6)$}. For comparison,
we also show the differences between the asymptotic and recursive solutions
minus the \mbox{$(\hz_0,\hz_1)=(2,6)$} matched PDF as black and red
dot-dashed curves, respectively. We see that the differences between any
two pairs of matched predictions are roughly in the range 
\mbox{$(-2,3)\cdot 10^{-4}$},
i.e.~at least a factor $25$ smaller than the recursive and the asymptotic 
solutions. While this statement progressively loses validity when moving
towards $\hz=8$ (where the asymptotic solution, which is the appropriate one
in this region, crosses zero), it also becomes less relevant, since indeed
all quantities of interest tend to become negligible in absolute value. 
Having said that, it is important to bear in mind that the dependence on 
the matching-function parameters is a genuine uncertainty that affects 
the matched predictions; plots such as those in fig.~\ref{fig:gamatch} 
help assess its size, and should be re-produced whenever new conditions
become relevant (specifically, for hard-scale values significantly different 
w.r.t.~those considered in fig.~\ref{fig:gamatch}). We finally point out that 
we have repeated the exercise by using $p=3$ and $p=4$; the overall 
uncertainties are similar to those obtained with $p=2$.

In summary, our best analytical results are obtained with the
matching formula of eq.~(\ref{matchsol}). In the case of the singlet
and the non-singlet, we employ eq.~(\ref{subtAz}) for the definition 
of the subtraction term, and a constant $G(z)$ function as in 
eq.~(\ref{Gzeqo}). In the case of the photon, the definition of 
the subtraction term is still given by eq.~(\ref{subtAz}) -- however,
there are more limited possibilities here, owing to eq.~(\ref{gaSReqA}).
The matched photon PDF does need a non-trivial matching function:
we adopt that of eq.~(\ref{Gdef}), with $\hz_0=2$, $\hz_1=6$, and $p=2$.

\section{Numerical and analytical predictions\label{sec:res}}
In this section we present our predictions for the PDFs, by computing
them both with the numerical code described in the first part of
sect.~\ref{sec:num}, and by evaluating the analytical formulae;
these are always the matched ones. We compare these two
classes of predictions, mutually validating them in the process.
Unless explicitly indicated, all results are NLL-accurate with
running $\aem$, and all have been obtained by setting $\mu_0=m$.

\begin{figure}[thb]
  \begin{center}
  \includegraphics[width=0.47\textwidth]{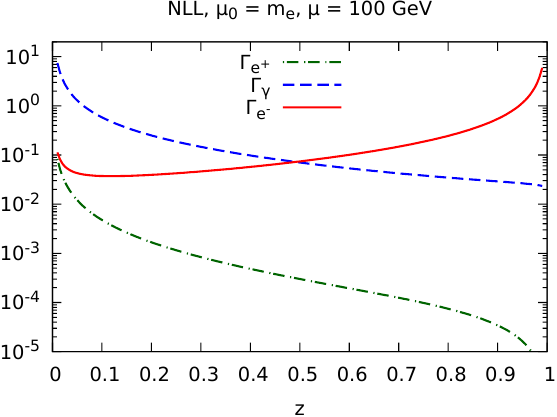}
$\phantom{aa}$
  \includegraphics[width=0.47\textwidth]{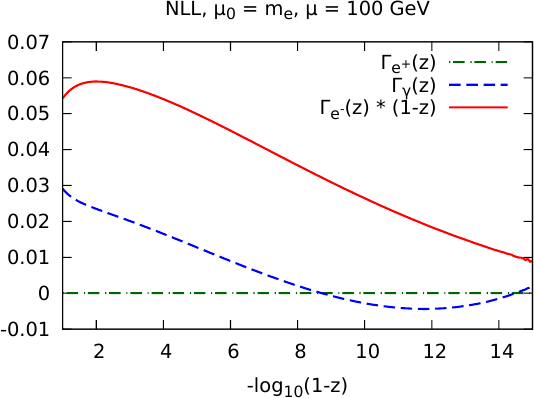}
\caption{\label{fig:allPDFs} 
Electron (solid red), photon (dashed blue), and positron PDFs (dot-dashed
green) PDFs at $\mu=100$~GeV. The electron PDF is multiplied by a 
factor $(1-z)$ in the plot on the right panel.
}
  \end{center}
\end{figure}

We begin by plotting in fig.~\ref{fig:allPDFs} the electron, photon,
and positron PDFs, computed at $\mu=100$~GeV with the numerical code. 
The left panel shows
these quantities in the full \mbox{$z\in (0,1)$} range, while the
right panel is a zoom to the large-$z$ region, where we consider
\mbox{$\hz\in (1,15)$} (see eq.~(\ref{hzdef}) for the definition
of $\hz$). Owing to the much faster growth of the electron 
PDFs in this region w.r.t.~that of the other two partons, we have 
multiplied this PDF by a factor equal to \mbox{$(1-z)$}, in order for 
all of the three curves to fit into the same layout. Figure~\ref{fig:allPDFs} 
renders it manifest that the production of heavy (relative to the
collider c.m.~energy) objects is dominated by the partonic 
lepton whose charge is the same as that of the particle lepton it stems 
from\footnote{The reader must bear in mind that all our results 
are obtained by assuming an electron particle. In the case of a positron
particle, the roles of the electron and positron partons are simply
reversed.} (in eq.~(\ref{master0}), one has the implicit constraint
\mbox{$\zp\zm\ge M^2/S$}, with $M$ and $\sqrt{S}$ the mass of the object
produced and the collider c.m.~energy, respectively). Note that from the right 
panel of fig.~\ref{fig:allPDFs}, given that the solid-red and dashed-blue 
curves are roughly of the same order, and that the former includes the 
$(1-z)$ factor, one can immediately see that the photon PDF is smaller
than the electron PDF by a number of orders of magnitude equal to the
value of $\hz$. Conversely, by producing lighter objects and/or by increasing 
the collider energy, the contribution(s) of the incoming photon(s) become(s) 
important.

In view of the smallness of the positron PDF as is documented in 
fig.~\ref{fig:allPDFs}, it is more convenient to present our findings in 
terms of the singlet and the non-singlet PDFs rather than by means of the 
electron and positron ones. This is what we shall do in the remainder
of this section.

\begin{figure}[thb]
  \begin{center}
  \includegraphics[width=0.47\textwidth]{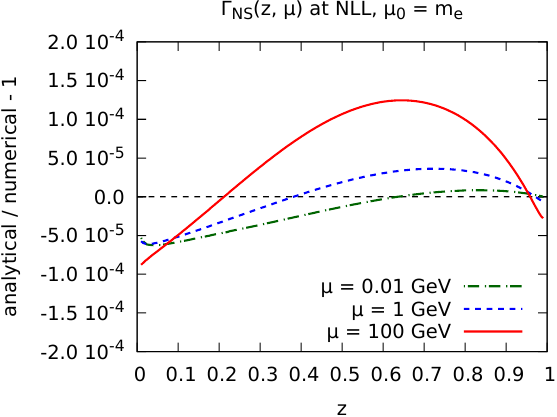}
$\phantom{aa}$
  \includegraphics[width=0.47\textwidth]{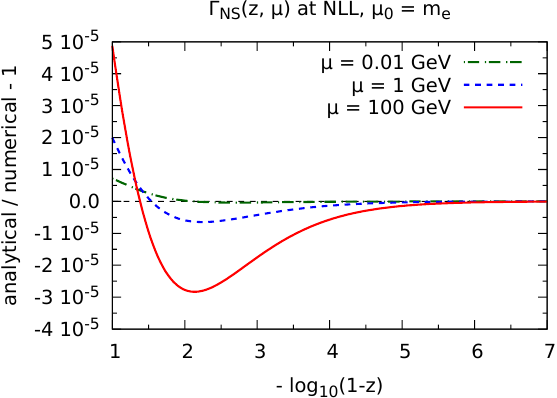}
\caption{\label{fig:anvsnumNS} 
Comparison between the numerical and analytical predictions for
the non-singlet, for three different hard-scale choices.
}
  \end{center}
\end{figure}

\begin{figure}[thb]
  \begin{center}
  \includegraphics[width=0.47\textwidth]{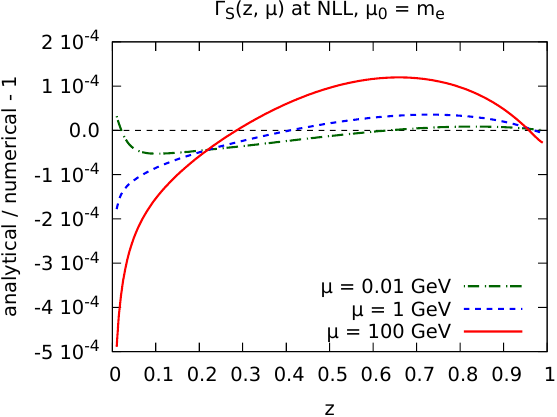}
$\phantom{aa}$
  \includegraphics[width=0.47\textwidth]{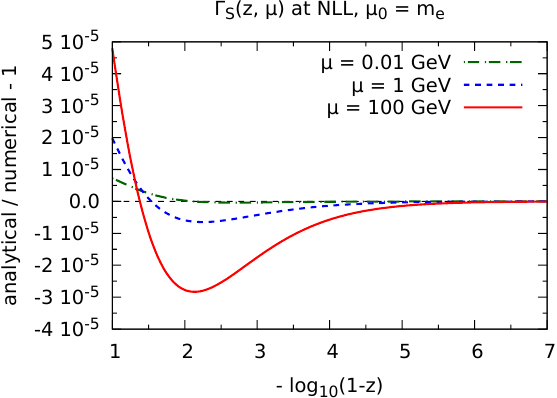}
\caption{\label{fig:anvsnumS} 
As in fig.~\ref{fig:anvsnumNS}, for the singlet.
}
  \end{center}
\end{figure}

\begin{figure}[thb]
  \begin{center}
  \includegraphics[width=0.47\textwidth]{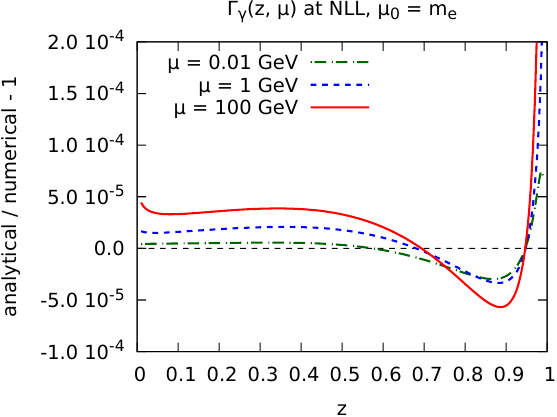}
$\phantom{aa}$
  \includegraphics[width=0.47\textwidth]{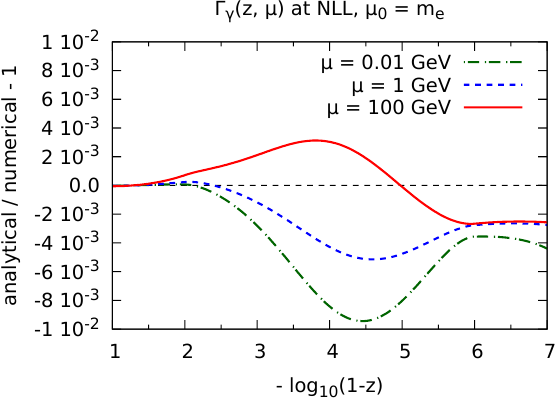}
\caption{\label{fig:anvsnumga} 
As in fig.~\ref{fig:anvsnumNS}, for the photon.
}
  \end{center}
\end{figure}

\begin{figure}[thb]
  \begin{center}
  \includegraphics[width=0.60\textwidth]{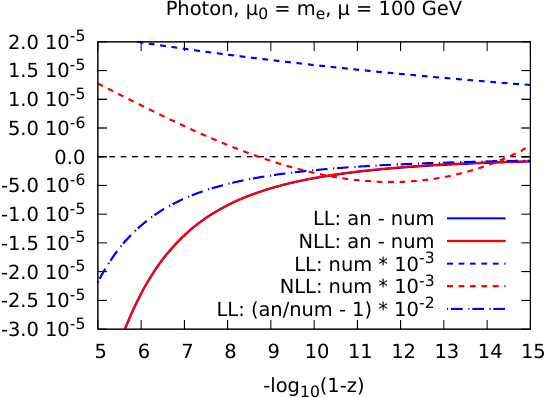}
\caption{\label{fig:gaasy} 
Behaviour of the photon PDF at very large $z$ values, where the analytical
and numerical predictions are compared. We have set $\mu=100$~GeV.
}
  \end{center}
\end{figure}

In order to establish the level of agreement between our numerical
and analytical predictions, we plot in figures~\ref{fig:anvsnumNS},
\ref{fig:anvsnumS}, and~\ref{fig:anvsnumga} the ratios of the latter
over the former, minus one, in the cases of the non-singlet, singlet,
and photon, respectively.
In each plot, there are three curves, corresponding to three different
choices of hard scales: $\mu=0.01$~GeV (dot-dashed green curves),
$\mu=1$~GeV (dashed blue curves), and $\mu=100$~GeV (solid red curves).
An overarching observation is that, in all of the cases bar for the
photon at large $z$'s (an exception to which we shall return later),
the $\mu=100$~GeV results are those which display the largest 
analytical-numerical disagreements. 
However, even in this worst-case scenario, the level of agreement
is excellent, being typically of the order of $10^{-5}$--$10^{-4}$
(relative); the largest disagreements are to be found at small $z$'s
in the case of the singlet (because of the presence of un-resummed
$\log z$ terms\footnote{Techniques to resum such logarithms exist,
see e.g.~refs.~\cite{Altarelli:2008aj,Ciafaloni:2003rd}.}). 
In keeping with the previous remark relevant to the hard-scale
dependence, the case of the photon at $z\simeq 1$ constitutes an exception:
from the right panel of fig.~\ref{fig:anvsnumga} we see that the analytical
and numerical predictions agree at the level of $10^{-3}$ ($10^{-2}$) at
$\mu=100$~GeV ($\mu=0.01$~GeV) for \mbox{$2\lesssim\hz\lesssim 6\,$};
furthermore, the behaviour at $\hz>6$ might seem to suggest that the
$z\to 1$ limits of the analytical and numerical computations are different.
We shall show in the following (see fig.~\ref{fig:gaasy}) that this is
in fact {\em not} the case. For the time being, the crucial thing to bear 
in mind is that, in the $z$ region we are discussing, the photon PDF
is very small in absolute value and, more importantly, smaller than the 
electron PDF by several orders of magnitude. Thus, even a relatively 
large discrepancy of 0.1--1\% between the numerical and analytical photon 
PDFs will be quite irrelevant. The general conclusion, which applies to
all partons, is therefore that the analytical formulae appear to be 
perfectly adequate, and can be employed in calculations of cross sections 
for phenomenological purposes.

\begin{figure}[thb]
  \begin{center}
  \includegraphics[width=0.47\textwidth]{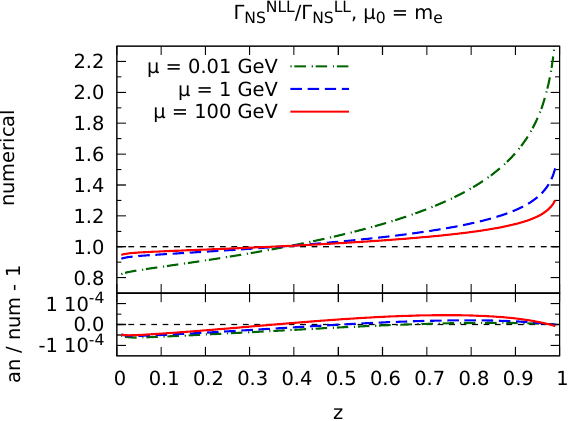}
$\phantom{aa}$
  \includegraphics[width=0.47\textwidth]{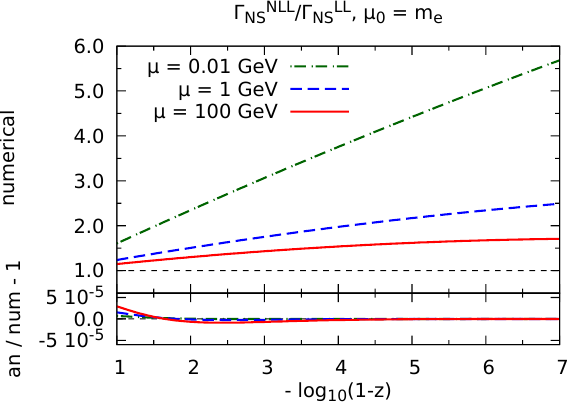}
\caption{\label{fig:NLLvsLLNS} 
Main frames: ratios of NLL over LL PDF, as computed with the numerical code,
for the non-singlet, and for three different hard-scale choices. Insets: 
ratio of the ratio shown in the main frame, over the same quantity computed 
analytically, minus one.
}
  \end{center}
\end{figure}
\begin{figure}[thb]
  \begin{center}
  \includegraphics[width=0.47\textwidth]{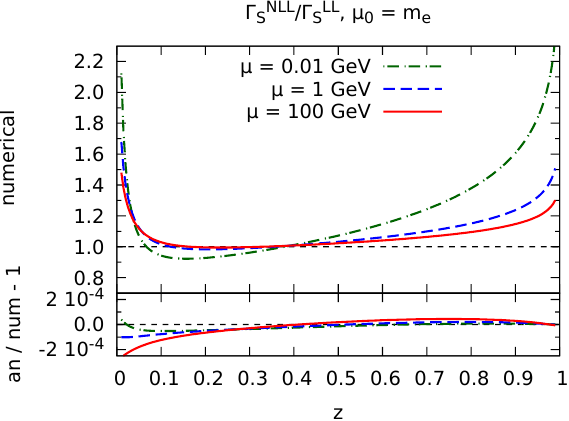}
$\phantom{aa}$
  \includegraphics[width=0.47\textwidth]{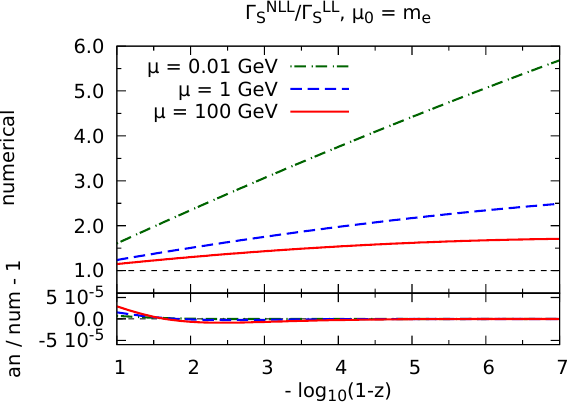}
\caption{\label{fig:NLLvsLLS} 
As in fig.~\ref{fig:NLLvsLLNS}, for the singlet.
}
  \end{center}
\end{figure}
\begin{figure}[thb]
  \begin{center}
  \includegraphics[width=0.47\textwidth]{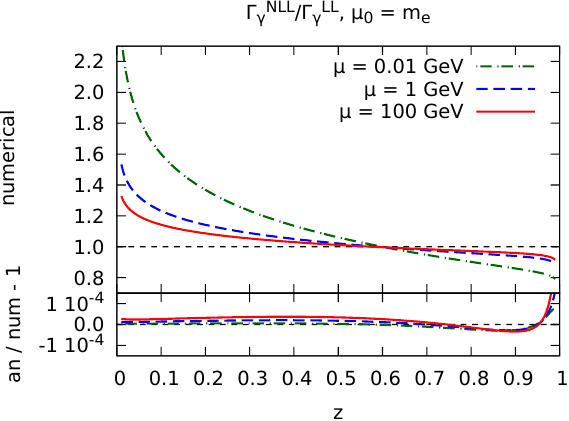}
$\phantom{aa}$
  \includegraphics[width=0.47\textwidth]{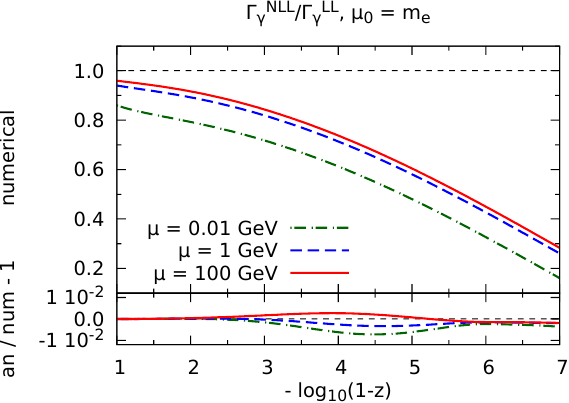}
\caption{\label{fig:NLLvsLLga} 
As in fig.~\ref{fig:NLLvsLLNS}, for the photon.
}
  \end{center}
\end{figure}
We now turn to assessing the effects on the PDFs of the NLL corrections, 
by comparing the NLL results with their LL counterparts.
While this will fully account for the predictions obtained here for
the first time, it is important to bear in mind that the PDFs are
unphysical quantities, and that beyond LL cancellations do occur
(in particular, in $\MSb$) between them and the short-distance 
cross sections. Thus, an increase or decrease by a factor $X$ 
of an NLL PDF w.r.t.~an LL one will most definitely not translate
into a corresponding increase or decrease of the NLO physical cross
section w.r.t.~its LO counterpart.

In the main frames of figs.~\ref{fig:NLLvsLLNS}, \ref{fig:NLLvsLLS}, 
and~\ref{fig:NLLvsLLga}, we plot the ratios of the NLL PDFs over the
LL ones, both computed with the numerical code. As was done previously,
all figures feature three curves, that correspond to different 
choices of hard scales; the same scale values, and the same graphical
patterns, are used here as in figs.~\ref{fig:anvsnumNS}--\ref{fig:anvsnumga}.
All the figures have an inset, which displays the double ratio (minus one):
\beq
\left.\frac{{\rm PDF}_{\sss\rm NLL}}{{\rm PDF}_{\sss\rm LL}}\right|_{\rm an}
\Bigg/
\left.\frac{{\rm PDF}_{\sss\rm NLL}}{{\rm PDF}_{\sss\rm LL}}\right|_{\rm num}
-1\,.
\eeq
The agreement between the numerical and analytical predictions is again
extremely good, especially at large $z$'s; once more, the photon in 
this region is the (relative) exception to that general rule, on which 
we shall comment later. The agreement becomes marginally worse with 
increasing $\mu$, but this effect is less evident w.r.t.~that in the case 
of the absolute predictions of figs.~\ref{fig:anvsnumNS}--\ref{fig:anvsnumga}.
Interestingly, the size of the NLL effects {\em decreases} with the hard
scale. This is particularly easy to understand in the large-$z$ region in
the case of the singlet (or non-singlet), since it can be directly read
from eq.~(\ref{NLLsol3run}). As was already remarked there, this behaviour
is driven by eq.~(\ref{Aexp}), which implies that: {\em a)} the coefficient
of the $\log(1-z)$ term is much larger than that of the $\log^2(1-z)$ term
up to extremely large values of $z$; {\em b)} such a coefficient, being
proportional to \mbox{$1/\aem(\mu)$}, decreases with $\mu$. These two
effects can clearly be seen in the main frames of the right panels of 
figs.~\ref{fig:NLLvsLLNS} and~\ref{fig:NLLvsLLS}, where the various
lines are almost straight ones, but relatively less so at larger values
of the hard scales. Keeping in mind the general observation made above
on the unphysical nature of the PDFs, we point out that the natural
applications of the quantities computed here involve scales that are large.

We now go back to commenting on the large-$z$ behaviour of the
photon PDF. We have already remarked in fig.~\ref{fig:anvsnumga} 
that such a PDF in this region is close to zero in absolute value,
and orders of magnitude smaller than its electron counterpart. On 
top of that, for the specific issue of the NLL vs LL results,
the comparison between eqs.~(\ref{gaNLLsol4run}) and~(\ref{gaLLsol4run})
(or between their over-simplified forms of eqs.~(\ref{ePDFgaasy})
and~(\ref{ePDFgaasyLL})) shows that, at variance with the case of
the electron (eqs.~(\ref{NLLsol3run}) and~(\ref{LLsol3run2})), the
NLL asymptotic photon PDF does not factorise the functional form
relevant to its LL version. Hence, larger differences in the matching 
region have to be expected between the NLL and LL photon PDF, which are
larger than those for the electron.

At the right end of the $z$ range in fig.~\ref{fig:NLLvsLLga} we see again 
the kind of pattern as in the same region of fig.~\ref{fig:anvsnumga}, which 
might cast doubts on the agreement between the $z\to 1$ limits of the 
analytical and numerical predictions. In order to address this concern, 
in fig.~\ref{fig:gaasy} we plot the photon PDF in a much more extended
$z$ range w.r.t.~what was done so far. 
The blue and red solid curves are the differences between
the analytical and numerical results computed at the LL and NLL, respectively;
the dashed curves of the same colours are the corresponding PDFs, multiplied
by an overall constant factor equal to $10^{-3}$; finally, the blue dot-dashed
curve is the rescaled ratio of the analytical over the numerical LL results, 
minus one, which can be sensibly computed owing to the fact that the LL PDF 
does not vanish for values of $z\ne 1$. Apart from the similarity between 
the LL and NLL differences, the key message of fig.~\ref{fig:gaasy} is that 
at $z\to 1$ the analytical and numerical predictions do tend to the same 
value, but in a much slower way w.r.t.~the case of the singlet/non-singlet.
In other words, the onset of the true asymptotic regime occurs at much 
larger $z$ values for the photon than for the singlet or non-singlet.
This needs not be surprising, owing to the mechanism that governs the 
asymptotic photon behaviour, as is documented in appendix~\ref{sec:asyph}. 
An improvement of the analytical large-$z$ PDF computed here would require
keeping all terms suppressed by powers of $N^{-2}$ in Mellin space, an 
extremely involved computation which is not justified in view of the 
smallness of the photon PDF in this region.

\section{Conclusions\label{sec:concl}}
In this paper we have computed the electron, positron, and photon PDFs 
relevant to an incoming unpolarised electron; the case of an incoming
positron is trivially obtained by charge conjugation. Our predictions 
include up to next-to-leading logarithmic (NLL) terms, and are obtained 
by evolving the initial conditions that have recently been 
calculated~\cite{Frixione:2019lga} at the next-to-leading order (NLO). 
We thus improve upon the long-standing leading-logarithmic (LL) PDFs of 
refs.~\cite{Skrzypek:1990qs,Skrzypek:1992vk,Cacciari:1992pz}; this is 
crucial to help achieve the high-precision results needed at future 
$\epem$ colliders. All of the calculations are perturbative in the QED 
coupling constant $\aem$, which by default we take as running.

The PDFs have been obtained by means of both numerical and analytical methods, 
which are shown to agree extremely well with each other (typically, at the 
$10^{-4}$ level or better for those $z$ values where the relevant PDFs
are not vanishingly small). As far as the analytical results are concerned,
they stem from an additive matching formula (eq.~(\ref{matchsol})),
which combines a prediction that is accurate to all orders in $\aem$ but 
only at $z\to 1$, with a prediction that is accurate up to $\ord(\aem^3)$ 
in the whole $z$ range; these are
referred to as the asymptotic and the recursive solutions, respectively.
The latter are thus called because they stem from recursive equations
(derived here for the first time at the NLL accuracy), whereby the
$\ord(\aem^k)$ contributions are obtained from the $\ord(\aem^p)$ 
ones (with $p<k$), the NLO initial conditions, and the Altarelli-Parisi
kernels. Although we have limited ourselves to considering $k\le 3$ in
this work, nothing prevents one from employing the recursive equations 
to achieve an even higher precision if need be.

The electron LL PDF is extremely peaked towards $z\to 1$, where it
features an integrable singularity. The NLL result has the same
qualitative behaviour, and in fact the $z\to 1$ singularity is even 
more pronounced than at the LL because of the presence of additional
\mbox{$\log(1-z)$} terms. While the photon LL PDF vanishes at $z\to 1$,
its NLL counterpart grows logarithmically there, thus exhibiting the
same enhanced growth at higher orders as the electron. However, one must 
bear in mind that PDFs are unphysical quantities: in particular, beyond LL 
they become dependent on the adopted collinear subtraction scheme. In this 
paper we have worked in $\MSb$, and some of the logarithms mentioned above
stem directly from this scheme choice; as such, they will cancel against
their counterparts in the short-distance cross sections, so as to have
scheme-independent predictions for physical observables.

The analytical knowledge of the PDFs is important to better understand
the details of QED collinear dynamics. However, in view of the
rapidity of the growth of the electron PDF at $z\to 1$, such a
knowledge is also crucial in the context of numerical computations,
because it gives one the possibility to adapt in a very tailored manner
the integration procedure, which would otherwise be hardly converging.

We finally point out that the PDFs of the photon (understood as an
incoming particle), and/or those of any polarised particle, can be dealt 
with similar techniques as those we have employed here.

\section*{Acknowledgments}
SF wishes to thank M.~Bonvini, S.~Catani, M.~Mangano, S.~Marzani, D.~Pagani, 
F.~Piccinini, G.~Ridolfi, H-S.~Shao, and B.~Ward for various conversations
during the course of this work, and the CERN TH division for hospitality.
V.B. acknowledges the support from the European Research Council (ERC) under 
the European Union's Horizon 2020 research and innovation program (grant
agreement No. 647981, 3DSPIN).

\appendix
\section{Results for the recursive solutions\label{sec:recres}}
In this appendix we report the results for the $\jLL_k$
and $\jNLL_k$ basis functions that appear in eq.~(\ref{PDFexpt}), which
we have computed for \mbox{$0\le k\le 3$} and \mbox{$0\le k\le 2$},
respectively, i.e.~up to $\ord(\aem^3)$. We write the actual
recursive solution that constitutes one of the main results of this
paper, and which we have used in our numerical studies, as follows:
\beq
\Gamma(z,\mu^2)=
\sum_{k=0}^{k_{\rm max}^{\sss\rm LL}} \frac{t^k}{k!}\,\jLL_k(z)
+\frac{\aem(t)}{2\pi}\sum_{k=0}^{k_{\rm max}^{\sss\rm NLL}} 
\frac{t^k}{k!}\,\jNLL_k(z)\,,
\label{PDFexptL}
\eeq
with
\beq
k_{\rm max}^{\sss\rm LL}=3\,,
\;\;\;\;\;\;\;\;
k_{\rm max}^{\sss\rm NLL}=2\,.
\label{ksmax}
\eeq
We also remind the reader that from eq.~(\ref{PDFexptL}) one can obtain
the solution in the case of non-running $\aem$, by replacing $\jLL_k$ with
$\iLL_k$ and $\jNLL_k$ with $\iNLL_k$, where:
\beqn
\iLL_k(z)&=&\jLL_k(z)\,,
\\
\iNLL_k(z)&=&\jNLL_k(z)\big[b_0\to 0\,,b_1\to 0\,,b_1/b_0\to 0\big]\,.
\eeqn
It is convenient, also in view of the matching with the large-$z$ solution, 
to present the results for the basis functions by writing them as follows:
\beqn
\jLL_k(z)&=&\jbLL_k(z)+\jhLL_k(z)\,,
\label{jbhLLdef}
\\
\jNLL_k(z)&=&\jbNLL_k(z)+\jhNLL_k(z)\,.
\label{jbhNLLdef}
\eeqn
By definition, $\jhLL_k$ and $\jhNLL_k$ collect all of the terms of
$\jLL_k$ and $\jNLL_k$, respectively, that vanish at $z=1$:
\beq
\lim_{z\to 1}\jhLL_k(z)=\lim_{z\to 1}\jhNLL_k(z)=0\,.
\label{jhdef}
\eeq
It then follows that $\jbLL_k$ and $\jbNLL_k$ include all contributions
that are either divergent (which then feature all the $\log^p(1-z)$ terms)
or equal to a non-null constant at $z=1$. Because of this, it is useful
to introduce the following auxiliary functions:
\beqn
\lbase_i(z)&=&\frac{\log^i(1-z)}{1-z}\,,\;\;\;\;\;\;\;\;i\ge 0\,,
\label{lbasedef}
\\
\qbase_i(z)&=&\log^i(1-z)\,,\;\;\;\;\;\;\;\;\;i\ge 0\,,
\label{qbasedef}
\eeqn
and write:
\beqn
\jbLL_k(z)&=&\sum_{i=0}^{i_{\rm max}^{\sss\rm LL}(k)}\Big[\,
b_{k,i}^{\sss\rm LL}\,\lbase_i(z)+
c_{k,i}^{\sss\rm LL}\,\qbase_i(z)\Big],
\;\;\;\;\;\;\;\;\;\;\;k\ge 1\,,
\label{jbLLkexp}
\\
\jbNLL_k(z)&=&
\sum_{i=0}^{i_{\rm max}^{\sss\rm NLL}(k)} \Big[\,
b_{k,i}^{\sss\rm NLL}\,\lbase_i(z)+
c_{k,i}^{\sss\rm NLL}\,\qbase_i(z)\Big],
\;\;\;\;\;\;\;\;k\ge 0\,.
\label{jbNLLkexp}
\eeqn
with:
\beqn
i_{\rm max}^{\sss\rm LL}(k)&=&k-1\,,
\\
i_{\rm max}^{\sss\rm NLL}(k)&=&k+1\,.
\eeqn
In addition to this, one must take into account that, at $\ord(\aem^0)$:
\beq
\jLL_0(z)=\jbLL_0(z)=\jhLL_0(z)=0\,.
\eeq
The contribution to $\Gamma(z)$ that does not vanish at $z\to 1$
is then written as follows:
\beq
\overline{\Gamma}(z,\mu^2)=
\sum_{k=0}^{k_{\rm max}^{\sss\rm LL}} \frac{t^k}{k!}\,\jbLL_k(z)
+\frac{\aem(t)}{2\pi}\sum_{k=0}^{k_{\rm max}^{\sss\rm NLL}} 
\frac{t^k}{k!}\,\jbNLL_k(z)\,.
\label{PDFexptLb}
\eeq
The expressions of the $b_{k,i}^{\sss\rm (N)LL}$ and 
$c_{k,i}^{\sss\rm (N)LL}$ coefficients for the non-singlet, singlet, 
and photon PDFs will be presented in appendix~\ref{sec:abJhLL} (LL results), 
and in appendix~\ref{sec:abJhNLL} (NLL results). The expressions of the
functions $\hat{J}^{\sss\rm (N)LL}(z)$ are lengthy (with some of them
receiving contributions that we have not computed analytically, as detailed
below), and not relevant to the matching; for these reasons, they are only 
reported in an ancillary file that will accompany the submission of this
paper to the {\tt arXiv}. We remind the reader that the 
recursive solutions are obtained by following the procedure
outlined in sect.~\ref{sec:rec}. Namely, one first computes the
$\JLL_k$ and $\JNLL_k$ functions, by employing eqs.~(\ref{JLLsol})
and~(\ref{JNLLsol}). These equations must be applied recursively, 
by working one's own way up in $k$ from the known $k=0$ results (given in
eqs.~(\ref{ILLSNSini})--(\ref{INLLgamini})). The expressions for
the Altarelli-Parisi kernels are taken from ref.~\cite{deFlorian:2016gvk}.
Finally, the $\jLL_k$ and $\jNLL_k$ functions are obtained by
derivation, according to eq.~(\ref{jfromJ}).
\begin{figure}[thb]
  \begin{center}
  \includegraphics[width=0.47\textwidth]{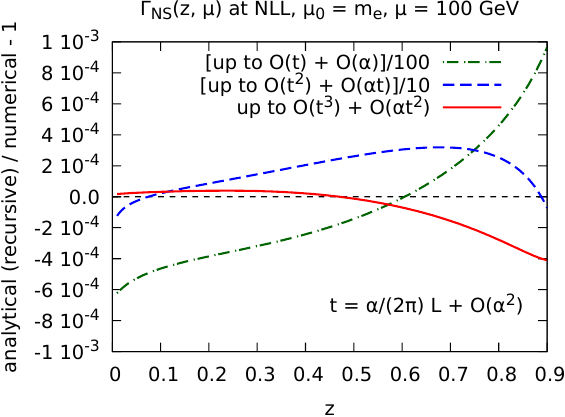}
$\phantom{aa}$
  \includegraphics[width=0.47\textwidth]{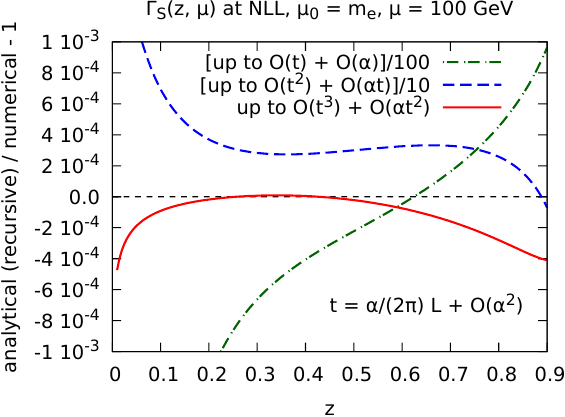}
$\phantom{aa}$
  \includegraphics[width=0.47\textwidth]{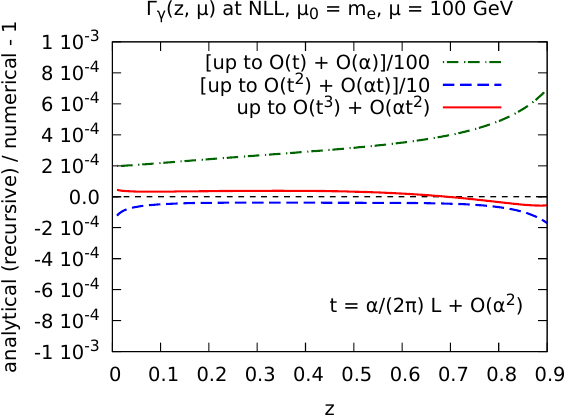}
\caption{\label{fig:conv} 
Agreement between recursive solutions of various accuracies, and the 
numerical predictions, for the non-singlet (top left panel), singlet 
(top right panel), and photon (bottom panel), for $\mu=100$~GeV. 
See the text for details.
}
  \end{center}
\end{figure}
In order to document the effect of increasing the number of terms
included in the recursive solutions, we plot in fig.~\ref{fig:conv} 
the ratio of the result of eq.~(\ref{PDFexptL}) over the numerical
predictions minus one; eq.~(\ref{PDFexptL}) is computed by setting:
\beq
k_{\rm max}^{\sss\rm NLL}=k_{\rm max}^{\sss\rm LL}-1\,,
\;\;\;\;\;\;\;\;
k_{\rm max}^{\sss\rm LL}=1,2,3\,.
\eeq
The ratios are displayed as green dot-dashed lines 
(\mbox{$k_{\rm max}^{\sss\rm LL}=1$}), blue dashed lines 
(\mbox{$k_{\rm max}^{\sss\rm LL}=2$}), and red solid lines 
(\mbox{$k_{\rm max}^{\sss\rm LL}=3$}). In order for the results 
to fit into the layout of the figures, the green and blue curves are 
multiplied by a constant factor equal to $10^{-2}$ and $10^{-1}$,
respectively. In keeping with what has been discussed in 
sect.~\ref{sec:res}, we see that our most accurate recursive predictions
(\mbox{$k_{\rm max}^{\sss\rm LL}=3$}) agree with the numerical results
at the level of a few $10^{-4}$ at the worst. Note that since here we
are dealing only with the recursive solutions we have limited ourselves
to plotting the PDFs in the range \mbox{$z\in (0,0.9)$} -- at the upper 
end of the range, the absence of the contribution from the asymptotic
solution starts to be felt. The new information stemming from
fig.~\ref{fig:conv} is that, if we had only computed either the first term or 
the first two terms in the sums of eq.~(\ref{PDFexptL}), the $\ord(10^{-4})$ 
agreement remarked above would actually have been roughly equal to, but 
generally worse than, $10^{-2}$ and $10^{-3}$, respectively. The figure 
also shows that, for any given accuracy of the recursive solution, the 
agreement with the numerical prediction marginally worsens towards $z\to 0$ 
in the case of the singlet, owing to the presence of $\log z$ terms which 
are not resummed.

In the course of the recursive procedure, we have found that some integrals 
relevant to $\JNLL_2$ (i.e.~the function associated with the $\ord(\aem t^2)$ 
term in the representation of the PDFs) are not easily computable analytically. 
We have therefore opted to limit ourselves to obtaining their $z\to 1$ 
leading terms analytically, while evaluating all of the remaining terms
numerically, so that the latter contribute only to $\jhNLL_2$
(we point out that an analogous strategy has already been adopted 
in ref.~\cite{Cacciari:1992pz}). More precisely, let us consider
the generic modified-convolution integral of eq.~(\ref{modconv}).
We distinguish two possibilities: either $g(x)$ is a plus distribution,
or it is an ordinary function. Notation-wise, these two cases are
written as follows:
\beqn
&&{\rm plus~distribution:}\phantom{aaaaaaa} 
g(x)=\Big[\hat{g}(x)\Big]_+\,,
\label{gdistr}
\\
&&{\rm ordinary~function:}\phantom{aaaaaa} 
g(x)\,.
\label{gfunc}
\eeqn
In the case of eq.~(\ref{gdistr}), we have:
\beq
\big[\hat{g}\big]_+\ootimes_z h = 
\left.\big[\hat{g}\big]_+\ootimes_z h\right|_{\rm end}+
\left.\big[\hat{g}\big]_+\ootimes_z h\right|_{\rm bulk}\,,
\eeq
where we have defined the endpoint and bulk contributions, respectively,
as follows:
\beqn
\left.\big[\hat{g}\big]_+\ootimes_z h\right|_{\rm end}&=&
-h(z)\int_0^z dx\,\hat{g}(x)\,,
\label{hgend}
\eeqn
\beqn
\left.\big[\hat{g}\big]_+\ootimes_z h\right|_{\rm bulk}&=&
\int_z^1 dx\,\hat{g}(x)\left[h\left(\frac{z}{x}\right)-h(z)\right]
\nonumber\\*&=&
\int_0^1 dy\,(1-z)\,\hat{g}\Big(1-(1-z)y\Big)\left[
h\!\left(\frac{z}{1-(1-z)y}\right)-h(z)\right]\,.\phantom{aa}
\label{hgbulk}
\eeqn
These equations can also be used in the simpler case of eq.~(\ref{gfunc}):
one simply sets the endpoint contribution equal to zero, and computes
eq.~(\ref{hgbulk}) by removing the subtraction term $h(z)$ and with the
formal replacement \mbox{$\hat{g}\to g$} there.

The endpoint contribution of eq.~(\ref{hgend}) is always computed 
analytically, and its results are included in $\jbNLL_2(z)$ and/or
$\jhNLL_k(z)$, according to the behaviour of $h(z)$ at $z\to 1$.
As far as eq.~(\ref{hgbulk}) is concerned, for the sake of the 
forthcoming discussion let us re-write it more compactly as follows:
\beq
F(z) = \int_0^1 dy\,f(y,z)\,.
\label{hgbulk2}
\eeq
If the integral in eq.~(\ref{hgbulk2}) were strongly convergent,
then we might obtain its contribution to the PDF (see eq.~(\ref{Fdef}))
by means of a derivation under the integral sign, namely:
\beq
-\frac{\partial F(z)}{\partial z} = 
-\int_0^1 dy\,\frac{\partial f(y,z)}{\partial z}\,.
\label{hgbulk2der}
\eeq
Unfortunately, the strong convergence of the integral is not guaranteed,
given that $F(z)$ in general is logarithmically divergent at $z\to 1$.
However, the contributions that are non vanishing at $z=1$ are also
easy to compute analytically; such computation can be carried out
directly at the differential level of eq.~(\ref{hgbulk2der}),
and stems from expanding the integrand on the r.h.s.~of that equation
in a series of $z$ around $1$. The latter must include all terms that
result in either a logarithmically-divergent or a constant non-null
term at $z\to 1$, which typically implies up to \mbox{$(1-z)^0$}
contributions. In this way we arrive at the following identity:
\beq
-\frac{\partial F(z)}{\partial z} = 
-\left[\frac{\partial F(z)}{\partial z}-
\left.\frac{\partial F(z)}{\partial z}\right|_{\rm asy}\right]
-\left.\frac{\partial F(z)}{\partial z}\right|_{\rm asy}\,,
\label{tmp1}
\eeq
with:
\beq
\left.\frac{\partial F(z)}{\partial z}\right|_{\rm asy}=
\int_0^1 dy\,\left.\frac{\partial f(y,z)}{\partial z}\right|_{\rm exp}\,,
\label{dFdzay}
\eeq
having denoted by \mbox{$\partial f/\partial z|_{\rm exp}$} the  
aforementioned series expansion.
The integral in eq.~(\ref{dFdzay}) is computed analytically, and its
result added to $\jbNLL_2$ (thus, given eq.~(\ref{jbNLLkexp}), it
contributes to \mbox{$c_{2,i}^{\sss\rm NLL}$} for some $i$, 
depending on $h(z)$; there are no contributions to 
\mbox{$b_{2,i}^{\sss\rm NLL}$}):
\beq
- \left.\frac{\partial F(z)}{\partial z}\right|_{\rm asy}
\;\longrightarrow\;\jbNLL_2(z)\,.
\label{Jnumasy}
\eeq
Conversely, the quantity in square brackets in eq.~(\ref{tmp1}), where 
the rightmost term is regarded as a regularising factor, is computed 
numerically\footnote{These are one-dimensional integrations of regularised
integrals: the routine {\tt gsl\_integration\_qag} of the GSL library is 
employed, which guarantees a fast and reliable convergence.}, and eventually 
included in $\jhNLL_2$:
\beq
- \left[ \frac{\partial F(z)}{\partial z}-
\left.\frac{\partial F(z)}{\partial z}\right|_{\rm asy} \right] \equiv
- \int_0^1 dy\,\left(\frac{\partial f(y,z)}{\partial z}-
\left.\frac{\partial f(y,z)}{\partial z}\right|_{\rm exp}\right)
\;\longrightarrow\;\jhNLL_2(z)\,.
\label{jNLL2num}
\eeq
We list here the pairs $\hat{g}$ (or $g$) and $h$ that we handle 
in the way we have just described:
\begingroup 
\allowdisplaybreaks
\begin{align}
\hat{g}_a(v)   &= \frac{1+v^2}{1-v}\,, &\quad
h_a(v)   &= \log^2(1-v) \log v\,, \\
\hat{g}_b(v)   &= \frac{1+v^2}{1-v}\,, &\quad
h_b(v)   &= \log(1-v)\,\Li{2}(v)\,, \\
\hat{g}_c(v)   &= \frac{1+v^2}{1-v}\,, &\quad
h_c(v)   &= \log^2(v)\,\log(1+v)\,, \\
\hat{g}_d(v)   &= \frac{1+v^2}{1-v}\,, &\quad
h_d(v)   &= \log(v)\,\log^2(1+v)\,, \\
\hat{g}_e(v)   &= \frac{1+v^2}{1-v}\,, &\quad
h_e(v)   &= \log(v)\,\Li{2}(-v)\,, \\
\hat{g}_f(v)   &= \frac{1+v^2}{1-v}\,, &\quad
h_f(v)   &= \log(1+v)\,\Li{2}(-v)\,, \\
\hat{g}_g(v) &= \frac{1+v^2}{1-v}\,, &\quad
h_g(v) &= \log(1+v)\,\Li{2}\left(\frac{1}{1+v}\right)\,, \\
\hat{g}_h(v)   &= \frac{1+v^2}{1-v}\,, &\quad
h_h(v)   &= \Li{3}(1-v)\,, \\
\hat{g}_i(v) &= \frac{1+v^2}{1-v}\,, &\quad
h_i(v) &= \Li{3}(-v)\,, \\
\hat{g}_j(v) &= \frac{1+v^2}{1-v}\,, &\quad
h_j(v) &= \Li{3}\left(\frac{1}{1+v}\right)\,, \\
\hat{g}_k(v) &= \frac{1+v^2}{1-v}\,\log(1-v)\,\log v\,, &\quad
h_k(v) &= \log(1-v)\,, \\
g_l(v) &= \frac{1+v^2}{1+v}\,\log^2 v\,, &\quad
h_l(v) &= \log(1-v)\,, \\
g_m(v) &= \frac{1+v^2}{1+v}\,\log v\,\log(1+v)\,, &\quad
h_m(v) &= \log(1-v)\,, \\
g_n(v) &= \frac{1+v^2}{1+v}\,\Li{2}(-v)\,, &\quad
h_n(v) &= \log(1-v)\,, \\
g_o(v) &= \frac{1}{v}\,, &\quad
h_o(v) &= \log v\,\log^2(1+v)\,, \\
g_p(v) &= \frac{1}{v}\,, &\quad
h_p(v) &= \log(1+v)\,\Li{2}(-v)\,, \\
g_q(v) &= \frac{1}{v}\,, &\quad
h_q(v) &= \log(1+v)\,\Li{2}\left(\frac{1}{1+v}\right)\,, \\
g_r(v) &= 1\,, &\quad
h_r(v) &= \Li{3}(1-v)\,, \\
g_s(v) &= \frac{1}{v}\,, &\quad
h_s(v) &= \Li{3}(1-v)\,, \\
g_t(v) &= 1\,, &\quad
h_t(v) &= \Li{3}\left(\frac{1}{1+v}\right)\,, \\
g_u(v) &= \frac{1}{v}\,, &\quad
h_u(v) &= \Li{3}\left(\frac{1}{1+v}\right)\,, \\
g_v(v) &= \frac{1}{v}\log^2(1-v)\,, &\quad
h_v(v) &= \log(1-v)\,.
\end{align}
\endgroup 
We stress again that each of these pairs will contribute to both
eq.~(\ref{Jnumasy}) and~(\ref{jNLL2num}). We denote generically
either of these contributions as follows:
\beq
\jnum{\rho}(z)\;\longleftrightarrow\;\big(\hat{g}_\rho,h_\rho\big)
\;\;\;\;{\rm or}\;\;\;\;\big(g_\rho,h_\rho\big)\,,
\;\;\;\;\;\;\;\;\rho=a,\ldots v\,.
\eeq
These will enter $\jbNLL_2(z)$ and $\jhNLL_2(z)$ as linear combinations
with identical coefficients (owing to eq.~(\ref{tmp1})), which however
do depend on the flavour structure. Explicitly:
\beqn
{\rm non-singlet:}~~~~&&
\sum_\rho w_{{\rm\sss NS},\rho}\,\jnum{\rho}(z)=
\label{jbNLL2numNS}
\\*&&\phantom{aaaa}
4\,\jnum{a} + 4\,\jnum{b} + 4\,\jnum{h} + 2\,\jnum{c}
+4\,\jnum{d}+ 4\,\jnum{e} +4\,\jnum{f} 
\nonumber\\*&&\phantom{aa}\!
- 4\,\jnum{g} - 4\,\jnum{i}+ 8\,\jnum{j}
- 4\,\jnum{k} - 2\,\jnum{l} + 8\,\jnum{m} + 8\,\jnum{n}\,,
\nonumber
\\
{\rm singlet:}~~~~&&
\sum_\rho w_{{\rm\sss S},\rho}\,\jnum{\rho}(z)=
\label{jbNLL2numS}
\\*&&\phantom{aaaa}
4\,\jnum{a} + 4\,\jnum{b} + 4\,\jnum{h} - 2\,\jnum{c} -4\,\jnum{d}
- 4\,\jnum{e} -4\,\jnum{f}
\nonumber \\*&&\phantom{aa}\!
+ 4\,\jnum{g} + 4\,\jnum{i} - 8\,\jnum{j} - 4\,\jnum{k} + 2\,\jnum{l} 
-8\,\jnum{m} 
\nonumber \\*&&\phantom{aa}\!
-8\,\jnum{n} - 24\,\NF\,\jnum{r}\,,
\nonumber
\\
{\rm photon:}~~~~&&
\sum_\rho w_{\gamma,\rho}\,\jnum{\rho}(z)=
\label{jbNLL2numga}
\\*&&\phantom{aa}
-8\,\jnum{o} -8\,\jnum{p} + 8\,\jnum{q} + 8\,\jnum{s}
+ 16\,\jnum{t} - 16\,\jnum{u} -4\,\jnum{v}\,.
\nonumber
\eeqn
The results of these linear combinations when the $\jnum{\rho}$
contributions are computed analytically as in eq.~(\ref{Jnumasy})
are the following:
\beqn
\sum_\rho w_{{\rm\sss NS},\rho}\,\jnum{\rho}(z)&=&
-\frac{2}{3}\pi^2 \log(1-z) + \frac{4}{3} \pi^2 + 10\log(2)^2\,,
\label{resjbNLL2numNS}
\\
\sum_\rho w_{{\rm\sss S},\rho}\,\jnum{\rho}(z)&=&
\frac{2}{3}\pi^2 \log(1-z) + 4\pi^2 - 10\log(2)^2\,,
\label{resjbNLL2numS}
\\
\sum_\rho w_{\gamma,\rho}\,\jnum{\rho}(z)&=&
-4 \log^3(1-z) + \frac{4}{3} \pi^2 \log(1-z)
+\frac{4}{3}\pi^2 \log(2) - 4 \log(2)^3 - 8 \zeta_3\,.
\nonumber\\*&&
\label{resjbNLL2numga}
\eeqn
As was anticipated, the results on the r.h.s.~of 
eqs.~(\ref{resjbNLL2numNS})--(\ref{resjbNLL2numga}) do not contribute
to any of the \mbox{$b_{2,i}^{\sss\rm NLL}$} coefficients, while they
enter the coefficients \mbox{$c_{2,1}^{\sss\rm NLL}$} and
\mbox{$c_{2,0}^{\sss\rm NLL}$} (singlet and non-singlet), and
\mbox{$c_{2,3}^{\sss\rm NLL}$}, \mbox{$c_{2,1}^{\sss\rm NLL}$},
and \mbox{$c_{2,0}^{\sss\rm NLL}$} (photon).

\subsection{LL coefficients\label{sec:abJhLL}}
In this appendix we report the results for the coefficients that
enter eq.~(\ref{jbLLkexp}); all of the coefficients that do not appear
below are understood to be equal to zero.

\vskip 0.2truecm
\noindent
$\bullet$~Non-singlet:
\beq
b_{{\rm\sss NS},\,1,0}^{\sss\rm LL} = 2\,,
\eeq
\beq
c_{{\rm\sss NS},\,1,0}^{\sss\rm LL} = -2\,,
\eeq
\beq
b_{{\rm\sss NS},\,2,1}^{\sss\rm LL} = 8\,,
\eeq
\beq
c_{{\rm\sss NS},\,2,1}^{\sss\rm LL} = -8\,,
\eeq
\beq
b_{{\rm\sss NS},\,2,0}^{\sss\rm LL} = 6\,,
\eeq
\beq
c_{{\rm\sss NS},\,2,0}^{\sss\rm LL} = -2\,,
\eeq
\beq
b_{{\rm\sss NS},\,3,2}^{\sss\rm LL} = 24\,,
\eeq
\beq
c_{{\rm\sss NS},\,3,2}^{\sss\rm LL} = -24\,,
\eeq
\beq
b_{{\rm\sss NS},\,3,1}^{\sss\rm LL} = 36\,,
\eeq
\beq
c_{{\rm\sss NS},\,3,1}^{\sss\rm LL} = -12\,,
\eeq
\beq
b_{{\rm\sss NS},\,3,0}^{\sss\rm LL} = \frac{27}{2} - 4 \pi^2\,,
\eeq
\beq
c_{{\rm\sss NS},\,3,0}^{\sss\rm LL} = \frac{9}{2} + 4 \pi^2\,.
\eeq

\vskip 0.2truecm
\noindent
$\bullet$~Singlet:
\beqn
b_{{\rm\sss S},\,k,i}^{\sss\rm LL}&=&b_{{\rm\sss NS},\,k,i}^{\sss\rm LL}
\;\;\;\;\;\;\;\;\forall\,k\,,i\,,
\\
c_{{\rm\sss S},\,k,i}^{\sss\rm LL}&=&c_{{\rm\sss NS},\,k,i}^{\sss\rm LL}
\;\;\;\;\;\;\;\;\forall\,k\,,i\,.
\eeqn

\vskip 0.2truecm
\noindent
$\bullet$~Photon:
\beq
c_{\gamma,\,1,0}^{\sss\rm LL} = 1\,,
\eeq
\beq
c_{\gamma,\,2,1}^{\sss\rm LL} = 2\,,
\eeq
\beq
c_{\gamma,\,2,0}^{\sss\rm LL} = \frac{3}{2} - \frac{2}{3}\NF\,,
\eeq
\beq
c_{\gamma,\,3,2}^{\sss\rm LL} = 4\,,
\eeq
\beq
c_{\gamma,\,3,1}^{\sss\rm LL} = 6 - \frac{4}{3}\NF\,,
\eeq
\beq
c_{\gamma,\,3,0}^{\sss\rm LL} = \frac{9}{4} - \frac{2}{3} \pi^2
- \NF + \frac{4}{9} \NF^2\,.
\eeq

\subsection{NLL coefficients\label{sec:abJhNLL}}
In this appendix we report the results for the coefficients that
enter eq.~(\ref{jbNLLkexp}). Note that these do already include
the r.h.s.~of eqs.~(\ref{resjbNLL2numNS})--(\ref{resjbNLL2numga}); 
all of the coefficients that do not appear below are understood 
to be equal to zero. We employ the following shorthand notation:
\beq
L_0=\log\frac{\mu_0^2}{m^2}\,.
\label{L0def}
\eeq

\vskip 0.2truecm
\noindent
$\bullet$~Non-singlet:
\beq
b_{{\rm\sss NS},\,0,1}^{\sss\rm NLL} = -4\,,
\eeq
\beq
c_{{\rm\sss NS},\,0,1}^{\sss\rm NLL} = 4\,,
\eeq
\beq
b_{{\rm\sss NS},\,0,0}^{\sss\rm NLL} = 2\,(\Lzero -1)\,,
\eeq
\beq
c_{{\rm\sss NS},\,0,0}^{\sss\rm NLL} = -2\,(\Lzero -1)\,,
\eeq
\beq
b_{{\rm\sss NS},\,1,2}^{\sss\rm NLL} = -12\,,
\eeq
\beq
c_{{\rm\sss NS},\,1,2}^{\sss\rm NLL} = 12\,,
\eeq
\beq
b_{{\rm\sss NS},\,1,1}^{\sss\rm NLL} = -14 + 8 \Lzero + 8 \pi b_0\,,
\eeq
\beq
c_{{\rm\sss NS},\,1,1}^{\sss\rm NLL} = 10 - 8 \Lzero - 8 \pi b_0\,,
\eeq
\beq
b_{{\rm\sss NS},\,1,0}^{\sss\rm NLL} = 1 - \frac{20}{9}\NF + 4 \pi b_0
- \frac{4 \pi b_1}{b_0} + \frac{4}{3}\pi^2 + \Lzero(6 - 4 \pi b_0)\,,
\eeq
\beq
c_{{\rm\sss NS},\,1,0}^{\sss\rm NLL} = -2 + \frac{32}{9}\NF - 4 \pi b_0 
+ \frac{4 \pi b_1}{b_0} - \frac{4}{3}\pi^2 + \Lzero(-2 + 4 \pi b_0)\,,
\eeq
\beq
b_{{\rm\sss NS},\,2,3}^{\sss\rm NLL} = -32\,,
\eeq
\beq
c_{{\rm\sss NS},\,2,3}^{\sss\rm NLL} = 32\,,
\eeq
\beq
b_{{\rm\sss NS},\,2,2}^{\sss\rm NLL} = 12 (-5 + 2 \Lzero + 4 \pi b_0)\,,
\eeq
\beq
c_{{\rm\sss NS},\,2,2}^{\sss\rm NLL} = -12 (-3 + 2 \Lzero + 4 \pi b_0)\,,
\eeq
\beq
b_{{\rm\sss NS},\,2,1}^{\sss\rm NLL} = -17 - \frac{160}{9} \NF
+ 56 \pi b_0 - \frac{32 \pi b_1}{b_0} + \frac{40}{3} \pi^2
- 16 \pi^2 b_0^2 - 4 \Lzero (-9 + 8 \pi b_0)\,,
\eeq
\beq
c_{{\rm\sss NS},\,2,1}^{\sss\rm NLL} = -7 + \frac{208}{9} \NF - 32 \pi b_0
+ \frac{32 \pi b_1}{b_0} - \frac{40}{3}\pi^2 + 16 \pi^2 b_0^2
+ 4 \Lzero(-3 + 8 \pi b_0)\,,
\eeq
\beqn
b_{{\rm\sss NS},\,2,0}^{\sss\rm NLL} &=&
9 - \frac{24 \pi b_1}{b_0} - 4 \pi b_0
+ 6 \pi^2 + 8 \pi^2 b_1 - 8 \pi^2 b_0^2
- \frac{16}{3} \pi^3 b_0
- 40 \zeta_3
\nonumber \\*&& 
+ \Lzero \left(\frac{27}{2} - 24 \pi b_0 - 4 \pi^2 + 8 \pi^2 b_0^2\right)
+ \NF \left(\frac{40 \pi b_0}{9} - \frac{2}{9} (33 + 4 \pi^2) \right),
\phantom{aa}
\eeqn
\beqn
c_{{\rm\sss NS},\,2,0}^{\sss\rm NLL} &=&
-4 - \frac{10}{3}\pi^2 + 8 \pi^2 b_0^2 + \frac{8 \pi b_1}{b_0} - 8 \pi^2 b_1
+ 14 \pi b_0 + \frac{16}{3} \pi^3 b_0 
\nonumber \\*&& 
+ \NF \left( \frac{22}{9} - \frac{64}{9} \pi b_0 + \frac{8}{9} \pi^2 \right)
+ \Lzero \left(\frac{9}{2} + 8 \pi b_0 + 4 \pi^2 - 8 \pi^2 b_0^2 \right) 
+ 40 \zeta_3\,.\phantom{aa}
\eeqn

\vskip 0.2truecm
\noindent
$\bullet$~Singlet:
\beqn
b_{{\rm\sss S},\,k,i}^{\sss\rm NLL}&=&b_{{\rm\sss NS},\,k,i}^{\sss\rm NLL}
\;\;\;\;\;\;\;\;\forall\,k\,,i\,,
\\
c_{{\rm\sss S},\,k,i}^{\sss\rm NLL}&=&c_{{\rm\sss NS},\,k,i}^{\sss\rm NLL}
\;\;\;\;\;\;\;\;\forall\,k\,,i\,.
\eeqn

\vskip 0.2truecm
\noindent
$\bullet$~Photon:
\beq
c_{\gamma,\,0,0}^{\sss\rm NLL} = (\Lzero -1)\,,
\eeq
\beq
c_{\gamma,\,1,2}^{\sss\rm NLL} = -3\,,
\eeq
\beq
c_{\gamma,\,1,1}^{\sss\rm NLL} = -7 + 2 \Lzero - \frac{4}{3} \NF\,,
\eeq
\beq
c_{\gamma,\,1,0}^{\sss\rm NLL} =
-4 + \NF \left(- \frac{26}{9} - \frac{2}{3}\Lzero \right)
+ 2 \pi b_0 - \frac{2 \pi b_1}{b_0}
+ \Lzero \left(\frac{3}{2} - 2 \pi b_0\right)\,,
\eeq
\beq
c_{\gamma,\,2,3}^{\sss\rm NLL} = -6\,,
\eeq
\beq
c_{\gamma,\,2,2}^{\sss\rm NLL} =
-\frac{37}{2} + 4 \Lzero - \frac{2}{3} \NF + 10 \pi b_0\,,
\eeq
\beqn
c_{\gamma,\,2,1}^{\sss\rm NLL} &=&
-\frac{37}{2} + \frac{8}{9}\NF^2
+ 18 \pi b_0 - \frac{8 \pi b_1}{b_0}
+ 2 \pi^2
\nonumber \\*&& 
+ \Lzero (6 - 8 \pi b_0)
- \frac{4}{3} \NF \left(5 + \Lzero -2 \pi b_0 \right)\,,
\eeqn
\beqn
c_{\gamma,\,2,0}^{\sss\rm NLL} &=&
- \frac{45}{8} + \left(\frac{52}{27} + \frac{4}{9} \Lzero \right)\NF^2
+ 4 \pi b_0 + \frac{11}{6}\pi^2 - 4 \pi^2 b_0^2 - \frac{6 \pi b_1}{b_0}
+ 4 \pi^2 b_1
\nonumber \\*&& 
+ \NF \left(
- \frac{23}{6} + \frac{40 \pi b_0}{9} + \frac{8 \pi b_1}{3 b_0}
+ \frac{2}{9} \pi^2 - \Lzero + \Lzero \frac{8 \pi b_0}{3}
\right)
\nonumber \\*&& 
+ \Lzero \left(\frac{9}{4} - 6 \pi b_0
- \frac{2}{3} \pi^2 + 4 \pi^2 b_0^2 \right)
- 6 \zeta_3\,.
\eeqn

\section{Asymptotic large-$z$ solution for photon PDF beyond leading $N$
\label{sec:asyph}}
In this appendix we consider the problem that has been anticipated
in sect.~\ref{sec:asyNLLs}, namely the improvement of the asymptotic
behaviour of the photon PDF given in eq.~(\ref{gaNLLsol3run}) stemming
from the inclusion of the off-diagonal elements in the evolution kernels.
In order to do this, we start from writing the $\ord(\aem)$ expressions
of the Altarelli-Parisi kernels as follows (see eq.~(\ref{APmatex})):
\beqn
\APmat_{{\rm\sss S},N}&=&\APmat_{{\rm\sss S},N}^{[0]}+
\frac{\aem(\mu)}{2\pi}\APmat_{{\rm\sss S},N}^{[1]}+\ord(\aem^2)
\\&\equiv&
\left(\APmat_{{\rm\sss S},N}^{[0,0]}+
\frac{1}{N}\APmat_{{\rm\sss S},N}^{[0,1]}+\ord\left(N^{-2}\right)\right)
\nonumber\\*&&\phantom{aaa}
+\frac{\aem(\mu)}{2\pi}
\left(\APmat_{{\rm\sss S},N}^{[1,0]}+
\frac{1}{N}\APmat_{{\rm\sss S},N}^{[1,1]}+\ord\left(N^{-2}\right)\right)
+\ord(\aem^2)\,,
\label{APNSgsubN}
\eeqn
having introduced, at each order in $\aem$, the leading- and subleading-$N$
contributions. They read as follows:
\beqn
\APmat_{{\rm\sss S},N}^{[0,0]}&=&
\left(
\begin{array}{cc}
-2\log\bN+2\lambda_0 & 0 \\
0 & -\frac{2}{3}\,\NF \\
\end{array}
\right),
\label{APnsg00}
\\
\APmat_{{\rm\sss S},N}^{[0,1]}&=&
\left(
\begin{array}{cc}
-1~~ & 2\NF \\
1 & 0\\
\end{array}
\right),
\label{APnsg01}
\\
\APmat_{{\rm\sss S},N}^{[1,0]}&=&
\left(
\begin{array}{cc}
\frac{20}{9}\NF\log\bN+\lambda_1 & 0 \\
0 & -\NF \\
\end{array}
\right),
\label{APnsg10}
\\
\APmat_{{\rm\sss S},N}^{[1,1]}&=&
\left(
\begin{array}{cc}
-4\log\bN+\frac{27+22\NF}{9}~~~~ & 
   2\NF\left(\log^2\bN+\frac{15-\pi^2}{6}\right) \\
-\log^2\bN+\frac{15+4\NF}{3}\log\bN-
\frac{64\NF+3(36+\pi^2)}{18} & 0 \\
\end{array}
\right).
\label{APnsg11}
\eeqn
Note that, by considering only eqs.~(\ref{APnsg00}) and~(\ref{APnsg10}),
one recovers eq.~(\ref{APmatasy2}). According to eq.~(\ref{matAPmell4}),
the Altarelli-Parisi kernels define the evolution kernel as follows:
\beq
\Mmat_N=\APmat_{{\rm\sss S},N}^{[0]}+\frac{\aem(\mu)}{2\pi}\left(
\APmat_{{\rm\sss S},N}^{[1]}-
\frac{2\pi b_1}{b_0}\,\APmat_{{\rm\sss S},N}^{[0]}\right)\,,
\label{matMNdef}
\eeq
whence one can write the evolution equation and its formal solution
as follows:
\beq
\frac{\partial\Eop_N(t)}{\partial t}=\Mmat_N(t)\Eop_N(t)
\;\;\;\;\;\;\Longrightarrow\;\;\;\;\;\;
\Eop_N(t)=\exp\left[\sum_{k=1}^\infty\Omega_{k,N}(t)\right]\,.
\label{EopMagExp}
\eeq
The solution in eq.~(\ref{EopMagExp}) is based on the so-called Magnus 
expansion~\cite{Magnus:1954} (see also ref.~\cite{Magnus:rev2008}), which 
is constructed solely in terms of the evolution kernel:
\beqn
\Omega_{1,N}(t)&=&\int_0^t dt_1\Mmat_N(t_1)\,,
\label{Omega1}
\\
\Omega_{2,N}(t)&=&\half\int_0^t dt_1\int_0^{t_1} dt_2
\Big[\Mmat_N(t_1),\Mmat_N(t_2)\Big]\,,
\label{Omega2}
\\
\Omega_{3,N}(t)&=&\ldots\,,
\label{Omega3}
\eeqn
with $\Omega_{k,N}(t)$ featuring $k$ instances of $\Mmat_N$, all appearing
in commutators. Thus, in the case of a one-dimensional flavour space or
of a diagonal evolution kernels, eq.~(\ref{EopMagExp}) is identical
to the solutions given in eq.~(\ref{Esol1}) and in sect.~\ref{sec:asyNLLs}.
As far as the singlet-photon sector is concerned, we can indeed 
recover the solutions we have found previously in terms of the quantity
introduced in this appendix. We define the leading-$N$ evolution
kernel:
\beq
\Mmat_N^{(0)}=\APmat_{{\rm\sss S},N}^{[0,0]}+\frac{\aem(\mu)}{2\pi}\left(
\APmat_{{\rm\sss S},N}^{[1,0]}-
\frac{2\pi b_1}{b_0}\,\APmat_{{\rm\sss S},N}^{[0,0]}\right)
\label{matM0Ndef}
\eeq
and denote by $\Eop_N^{(0)}(t)$ the corresponding evolution operator.
Thus:
\beq
\frac{\partial\Eop_N^{(0)}(t)}{\partial t}=\Mmat_N^{(0)}(t)\Eop_N^{(0)}(t)
\;\;\;\;\;\;\Longrightarrow\;\;\;\;\;\;
\Eop_N^{(0)}(t)=
\left(
\begin{array}{cc}
E_{\Sigma\Sigma,N}^{(0)} & 0\\
0 & E_{\gamma\gamma,N}^{(0)} \\
\end{array}
\right),
\label{Eop0}
\eeq
where:
\beqn
E_{\Sigma\Sigma,N}^{(0)}&=&\exp\left[-\xi_1\log\bN+\hat{\xi}_1\right],
\label{Eop0SS}
\\
E_{\gamma\gamma,N}^{(0)}&=&\exp\left[-\frac{2\NF}{3}t-
\frac{\aemmu-\aemz}{4\pi^2 b_0}\NF\Big(1-\frac{4\pi b_1}{3b_0}\Big)\right]
\label{Eop0gg1}
\\*&=&
\left(\frac{\aemz}{\aemmu}\right)^\frac{\NF}{3\pi b_0}
\exp\left[-\frac{\aemmu-\aemz}{4\pi^2 b_0}\NF
\Big(1-\frac{4\pi b_1}{3b_0}\Big)\right]
\;\stackrel{{\rm QED}}{\longrightarrow}\;
\frac{\aemz}{\aemmu}\,.
\label{Eop0gg2}
\eeqn
Equation~(\ref{Eop0SS}) coincides with eq.~(\ref{ENxi1}), while
eq.~(\ref{Eop0gg1}) coincides with eq.~(\ref{Eggsol}), as they should.
This is not immediately apparent in the case of eq.~(\ref{Eop0gg1})
since there, at variance with what has been done in eq.~(\ref{Eggsol}),
we have not used the simplifications induced by the explicit expressions
of the QED $\beta$-function coefficients (see eq.~(\ref{b0b1})). This
is useful when one considers the limit of non-running $\alpha$ of the
formulae presented here. An expression equivalent to eq.~(\ref{Eop0gg1}),
as well as the QED ``limit'' of both, is given in eq.~(\ref{Eop0gg2}).

We stress that the case of non-running $\aem$ is problematic, as it might
lead to inconsistencies. By switching off the running, one effectively
neglects bubble-diagram contributions which are exactly the same as
those that lead to the $\gamma\gamma$ entries in eqs.~(\ref{APnsg00})
and~(\ref{APnsg10}). In this paper we ignore such potential inconsistencies,
but then we need to carefully distinguish the $\gamma\gamma$ contributions
to the Altarelli-Parisi kernels (which we always parametrise by means
of $\NF$) from those to the QED $\beta$ function (which we parametrise
by means of the $\beta$-function coefficients $b_i$). We shall return
to this point with one explicit example later in this appendix
(see eqs.~(\ref{faemrepl1}) and~(\ref{faemrepl2})).

In order to improve on the leading-$N$ results, we shall introduce the
subleading-$N$ contributions to the evolution kernel, and treat them
as a perturbation to the solution of eq.~(\ref{Eop0}). This entails
writing:
\beq
\Mmat_N=\Mmat_N^{(0)}+\frac{1}{N}\Mmat_N^{(1)}
\;\;\;\;\;\;\Longrightarrow\;\;\;\;\;\;
\Eop_N(t)=\Eop_N^{(0)}(t)\Eop_N^{(1)}(t)\,,
\label{Eopfull}
\eeq
having defined:
\beq
\Mmat_N^{(1)}=\APmat_{{\rm\sss S},N}^{[0,1]}+\frac{\aem(\mu)}{2\pi}\left(
\APmat_{{\rm\sss S},N}^{[1,1]}-
\frac{2\pi b_1}{b_0}\,\APmat_{{\rm\sss S},N}^{[0,1]}\right).
\label{matM1Ndef}
\eeq
By replacing eq.~(\ref{Eopfull}) into eq.~(\ref{EopMagExp}), one arrives
at the evolution equation for the operator $\Eop_N^{(1)}(t)$:
\beq
\frac{\partial\Eop_N^{(1)}(t)}{\partial t}=
\widehat{\Mmat}_N^{(1)}(t)\Eop_N^{(1)}(t)\,,
\;\;\;\;\;\;\;\;
\widehat{\Mmat}_N^{(1)}(t)=\frac{1}{N}\left(\Eop_N^{(0)}(t)\right)^{-1}
\Mmat_N^{(1)}(t)\Eop_N^{(0)}(t)\,.
\label{M1Nevol}
\eeq
Equation~(\ref{M1Nevol}) can be solved as is written in eq.~(\ref{EopMagExp}), 
by constructing the $\Omega_{k,N}(t)$ terms according to 
eqs.~(\ref{Omega1})--(\ref{Omega3}) with 
\mbox{$\Mmat_N\to\widehat{\Mmat}_N^{(1)}$} there. We then observe
that \mbox{$\Omega_{k,N}\propto 1/N^k$}, and thus for consistency
with eq.~(\ref{APNSgsubN}) we are allowed to discard all contributions
with $k\ge 2$. Therefore:
\beq
\Eop_N^{(1)}(t)=\exp\Big[\Omega_{1,N}(t)\Big]+\ord\left({1/N^2}\right)=
I+\int_0^t dt_1\widehat{\Mmat}_N^{(1)}(t_1)+\ord\left({1/N^2}\right)\,.
\label{Eop1sol1}
\eeq
In spite of these simplifications, the integral on the r.h.s.~of
eq.~(\ref{Eop1sol1}) features contributions of the type
\mbox{$\exp(at_1)\exp(\exp(bt_1))$} for certain $a$ and $b$, where
the functional dependence $\exp(\exp(bt_1))$ stems for the dependence
on $t_1$ of $\aemmu$ in $\Eop_N^{(0)}$. Apart from rendering the $t_1$
integral in eq.~(\ref{Eop1sol1}) non trivial, this will also induce
functional forms in the $N$-space whose analytical inverse Mellin
transforms will be extremely hard to compute. We shall therefore 
resort to simplifying the expression of $\Eop_N^{(0)}$, by linearising
the dependence on $t_1$ of $\aemmu$ there. This implies that, as an
evolution kernel, we shall use what follows:
\beq
\widehat{\Mmat}_N^{(1,L)}(t)=\widehat{\Mmat}_N^{(1)}(t)
\Bigg[\Eop_N^{(0)}(t)\;\longrightarrow\;\Eop_N^{(0,L)}(t)\Bigg],
\eeq
where:
\beq
\Eop_N^{(0,L)}(t)=
\left(
\begin{array}{cc}
E_{\Sigma\Sigma,N}^{(0,L)} & 0\\
0 & E_{\gamma\gamma,N}^{(0,L)} \\
\end{array}
\right),
\eeq
whose expression can be obtained from eqs.~(\ref{Eop0SS}) and~(\ref{Eop0gg1})
after the linearisation introduced above. Thus:
\beqn
E_{\Sigma\Sigma,N}^{(0,L)}&=&\exp\left[\left(-\xi_{1,0}\log\bN
+\hat{\xi}_{1,0}\right)t\,\right]\,,
\label{Eop0SSL}
\\
E_{\gamma\gamma,N}^{(0,L)}&=&
\exp\left[-\left(\frac{2\NF}{3}+\chi_{1,0}\right)t\,\right]\,.
\label{Eop0gg1L}
\eeqn
In equation~(\ref{Eop0SSL}) we have introduced the quantities $\xi_{1,0}$
and $\hat{\xi}_{1,0}$ which we have defined as follows:
\beq
\xi_1=\xi_{1,0}\,t+\ord(t^2)\,,
\;\;\;\;\;\;\;\;\;\;
\hat{\xi}_1=\hat{\xi}_{1,0}\,t+\ord(t^2)\,,
\label{hx10def}
\eeq
with $\xi_1$ and $\hat{\xi}_1$ given in eqs.~(\ref{xiR1def})
and~(\ref{chi1Rdef}). By means of an explicit computations from the
latter two equations we obtain:
\beqn
\xi_{1,0}&=&2\left[1-\frac{\aemz}{\pi}\left(\frac{5}{9}\NF+\frac{\pi b_1}{b_0}
\right)\right],
\label{xi10def}
\\
\hat{\xi}_{1,0}&=&
\frac{3}{2}\left[1+\frac{\aemz}{\pi}\left(\frac{\lambda_1}{3}\,
-\frac{\pi b_1}{b_0}\right)\right].
\label{hxi10def}
\eeqn
As far as eq.~(\ref{Eop0gg1L}) is concerned, its expression stems 
from that of eq.~(\ref{Eop0gg1}); in particular:
\beq
-\frac{\aemmu-\aemz}{4\pi^2 b_0}\NF\Big(1-\frac{4\pi b_1}{3b_0}\Big) =
-\chi_{1,0}\,t+\ord(t^2)\,,
\eeq
from whence:
\beq
\chi_{1,0}=\frac{\aemz}{2\pi}\NF\Big(1-\frac{4\pi b_1}{3b_0}\Big)
\;\stackrel{{\rm QED}}{\longrightarrow}\;0\,.
\label{chi10def}
\eeq
In summary, the evolution operator we shall use is the following:
\beq
\Eop_N(t)=\Eop_N^{(0,L)}(t)
\left(I+\int_0^t dt_1\widehat{\Mmat}_N^{(1,L)}(t)\right).
\label{EopNL}
\eeq
Having established that the asymptotic solutions presented in
sect.~\ref{sec:asy} are perfectly adequate for the case of the 
singlet, we shall now focus on the implications of eq.~(\ref{EopNL})
on the photon PDF. We obtain:
\beq
\ePDF{\gamma}(z)=
M^{-1}\Big[\big(\Eop_N(t)\big)_{\gamma\Sigma}\,\Gamma_{{\rm\sss S},0,N}\Big]
+M^{-1}\Big[\big(\Eop_N(t)\big)_{\gamma\gamma}\,\Gamma_{\gamma,0,N}\Big],
\label{gaPDFoffd}
\eeq
with $\Gamma_{{\rm\sss S},0,N}$ and $\Gamma_{\gamma,0,N}$ the $N$-space
expressions of the singlet and photon initial conditions, respectively.
These can be obtained from eqs.~(\ref{G0sol})--(\ref{Gpossol2}):
\beqn
\Gamma_{{\rm\sss S},0,N}&=&1+\frac{\aemz}{2\pi}
\left(\Fzero+\Fone\log\bN+\Ftwo\log^2\bN\right)+\ord\left(N^{-1}\right)\,,
\label{asyelectN}
\\
\Gamma_{\gamma,0,N}&=&\ord\left(N^{-1}\right)\,,
\label{asygammaN}
\eeqn
where:
\beqn
\Fzero&=&2-\frac{\pi^2}{3}+\frac{3}{2}\Lzero\,,
\label{F0def}
\\
\Fone&=&2\left(1-\Lzero\right)\,,
\label{F1def}
\\
\Ftwo&=&-2\,.
\label{F2def}
\eeqn
Let us start by considering the contribution of the first term on the
r.h.s.~of eq.~(\ref{gaPDFoffd}). With a straightforward, if tedious,
computation we obtain what follows:
\beq
\big(\Eop_N(t)\big)_{\gamma\Sigma}\,\Gamma_{{\rm\sss S},0,N}
\;\stackrel{N\to\infty}{\longrightarrow}\;E_{\gamma\gamma,N}^{(0,L)}\,
\frac{1}{N}\sum_{j=1}^4 \bN^{-\kappa_j}\frac{\sum_{i=0}^4 x_i^{(j)}\log^i\bN}
{y_0^{(j)}+y_1^{(j)}\log\bN}\,,
\label{gaPDFj14}
\eeq
with \mbox{$\{x_0^{(j)},\ldots x_4^{(j)},y_0^{(j)},y_1^{(j)}\}$}
four sets of $N$-independent quantities, whose specific forms are
unimportant here. For any given $j$, the five terms in the numerators
on the r.h.s.~of eq.~(\ref{gaPDFj14}) can be re-expressed {\em algebraically}
(i.e.~without any approximations) in terms of the corresponding
denominators. In this way, one arrives at the following forms (note
that $E_{\gamma\gamma,N}^{(0,L)}$ is independent of $N$):
\beqn
\ePDF{\gamma,j}(z)&=&M^{-1}\left[\frac{1}{N}
\bN^{-\kappa_j}\frac{\sum_{i=0}^4 x_i^{(j)}\log^i\bN}
{y_0^{(j)}+y_1^{(j)}\log\bN}\,,\right]\,,
\;\;\;\;\;\;\;\;\;\;j=1,2,3,4\,,
\nonumber
\\*&\equiv&
\sum_{i=1}^5 
R_i\Big(\Conej,\Ctwoj,\Cthreej,\DjCthree\big/\DjCtwo,\DjCtwo\Big)\,
\invm_i\Big(z;\kappa_j,\DjCtwo,\DjCthree\Big)\,.
\label{gaPDFj14res}
\eeqn
Here, we have introduced the inverse Mellin transforms relevant to
eq.~(\ref{gaPDFj14}) which are linearly independent from each other,
namely\footnote{Here and elsewhere, some quantities are denoted to 
depend on parameters which do not actually enter their functional
forms. This allows one to write eq.~(\ref{gaPDFj14res}) in a compact,
and formally correct, way.}:
\beqn
M^{-1}\left[\frac{\bN^{-\kappa}}{N}\frac{1}{\dendpar+\dencpar\log\bN}\right]
\;\;&\stackrel{z\to 1}{\longrightarrow}&\;\;
\invm_1(z;\kappa,\dencpar,\dendpar)\,,
\label{Mmoinv1}
\\
M^{-1}\left[\frac{\bN^{-\kappa}}{N}\log^p\bN\right]
\;\;&\stackrel{z\to 1}{\longrightarrow}&\;\;
\invm_{p+2}(z;\kappa,\dencpar,\dendpar)\,,
\;\;\;\;\;\;\;\;
p=0,1,2,3\,.\phantom{aaa}
\label{Mmoinv25}
\eeqn
Explicit computations give:
\beqn
\invm_1(z;\kappa,\dencpar,\dendpar)&=&
\frac{e^{-\gE\kappa}(1-z)^\kappa}{\Gamma(1+\kappa)}\Bigg(
\frac{1}{\dendpar-\dencpar\log(1-z)}
-\frac{(\pi^2\kappa-6\zeta_3\kappa^2)\dencpar}{6(\dendpar-\dencpar\log(1-z))^2}
\nonumber\\*&&
\phantom{\frac{e^{-\gE\kappa}(1-z)^\kappa}{\Gamma(1+\kappa)}\Bigg(}
-\frac{(30\pi^2-360\zeta_3\kappa+\pi^4\kappa^2)\dencpar^2}
{180(\dendpar-\dencpar\log(1-z))^3}
\Bigg)\,,
\label{Mmoinv1res}
\\
\invm_2(z;\kappa,\dencpar,\dendpar)&=&
\frac{e^{-\gE\kappa}(1-z)^\kappa}{\Gamma(1+\kappa)}\,,
\label{Mmoinv2res}
\\
\invm_3(z;\kappa,\dencpar,\dendpar)&=&
\frac{e^{-\gE\kappa}(1-z)^\kappa}{\Gamma(1+\kappa)}\left(
-\log(1-z)+\frac{\pi^2\kappa}{6}-\zeta_3\kappa^2\right)\,,
\label{Mmoinv3res}
\\
\invm_4(z;\kappa,\dencpar,\dendpar)&=&
\frac{e^{-\gE\kappa}(1-z)^\kappa}{\Gamma(1+\kappa)}\Bigg(
\log(1-z)^2-\frac{\pi^2}{6}
+\kappa\Big(\!-\frac{\pi^2}{3}\log(1-z)+2\zeta_3\Big)
\nonumber\\*&&
\phantom{\frac{e^{-\gE\kappa}(1-z)^\kappa}{\Gamma(1+\kappa)}\Bigg(}
+\kappa^2\Big(2\zeta_3\log(1-z)-\frac{\pi^4}{180}\Big)
\Bigg)\,,
\label{Mmoinv4res}
\\
\invm_5(z;\kappa,\dencpar,\dendpar)&=&
\frac{e^{-\gE\kappa}(1-z)^\kappa}{\Gamma(1+\kappa)}\Bigg(
-\log(1-z)^3+\frac{\pi^2}{2}\log(1-z)-2\zeta_3
\label{Mmoinv5res}
\\*&&
\phantom{\frac{e^{-\gE\kappa}(1-z)^\kappa}{\Gamma(1+\kappa)}\Bigg(}
+\kappa\Big(\frac{\pi^2}{2}\log(1-z)^2-6\zeta_3\log(1-z)-\frac{\pi^4}{60}\Big)
\nonumber
\\*&&
\phantom{aaaaaa}
+\kappa^2\Big(-3\zeta_3\log(1-z)^2+\frac{\pi^4}{60}\log(1-z)
+\frac{3}{2}\pi^2\zeta_3-12\zeta_5\Big)
\Bigg)\,,
\nonumber
\eeqn
where, consistently with eqs.~(\ref{Mmoinv1}) and~(\ref{Mmoinv25}),
in eqs.~(\ref{Mmoinv1res})--(\ref{Mmoinv5res}) some terms that vanish
at $z\to 1$ have not been included. This is of course arbitrary to
some extent, and the logic we have followed is that of keeping those
terms which, when expanded in series, either contribute to the same 
monomials $t^n$ and $\aem t^n$ as the recursive solutions considered 
in this paper, or have the same power of $\kappa$ as the former ones.
On top of this, one has the special case of eq.~(\ref{Mmoinv1res}) which
has the structure of a series in 
\mbox{$\dencpar^{k-1}(\dendpar-\dencpar\log(1-z))^{-k}$}. When $z\to 1$,
these terms are progressively more suppressed with increasing $k$.
Unfortunately, this hierarchy is not valid at intermediate $z$'s;
in fact, for the values of $\dencpar$ and $\dendpar$ relevant to
our computation there is a singularity at $z\simeq 0.65$ which
is dominated by increasingly large values of $k$. This is what prevents
the asymptotic solution of the photon PDF from being well-behaved
in all of the $z$ range, at variance with its electron counterpart.
This has significant implications for the matching, which are discussed
in sect.~\ref{sec:match}. A solution to this problem would be that of
resumming the series on the r.h.s.~of eq.~(\ref{Mmoinv1res}); we have
computed its first seven coefficients, but have not been able to identify
the corresponding generating function. Numerically, the use of all of 
these seven contributions instead of the three reported in 
eq.~(\ref{Mmoinv1res}) does not change the behaviour at large $z$'s,
and does not improve that at intermediate $z$'s.

The $R_i$ functions that appear in eq.~(\ref{gaPDFj14res}) are:
\beqn
\allowdisplaybreaks
\rrdo(\Cthree,\Cfour,\Cfive,\doc,\cpar)&=&
\Big(\Cfive-\doc\Cfour+\doc^2\Cthree\Big)
\nonumber\\*&&\phantom{aaa}
\times\left[1+\aemzotpi\Big(\Fzero-\doc\Fone+\doc^2\Ftwo\Big)\right],
\label{R1def}
\\
\rrcz(\Cthree,\Cfour,\Cfive,\doc,\cpar)&=&
\frac{1}{\cpar}\big(\Cfour-\doc\Cthree\big)
\nonumber\\*&+&
\aemzotpi\frac{1}{\cpar}\Big(\Cfour\Fzero+\Cfive\Fone-
\doc\big(\Cthree\Fzero+\Cfour\Fone+\Cfive\Ftwo\big)
\nonumber\\*&&\phantom{\aemzotpi\frac{1}{\cpar}}
+\doc^2\big(\Cthree\Fone+\Cfour\Ftwo\big)-\doc^3\Cthree\Ftwo\Big),
\label{R2def}
\\
\rrno(\Cthree,\Cfour,\Cfive,\doc,\cpar)&=&
\frac{\Cthree}{\cpar}+\aemzotpi\frac{1}{\cpar}\Big(\Cthree\Fzero+
\Cfour\Fone+\Cfive\Ftwo
\nonumber\\*&&\phantom{aaaa\aemzotpi\frac{1}{\cpar}}
-\doc\big(\Cthree\Fone+\Cfour\Ftwo\big)+\doc^2\Cthree\Ftwo\Big),
\label{R3def}
\\
\rrnt(\Cthree,\Cfour,\Cfive,\doc,\cpar)&=&
\aemzotpi\frac{1}{\cpar}\Big(\Cthree\Fone+\Cfour\Ftwo-\doc\Cthree\Ftwo\Big),
\label{R4def}
\\
\rrnth(\Cthree,\Cfour,\Cfive,\doc,\cpar)&=&
\aemzotpi\frac{\Cthree}{\cpar}\Ftwo\,,
\label{R5def}
\eeqn
where the $\Fis$ constants are given in eqs.~(\ref{F0def})--(\ref{F2def}).
Equations~(\ref{R1def})--(\ref{R5def}) must be evaluated as indicated in 
eq.~(\ref{gaPDFj14res}), with parameters:
\beqn
\DoneCtwo&=&\xi_{1,0}\,,
\label{D1C2}
\\
\DoneCthree&=&-\left(\frac{2\NF}{3}+2\pi b_0+
\hat{\xi}_{1,0}+\chi_{1,0}\right)\,,
\\
\Coneone&=&\frac{\aemz}{2\pi}\exp\big(\!-\!\DoneCthree t\big)\,,
\\
\Ctwoone&=&-\frac{\aemz}{2\pi}\left(5+\frac{4\NF}{3}\right)
\exp\big(\!-\!\DoneCthree t\big)\,,
\\
\Cthreeone&=&\frac{\aemz}{2\pi}\left(6+\frac{\pi^2}{6}+\frac{32\NF}{9}
+\frac{2\pi b_1}{b_0}\right)\exp\big(\!-\!\DoneCthree t\big)\,,
\\
\DonebCtwo&=&\DoneCtwo\,,
\\
\DonebCthree&=&\DoneCthree\,,
\\
\Conetwo&=&-\frac{\aemz}{2\pi}\,,
\\
\Ctwotwo&=&\frac{\aemz}{2\pi}\left(5+\frac{4\NF}{3}\right)\,,
\\
\Cthreetwo&=&-\frac{\aemz}{2\pi}\left(6+\frac{\pi^2}{6}+\frac{32\NF}{9}
+\frac{2\pi b_1}{b_0}\right)\,,
\eeqn
\beqn
\DtwoCtwo&=&\DoneCtwo\,,
\\
\DtwoCthree&=&-\left(\frac{2\NF}{3}+
\hat{\xi}_{1,0}+\chi_{1,0}\right)\,,
\\
\Conethree&=&0\,,
\\
\Ctwothree&=&0\,,
\\
\Cthreethree&=&-\exp\big(\!-\!\DtwoCthree t\big)\,,
\\
\DtwobCtwo&=&\DtwoCtwo\,,
\\
\DtwobCthree&=&\DtwoCthree\,,
\\
\Conefour&=&0\,,
\\
\Ctwofour&=&0\,,
\\
\Cthreefour&=&1\,,
\label{C34}
\eeqn
and:
\beqn
k_j&=&\xi_{1,0}\,t\,,\phantom{aaaaaaa}j=1,3\,,
\label{kappa13}
\\
k_j&=&0\,,\phantom{aaaaaaaaaa}j=2,4\,.
\label{kappa24}
\eeqn
We next consider the contribution of the second term on the
r.h.s.~of eq.~(\ref{gaPDFoffd}). Owing to eq.~(\ref{EopNL}), to the
$1/N$ suppression implicit in \mbox{$\widehat{\Mmat}_N^{(1,L)}$},
and to eq.~(\ref{asygammaN}), it is immediate to see that this
contribution, up to terms vanishing in the $z\to 1$ limit, is
identical to that of eq.~(\ref{gaNLLsol3run}), bar for an
$\aemz/\aemmu$ prefactor that here needs to be written according
to eq.~(\ref{Eop0gg1L}). Thus, by introducing the quantity:
\beqn
\ePDF{\gamma,5}(z)&=&
\frac{\aemz}{2\pi}\,\frac{1+(1-z)^2}{z}\left(
\log\frac{\muz^2}{m^2}-2\log z-1\right)\,,
\label{gaNLLsol3run2}
\eeqn
we can write the sought large-$z$ expression of the photon PDFs
in a compact form:
\beqn
\ePDF{\gamma}(z)&=&
\exp\left[-\left(\frac{2\NF}{3}+\chi_{1,0}\right)t\right]
\sum_{j=1}^5\ePDF{\gamma,j}(z)\,,
\label{gaNLLsol4run}
\eeqn
with $\ePDF{\gamma,j}(z)$ given in eq.~(\ref{gaPDFj14res}) for
$j\le 4$ and in eq.~(\ref{gaNLLsol3run2}) for $j=5$.

The results presented above allow one to obtain their counterparts
in the case of non-running $\aem$, by means of the following formal 
replacements (see eq.~(\ref{tina})):
\beqn
&&t\;\longrightarrow\;\frac{\eta_0}{2}\,,
\;\;\;\;\;\;\;\;
\chi_{1,0}\,t\;\longrightarrow\;
\frac{\aem}{2\pi}\,\frac{\eta_0}{2}\,\NF\,,
\label{faemrepl1}
\\*&&
b_0\;\longrightarrow\;0\,,
\;\;\;\;\;\;\;\;
b_1\;\longrightarrow\;0\,,
\;\;\;\;\;\;\;\;
b_1/b_0\;\longrightarrow\;0\,,
\label{faemrepl2}
\eeqn
with $\eta_0$ defined in eq.~(\ref{eta0def}). This procedure is
consistent with its analogue relevant to the recursive solutions
(see sect.~\ref{sec:rec} and appendix~\ref{sec:recres}). We can
also see that, by using the replacements above in the expression
for \mbox{$E_{\gamma\gamma,N}^{(0)}$} given in eq.~(\ref{Eop0gg1}),
one obtains the same result as one would have directly read from
the solution for the evolution operator relevant to the 
case of non-running $\aem$ (eq.~(\ref{Esol1ex}), with 
\mbox{$\aem(\mu)\to\aem$} and \mbox{$b_0\to 0$} there).
We observe that this would not have happened if one had used
eq.~(\ref{Eop0gg2}) instead of eq.~(\ref{Eop0gg1}), in spite of
these two equations being identical in QED. In other words, the
replacements in eqs.~(\ref{faemrepl1}) and~(\ref{faemrepl2}) might
lead to an incorrect result in the limit of non-running $\aem$ if
applied to an expression that contains two values of $\aem$ computed
at different scales; when this is the case, one must first express
one of such $\aem$ values in terms of the other one, and of $t$.
That being said, we point out again that the limit of non-running
$\aem$ must be interpreted with some care (see the comments that
follow eq.~(\ref{Eop0gg2})).

When not considering the case of non-running $\aem$, one can
re-expressed the exponential prefactors in eq.~(\ref{gaNLLsol4run}) 
and in eqs.~(\ref{D1C2})--(\ref{C34}), and their combinations,
in simpler ways, namely:
\beqn
\exp\big(\!-\!\DoneCthree t\big)\,
\exp\left[-\left(\frac{2\NF}{3}+\chi_{1,0}\right)t\right]&=&
\frac{\aemmu}{\aemz}\,e^{\hat{\xi}_{1,0}t}\;\longrightarrow\;
\frac{\aemmu}{\aemz}\,e^{\hat{\xi}_1}\,,
\label{simpID1}
\\
\exp\big(\!-\!\DtwoCthree t\big)\,
\exp\left[-\left(\frac{2\NF}{3}+\chi_{1,0}\right)t\right]&=&
e^{\hat{\xi}_{1,0}t}\;\longrightarrow\;
e^{\hat{\xi}_1}\,,
\label{simpID2}
\\
\exp\left[-\left(\frac{2\NF}{3}+\chi_{1,0}\right)t\right]&=&
\frac{\aemz}{\aemmu}\,.
\label{simpID3}
\eeqn
Two observations are in order. Firstly, the expressions on the
r.h.s.'s of eqs.~(\ref{simpID1}) and~(\ref{simpID2}) factorise
in the functions $R_i$, owing to the linearity of the latter 
w.r.t.~$\Conej$, $\Ctwoj$, and $\Cthreej$. Secondly, the replacements
on the rightmost sides of eqs.~(\ref{simpID1}) and~(\ref{simpID2}) stem
from eq.~(\ref{hx10def}); they are not mandatory, but are consistent
with the linearisation simplifications made when solving the evolution
equations. For scales of the order of up to a few hundred GeV's, in practice
they do not induce any significant numerical differences. With the
same arguments, in eq.~(\ref{gaPDFj14res}) one can also perform 
the replacements:
\beq
k_j=\xi_{1,0}\,t\;\;\longrightarrow\;\;\xi_1
\phantom{aaaaaaa}j=1,3\,,
\eeq
again from eq.~(\ref{hx10def}).

Equation~(\ref{gaNLLsol4run}) is the asymptotic solution that emerges 
from solving the evolution equation by keeping the dominant off-diagonal
terms in the Altarelli-Parisi kernels. As we shall discuss in
appendix~\ref{sec:lzexp}, it shares with its singlet and non-singlet
counterparts the nice property that its perturbative expansion lead
to the same coefficients as those of the recursive solutions (for 
certain classes of basis functions in the $z$ space). However, its
functional form is rather involved, but it is fortunately possible
to simplify it, by keeping only the truly dominant terms in the $z\to 1$
limit at each order in $\aem$. In order to do so one starts by observing 
that, in such a limit, one has:
\beqn
\invm_i(z;\kappa,\dencpar,\dendpar)\;\;
&\stackrel{z\to 1}{\longrightarrow}&\;\;0\,,
\label{inviklim}
\\
\invm_1(z;0,\dencpar,\dendpar)\;\;
&\stackrel{z\to 1}{\longrightarrow}&\;\;0\,,
\\
\invm_2(z;0,\dencpar,\dendpar)\;\;
&\stackrel{z\to 1}{\longrightarrow}&\;\;1\,,
\\
\invm_3(z;0,\dencpar,\dendpar)\;\;
&\stackrel{z\to 1}{\longrightarrow}&\;\;-\log(1-z)\,,
\\
\invm_4(z;0,\dencpar,\dendpar)\;\;
&\stackrel{z\to 1}{\longrightarrow}&\;\;\log^2(1-z)\,,
\\
\invm_5(z;0,\dencpar,\dendpar)\;\;
&\stackrel{z\to 1}{\longrightarrow}&\;\;
\frac{\pi^2}{2}\log(1-z)-\log^3(1-z)\,,
\eeqn
for any values of $\dencpar$ and $\dendpar$. Because of eqs.~(\ref{kappa13}) 
and~(\ref{kappa24}), eq.~(\ref{inviklim}) implies that only the $j=2$ and 
$j=4$ contributions to eq.~(\ref{gaNLLsol4run}) govern the divergent 
behaviour of $\ePDF{\gamma}(z)$ at $z\to 1$.
A simple computation then leads to the following result:
\beq
\ePDF{\gamma}(z)\;\stackrel{z\to 1}{\longrightarrow}\;
\frac{\aemz^2}{\aemmu}\,\frac{3}{2\pi\xi_{1,0}}\,\log(1-z)
-\frac{\aemz^3}{\aemmu}\,\frac{1}{2\pi^2\xi_{1,0}}\,\log^3(1-z)\,.
\label{ePDFgaasy}
\eeq
There is a certain similarity between eq.~(\ref{ePDFgaasy}) and 
eq.~(\ref{NLLsol3run}) which is worth stressing. In particular,
the dominant term at $z\to 1$ in both equations (proportional to
$\log(1-z)^3$ and $\log(1-z)^2$, respectively) is suppressed w.r.t.~the
subdominant one ($\log(1-z)$ in both cases) by a factor proportional
to $\aem$ (owing to eq.~(\ref{Aexp}) for eq.~(\ref{NLLsol3run})).
This implies that numerically the onset of the behaviour driven
by the most divergent terms occurs only at $z$ values which are
exceedingly large, and in fact hardly relevant to any phenomenological 
applications -- we have commented further on this
point in sect.~\ref{sec:res}.

Equation~(\ref{gaNLLsol4run}) simplifies considerably when one 
retains only the LL terms. A direct calculation leads to the following
result:
\beq
\ePDF{\gamma}(z)=
-e^{\hat{\xi}_0}\,\invm_1\!\left(z;\xi_0,D_1^{(0)},D_2^{(0)}\right)
+\frac{\aemz}{\aemmu}\,\invm_1\!\left(z;0,D_1^{(0)},D_2^{(0)}\right)\,,
\label{gaLLsol4run}
\eeq
with $\xi_0$ and $\hat{\xi}_0$ defined in eq.~(\ref{xiR0chi0Rdef}), and:
\beqn
D_1^{(0)}&=&2\,,
\\
D_2^{(0)}&=&-\frac{2\NF}{3}-\frac{3}{2}\,.
\eeqn
We point out that, consistently with the results of appendix~\ref{sec:abJhLL},
the LL photon PDF is of $\ord(t)$ (i.e.~it does vanish with $\aem\to 0$): the 
two terms on the r.h.s.~of eq.~(\ref{gaLLsol4run}) cancel each other at $t=0$.
From eq.~(\ref{Mmoinv1res}), we also see that the LL-accurate photon PDF of 
eq.~(\ref{gaLLsol4run}) vanishes in the $z\to 1$ limit:
\beq
\ePDF{\gamma}(z)\;\stackrel{z\to 1}{\longrightarrow}\;0\,.
\label{ePDFgaasyLL}
\eeq
By comparing eqs.~(\ref{ePDFgaasy}) and~(\ref{ePDFgaasyLL}) we observe
that the photon PDF has a behaviour analogous to that of the electron
PDF, namely that its NLL form grows faster than its LL counterpart at
$z\to 1$; to a good extent, this is an artifact of the $\MSb$ scheme.

We finally point out that eq.~(\ref{gaLLsol4run}) can be directly
obtained from solving the evolution equation of eq.~(\ref{EopMagExp}),
by using there:
\beq
\Mmat_N=\APmat_{{\rm\sss S},N}^{[0,0]}+
\frac{1}{N}\APmat_{{\rm\sss S},N}^{[0,1]}\,.
\label{MatNLL}
\eeq
Since the kernel of eq.~(\ref{MatNLL}) is independent of $t$,
eq.~(\ref{EopMagExp}) can be simply solved by diagonalisation.
After that, one multiplies the results by the LO initial conditions,
and performs the inverse Mellin transform. The fact that by doing so
one recovers eq.~(\ref{gaLLsol4run}) is a rather powerful check
on the procedure adopted in this appendix.

\section{Expansion of large-$z$ solutions\label{sec:lzexp}}
In view of the matching between the asymptotic large-$z$ solutions and 
the recursive solutions, it is useful to consider the expansion 
of the former ones in a series of $\aem$; this will also 
allow us to perform some consistency checks on them.
We can formally write the result of such an expansion for the NLL-accurate,
running-$\aem$ solutions of eqs.~(\ref{NLLsol3run}) and~(\ref{gaNLLsol4run})
in the same way as in eqs.~(\ref{PDFexptL}) and~(\ref{PDFexptLb}), 
namely\footnote{See footnote~\ref{ft:types} for what concerns the cases 
of solutions of different accuracy.}:
\beq
\overline{\Gamma}(z,\mu^2)=
\sum_{k=0}^{k_{\rm max}^{\sss\rm LL}} \frac{t^k}{k!}\,\KLL_k(z)
+\frac{\aem(t)}{2\pi}\sum_{k=0}^{k_{\rm max}^{\sss\rm NLL}} 
\frac{t^k}{k!}\,\KNLL_k(z)\,.
\label{lzPDFexpt}
\eeq
As the notation with an overline suggests, we only take into account
contributions that do no vanish at $z\to 1$.
We point out that we consider the expansion up to $\ord(\aem^3)$,
i.e.~we use the values in eq.~(\ref{ksmax}), for the sole reason of
consistency with what has been done for the recursive solutions in
appendix~\ref{sec:recres}.
The flavour structure of eq.~(\ref{lzPDFexpt}) is the same as
that in eqs.~(\ref{PDFasydef1}) and~(\ref{PDFasydef2}), and can therefore
be accounted for by $\KLL_k$ and $\KNLL_k$, precisely as is the case of the 
$\jLL_k$ and $\jNLL_k$ functions for the recursive solutions; in practice
we shall omit flavour indices here, in order to simplify the notation,
since no confusion is possible. In fact, one must bear in mind that the 
large-$z$ solutions of the singlet and non-singlet PDFs coincide, and that 
the one of the photon has a functional behaviour significantly different 
from the former two. Therefore we shall first deal with the singlet 
non-singlet cases together, and with that of the photon afterwards.

\subsection{Singlet and non-singlet\label{sec:SNSasyexp}}
When expanding eq.~(\ref{NLLsol3run}) to obtain $\KLL_k(z)$ and $\KNLL_k(z)$,
one can simply use the explicit expressions of $\xi_1$ and $\hat{\xi}_1$
in eqs.~(\ref{xiR1def}) and~(\ref{chi1Rdef}), respectively, and then
consider the Taylor series in $t$ and $\aem$. However, this procedure
cannot possibly give a correct answer at $z=1$, since $\Gamma(z)$
diverges there order by order, with non-integrable singularities.
In order to properly take such an endpoint contribution
into account, all $z$-dependent terms in $\Gamma(z)$ must be
regarded as distributions, rather than as regular functions. By doing so,
one can exploit the following identities:
\beq
\frac{\log^p(1-z)}{(1-z)^{1-\kappa}}=
\frac{(-1)^p\,\Gamma(1+p)}{\kappa^{1+p}}\,\delta(1-z)+
\sum_{i=0}^\infty \frac{\kappa^i}{\Gamma(1+\kappa)}\,\Lbase_{i+p}(z)\,,
\;\;\;\;\;\;\;\;p\ge 0\,,
\label{lpomzexp1}
\eeq
for any $\kappa$, and where:
\beq
\Lbase_i(z)=\Big[\lbase_i(z)\Big]_+\equiv
\left[\frac{\log^i(1-z)}{1-z}\right]_+,\;\;\;\;\;\;\;\;
i\ge 0\,,
\label{Llbasedef}
\eeq
having introduced $\lbase_i(z)$ in eq.~(\ref{lbasedef}).
By using eq.~(\ref{lpomzexp1}) in eq.~(\ref{NLLsol3run}) with
$\kappa=\xi_1$, and by subsequently expanding in $t$ and $\aem$, one
determines $\KLL_k(z)$ and $\KNLL_k(z)$. Because of the structure of
eq.~(\ref{lpomzexp1}), it is clear that the latter two quantities can
be expressed as linear combinations of the $\Lbase_i(z)$ distributions
and of Dirac delta's, namely:
\beqn
\KLL_k(z)&=&A_k^{\sss\rm LL}\delta(1-z)+
\left(1-\delta_{k0}\right)
\sum_{i=0}^{i_{\rm max}^{\sss\rm LL}(k)} B_{k,i}^{\sss\rm LL}\,\Lbase_i(z)\,,
\;\;\;\;\;\;\;\;\;\;\;k\ge 0\,,
\label{KLLkexp}
\\
\KNLL_k(z)&=&A_k^{\sss\rm NLL}\delta(1-z)+
\sum_{i=0}^{i_{\rm max}^{\sss\rm NLL}(k)} B_{k,i}^{\sss\rm NLL}\,\Lbase_i(z)\,,
\;\;\;\;\;\;\;\;\;\;\;\phantom{1-\delta_{k0}}k\ge 0\,.
\label{KNLLkexp}
\eeqn
Equations~(\ref{KLLkexp}) and~(\ref{KNLLkexp}) are by construction valid
for any $z$, including $z=1$, and so is eq.~(\ref{lzPDFexpt}). The $z=1$
contribution will be used in the following, but is not relevant for the
matching procedure. For the latter, $\overline{\Gamma}(z)$ will be
considered only with $z<1$, and thus becomes an ordinary function.
Its form can be read directly from eq.~(\ref{lzPDFexpt}), and is
as follows:
\beq
\overline{\Gamma}(z,\mu^2)=
\sum_{k=0}^{k_{\rm max}^{\sss\rm LL}} \frac{t^k}{k!}\,\kLL_k(z)
+\frac{\aem(t)}{2\pi}\sum_{k=0}^{k_{\rm max}^{\sss\rm NLL}} 
\frac{t^k}{k!}\,\kNLL_k(z)\,,
\label{lzPDFexptf}
\eeq
where:
\beqn
\kLL_k(z)&=&\KLL_k(z)\Big[A_k^{\sss\rm LL}\to 0\,,\;
\Lbase_i(z)\to\lbase_i(z)\Big],
\\
\kNLL_k(z)&=&\KNLL_k(z)\Big[A_k^{\sss\rm NLL}\to 0\,,\;
\Lbase_i(z)\to\lbase_i(z)\Big].
\eeqn
Note the strict similarity between eqs.~(\ref{lzPDFexptf}) 
and~(\ref{PDFexptLb}). This has to be expected, since both of these
expressions are $\ord(\aem^3)$ approximations of the PDF, that retain 
either some (eq.~(\ref{lzPDFexptf})) or all (eq.~(\ref{PDFexptLb}))
of the terms that are singular for $z\to 1$. 

We have determined the coefficients $A_k^{\sss\rm LL}$ and 
$B_{k,i}^{\sss\rm LL}$ for $k\le 3$, and $A_k^{\sss\rm NLL}$ and
$B_{k,i}^{\sss\rm NLL}$ for $k\le 2$, by means of a direct computation.
The results for $k=0$ are particularly interesting since, in view
of eq.~(\ref{lzPDFexpt}), they must be related to the initial
conditions of eqs.~(\ref{G0sol}) and~(\ref{G1sol2}). We have
obtained:
\beqn
A_0^{\sss\rm LL}&=&1\,,
\label{a0LL}
\\
A_0^{\sss\rm NLL}&=&2+\frac{3}{2}\,L_0\,,
\label{a0NLL}
\\
B_{0,0}^{\sss\rm NLL}&=&2\left(L_0-1\right)\,,
\label{b00NLL}
\\
B_{0,1}^{\sss\rm NLL}&=&-4\,,
\label{b01NLL}
\eeqn
where $L_0$ has been defined in eq.~(\ref{L0def}).
With the result of eq.~(\ref{a0LL}), $\kLL_0(z)$ is indeed identical 
to eq.~(\ref{G0sol}). However, by replacing the results of
eqs.~(\ref{a0NLL})--(\ref{b01NLL}) into eq.~(\ref{KNLLkexp}),
$\kNLL_0(z)$ turns out {\em not} to coincide with $\ePDF{\lm}^{[1]}(z)$
of eq.~(\ref{G1sol2}). This is hardly surprising: when working in the 
large-$z$ region, one is entitled to set $z=1$ in all of the polynomial 
terms that appear in the numerators. Therefore, while 
$\kNLL_0(z)$ should not necessarily be equal to $\ePDF{\lm}^{[1]}(z)$,
it {\em must} be equal to the $z\to 1$ asymptotic form of the latter --
if that were not the case, the large-$z$ solution would not be compatible
with the initial conditions from which it supposedly originates.
In order to obtain the asymptotic expression of the initial condition,
one cannot set $z=1$ in all of the numerators of the latter right away,
since $\ePDF{\lm}^{[1]}(z)$ is not an ordinary function, but a distribution.
Before doing so, one must first pull out the \mbox{$1+z^2$} factors from the 
plus distributions in eq.~(\ref{G1sol2}). This can be done by exploiting
the following identities:
\beqn
\frac{1+z^2}{(1-z)_+}&=&\left(\frac{1+z^2}{1-z}\right)_+
-\frac{3}{2}\,\delta(1-z)\,,
\label{pluscan}
\\
(1+z^2)\lppdistr{1-z}{+}&=&
\left(\frac{1+z^2}{1-z}\log(1-z)\right)_+
+\frac{7}{4}\,\delta(1-z)\,.\phantom{aaaa}
\label{Lpluscan}
\eeqn
After having done this, one can finally let \mbox{$1+z^2\to 2$} in
the numerators. It is a matter of simple algebra to show that this
procedure leads to the expected result:
\beq
\ePDF{\lm}^{[1]}(z)\;\stackrel{z\to 1}{\longrightarrow}\;\kNLL_0(z)\,.
\eeq
In summary, we have thus proven that the solution of eq.~(\ref{NLLsol3run}) 
embeds the initial conditions of eqs.~(\ref{G0sol}) and~(\ref{G1sol2}). 

We conclude this appendix by reporting the results for the coefficients
with $k>0$. We have obtained what follows:
\beqn
A_1^{\sss\rm LL}&=&\frac{3}{2}\,,
\\
A_1^{\sss\rm NLL}&=&\frac{27}{8}+\frac{\pi^2}{6}-2\zeta_3-
4\pi b_0-\frac{3\pi b_1}{b_0}
-\frac{\NF}{18}(3+4\pi^2)
\nonumber\\*
&+&\left(\frac{9}{4}-\frac{2\pi^2}{3}-3\pi b_0\right)L_0 
\\
A_2^{\sss\rm LL}&=&\frac{9}{4}-\frac{2\pi^2}{3}\,,
\\
A_2^{\sss\rm NLL}&=&\frac{45}{8}
+\left(8 b_0^2+6 b_1-\frac{5}{6}\right)\pi^2
+\frac{4\pi^4}{45}-\left(22-20\pi b_0\right)\zeta_3
-\left(\frac{51}{4}+\frac{5\pi^2}{3}\right)\pi b_0
\nonumber\\*
&-&\left(9-\frac{8\pi^2}{3}\right)\frac{\pi b_1}{b_0}
-\NF\left(\frac{1}{2}-\frac{22\pi^2}{27}
-\frac{\pi b_0}{3}-\frac{4\pi^3 b_0}{9}\right)
\nonumber\\*
&+&\left(\frac{27}{8}-3\pi^2-9\pi b_0+6\pi^2 b_0^2+
\frac{8\pi^3 b_0}{3}+16\zeta_3\right)L_0\,,
\\
A_3^{\sss\rm LL}&=&\frac{27}{8}-3\pi^2+16\zeta_3\,,
\eeqn
and:
\beqn
B_{k,i}^{\sss\rm LL}&=&
b_{{\rm\sss S},\,k,i}^{\sss\rm LL}=
b_{{\rm\sss NS},\,k,i}^{\sss\rm LL}\,,
\label{bLLlim}
\\
B_{k,i}^{\sss\rm NLL}&=&
b_{{\rm\sss S},\,k,i}^{\sss\rm NLL}=
b_{{\rm\sss NS},\,k,i}^{\sss\rm NLL}\,,
\label{bNLLlim}
\eeqn
with $b_{{\rm\sss S},\,k,i}^{\sss\rm LL}$ and
$b_{{\rm\sss NS},\,k,i}^{\sss\rm LL}$ given in 
appendix~\ref{sec:abJhLL}, and $b_{{\rm\sss S},\,k,i}^{\sss\rm NLL}$ 
and $b_{{\rm\sss NS},\,k,i}^{\sss\rm NLL}$ in appendix~\ref{sec:abJhNLL}.
We point out that eqs.~(\ref{bLLlim}) and~(\ref{bNLLlim}) hold for all
values of $k$ and $i$ we have considered here. This is remarkable, because
it tells one that with the expressions obtained in this paper all of the
\mbox{$\log^p(1-z)/(1-z)$} terms in the PDF are the same regardless of 
whether one obtains them from the recursive solution, or by expanding the
asymptotic solution. In general, one expects the logarithms from the latter 
to coincide with those of the former only for the larger values of $p$ at
any given $k$. The result obtained here ultimately stems from keeping some
formally subleading contributions in the procedure of sect.~\ref{sec:asyNLL};
in particular, it is important that the numerators in eqs.~(\ref{Ipdef})
and~(\ref{ILdef}) be \mbox{$1+x^2$} rather than $2$ (which would be a
perfectly fine choice in the asymptotic region)\footnote{\label{ft:asy}
It turns out that the use of \mbox{$1+x^2$} is essential in the determination
of the endpoint contributions in the plus distributions of eqs.~(\ref{Ipdef})
and~(\ref{ILdef}), which in turn induce (some of) the $z$-independent terms
in eqs.~(\ref{Ipi3}) and~(\ref{ILi3}). Conversely, away from the endpoints
the replacement of \mbox{$1+x^2$} with $2$ leads to power-suppressed
terms at $z\to 1$.}.

\subsection{Photon\label{sec:gaasyexp}}
In the case of the photon one needs to employ eq.~(\ref{gaNLLsol4run}).
We start by observing that the Taylor series in $t$ and $\aem$ of such
a quantity leads order by order to integrable singularities; as expected,
there is therefore no endpoint contribution, and the expansion of
the large-$z$ solution can be expressed in terms of ordinary functions.
Before turning to the explicit form of the latter, we point out that the 
$t^0$ term in the expansion of eq.~(\ref{gaNLLsol4run}) is equal 
to \mbox{$\ePDF{\gamma,5}(z)$}, since the contributions of the
\mbox{$\ePDF{\gamma,j}(z)$} with $j\le 4$ terms mutually cancel
(that of $j=1$ ($j=3$) against that of $j=2$ ($j=4$)). One thus
recovers the initial conditions of eqs.~(\ref{G0sol}) and~(\ref{Ggesol2}),
which is a first consistency check on eq.~(\ref{gaNLLsol4run}).
We now write the expansion of the large-$z$ photon PDF in the
same way as was done in eq.~(\ref{lzPDFexptf}), but with the
$K_k$ functions defined as follows:
\beqn
\kLL_k(z)&=&\left(1-\delta_{k0}\right)
\sum_{i=0}^{i_{\rm max}^{\sss\rm LL}(k)} C_{k,i}^{\sss\rm LL}\,\qbase_i(z)\,,
\;\;\;\;\;\;\;\;\;\;\;k\ge 0\,,
\label{gkLLkexp}
\\
\kNLL_k(z)&=&
\sum_{i=0}^{i_{\rm max}^{\sss\rm NLL}(k)} C_{k,i}^{\sss\rm NLL}\,\qbase_i(z)\,,
\;\;\;\;\;\;\;\;\;\;\;\phantom{1-\delta_{k0}}k\ge 0\,.
\label{gakNLLkexp}
\eeqn
having introduced the $\qbase_i(z)$ functions in eq.~(\ref{qbasedef}).
It is a matter of algebra to arrive at the final results:
\beqn
C_{k,i}^{\sss\rm LL}&=&
c_{\gamma,\,k,i}^{\sss\rm LL}\,,
\label{cLLlim}
\\
C_{k,i}^{\sss\rm NLL}&=&
c_{\gamma,\,k,i}^{\sss\rm NLL}\,,
\label{cNLLlim}
\eeqn
with $c_{\gamma,\,k,i}^{\sss\rm LL}$ given in appendix~\ref{sec:abJhLL}, 
and $c_{\gamma,\,k,i}^{\sss\rm NLL}$ in appendix~\ref{sec:abJhNLL}.
As was the case for their singlet and non-singlet counterparts
(eqs.~(\ref{bLLlim}) and~(\ref{bNLLlim})), eqs.~(\ref{cLLlim}) 
and~(\ref{cNLLlim}) have the property of holding for all of the
$k$ and $i$ values considered here. Thus, the same remarks done
previously are valid here as well (with the obvious exception that
they apply to the \mbox{$\log^p(1-z)$} terms rather than to the 
\mbox{$\log^p(1-z)/(1-z)$} ones relevant to the singlet and non-singlet
cases).

\section{Alternative $z$-space derivation of asymptotic large-$z$ 
solutions\label{sec:lzz}}
In this appendix we show how some of the asymptotic results
of sect.~\ref{sec:asy} can be obtained directly in configuration
space, that is without resorting to Mellin-space techniques, and
thus providing one with a cross-check on the results of the latter.
We have considered this alternative procedure starting from a couple
of simplifying assumptions\footnote{We do not make any claims as to whether
this $z$-space approach remains viable if either of these assumptions is
relaxed.}: namely, we only deal with the non-singlet case, and we neglect 
the running of $\aem$. We point out that this method has already been
used to obtain the LL solution of eq.~(\ref{LLsol3}) -- see 
e.g.~ref.~\cite{Altarelli:1981ax}. Here, we extend it to the NLL accuracy.

In essence, the procedure works as follows. One makes an ansatz for
the $z$-space functional form of $\Gamma(z,\mu^2)$, where the $\mu^2$
dependence is parametrised by unknown functions. The PDF evolution equations,
simplified in the $z\to 1$ limit, are then turned into differential
equations for such unknown functions, where the independent variable 
is $\mu^2$. By solving these equations, one is left with arbitrary integration
constants, whose values are finally determined by matching the solutions
to the known PDF initial conditions.

In order to proceed, we start by observing that the assumption of 
non-running $\aem$ implies that the dependence on $\mu^2$ can be
entirely parametrised by means of the quantity $\eta_0$, introduced
in eq.~(\ref{eta0def}); thus, we shall use the latter as our independent
variable. At the LL, this implies that the evolution equation of 
eq.~(\ref{APeqns}) reads as follows:
\beq
\frac{d}{d\eta_0} \GammaLL(z,\eta_0) = \frac{1}{2}\,
P^{[0]} \otimes_z \GammaLL(\eta_0)\,.
\label{APexplLL}
\eeq
For the computation of the convolution integral on the r.h.s.~of 
eq.~(\ref{APexplLL}) we approximate the first-order non-singlet 
Altarelli-Parisi kernel in the large-$z$ region as follows:
\beq
P^{[0]}(z)\;\stackrel{z \to 1}{\longrightarrow}\;
2\left(\frac{1}{1-z}\right)_+ + 2\lambda_0\,\delta(1-z)\,,
\label{P0asyxspace}
\eeq
which is the analogue of eq.~(\ref{P0asy2}). The parameter $\lambda_0$ 
has been defined in eq.~(\ref{bNdef}), and its value stems from the 
exact form of the denominator of the splitting kernels, \mbox{$1+z^2$};
thus, eq.~(\ref{P0asyxspace}) is fully consistent with what is observed
in footnote~\ref{ft:asy}. We now make the following ansatz for the
functional form of the LL PDF that appear in eq.~(\ref{APexplLL}):
\beq
\GammaLL(z,\eta_0)= b(\eta_0)\,(1-z)^{a(\eta_0)}\,.
\label{GammaxdepLL}
\eeq
By replacing eq.~(\ref{GammaxdepLL}) into eq.~(\ref{APexplLL}), and
by using eq.~(\ref{P0asyxspace}), the convolution integral has two trivial
contributions, induced by the $\delta(1-z)$ and by the subtraction term of 
the plus distribution (integrated in the $(0,z)$ range) in 
eq.~(\ref{P0asyxspace}). The non-trivial part of the convolution 
integral can also be easily computed in the $z\to 1$ limit, to read:
\beqn
\int_z^1  \frac{d x}{1-x} &&\left[
  \frac{1}{x} \left( 1-\frac{z}{x} \right)^{a(\eta_0)}
  - \left( 1-z \right)^{a(\eta_0)} \right]
\nonumber \\*&& \quad
\stackrel{z \to 1}{\simeq}
- (1 - z)^{a(\eta_0)} \Big[ \psi_0(a(\eta_0)+1) + \gE \Big]\,.
\label{intomz}
\eeqn
Thus, both sides of eq.~(\ref{APexplLL}) are linear combinations of
two terms, whose dependence on $z$ is equal to \mbox{$(1-z)^{a(\eta_0)}$} 
and to \mbox{$(1-z)^{a(\eta_0)}\log(1-z)$}, respectively. By equating
the coefficients of such terms one finally arrives at the sought
differential equations:
\beqn 
\frac{d}{d\eta_0} a(\eta_0) &=& 1\,, \\
\frac{d}{d\eta_0} b(\eta_0) &=& b(\eta_0) \Big[
-\big(\psi_0(a(\eta_0)+1) + \gE\big) + \lambda_0 \Big]\,.
\eeqn
The solutions of these are:
\beqn 
a(\eta_0) &=& \eta_0 + a_0\,,
\label{aLL}\\
b(\eta_0) &=& b_0\,\frac{e^{(\lambda_0 - \gE)\eta_0}}
{\Gamma(\eta_0 + a_0 + 1)}\,.
\label{bLL}
\eeqn
The quantities $a_0$ and $b_0$ are arbitrary integration constants, which
can be determined by observing that, in the limit \mbox{$\eta_0\to0$}, 
$\GammaLL(z,\eta_0)$ must be equal to the initial condition of 
eq.~(\ref{G0sol}). By imposing such an equality we obtain:
\beq 
a_0 = -1\,,\;\;\;\;\;\;\;\;
b_0 = 1\,.
\label{initcondeqdiffLL}
\eeq
It then becomes appartent that eq.~(\ref{GammaxdepLL}), supplemented with 
eqs.~(\ref{aLL}), (\ref{bLL}), and~(\ref{initcondeqdiffLL}), coincides 
with eq.~(\ref{LLsol3}).

The procedure outlined so far can now be extended to the NLL. We write
the analogue of eq.~(\ref{APexplLL}) as follows:
\beq
\frac{d}{d\eta_0} \GammaNLL(z,\eta_0) =
\frac{1}{2}
\left(P^{[0]} + \aemotpi P^{[1]} \right)
\otimes_z
\GammaNLL\left(\eta_0\right)\,,
\label{APexpl}
\eeq
with the second-order non-singlet  Altarelli-Parisi kernel 
approximated in the large-$z$ as follows:
\beq
P^{[1]}(z)\;\stackrel{z \to 1}{\longrightarrow}\;
-\frac{20}{9}\NF \left( \frac{1}{1-z} \right)_+
+ \lambda_1\,\delta(1-z)\,.
\label{P1asyxspace}
\eeq
Equation~(\ref{P1asyxspace}) is the $z$-space analogue of eq.~(\ref{P1asy2}),
with $\lambda_1$ defined in eq.~(\ref{lambda1}). We also need to replace
our LL ansatz of eq.~(\ref{GammaxdepLL}) with one that is appropriate
at the NLL, namely:
\beqn
\GammaNLL(z,\eta_0) &=& (1-z)^{a(\eta_0)}
\nonumber \\*&& \phantom{\times}
\times
\left\{b(\eta_0) + \frac{\aem}{\pi}
\Big[c(\eta_0) +d(\eta_0)\log(1-z)
  +e(\eta_0)\log^2(1-z) \Big]\right\}.\phantom{aaa}
\label{GammaxdepNLL}
\eeqn
The physical motivation of eq.~(\ref{GammaxdepNLL}) is the following.
Firstly, one observes that $P^{[0]}$ and $P^{[1]}$ have the same functional
large-$z$ behaviours. Secondly, we have seen that at the LL the convolution
of the evolution kernel with the r.h.s.~of eq.~(\ref{GammaxdepLL}) either
leaves the functional form of the latter unchanged, or it multiplies it
by a $\log(1-z)$ term. Therefore, since the $\ord(\aem)$ contribution
to the PDF initial condition in the $\MSb$ scheme, eq.~(\ref{G1sol2}), 
contains logarithmic terms up to the first power, its convolution with
the evolution kernel either leave those unchanged, or it increases their
powers by one unity.

As was the case at the LL, the convolution of the r.h.s.~of
eq.~(\ref{GammaxdepNLL}) with the Altarelli-Parisi kernels features
a few trivial contributions, due to the endpoints, and some non-trivial
ones, which can nevertheless be readily computed. Among the latter,
we find again eq.~(\ref{intomz}), and:
\beqn
\int_z^1 && \frac{d x}{1-x} \left[
  \frac{1}{x} \left( 1-\frac{z}{x} \right)^{a(\eta_0)} \log
  \left( 1-\frac{z}{x} \right)
  - \left( 1-z \right)^{a(\eta_0)} \log \left( 1-z \right) \right]
\nonumber \\*&& \quad
\stackrel{z \to 1}{\simeq}
- (1 - z)^{a(\eta_0)}
\Big\{\log (1 - z) [\psi_0(a(\eta_0)+1) + \gE]
+ \psi_1(a(\eta_0)+1)\Big\}\,,\phantom{aaa}
\\
\int_z^1 && \frac{d x}{1-x} \left[
  \frac{1}{x} \left( 1-\frac{z}{x} \right)^{a(\eta_0)} \log^2
  \left( 1-\frac{z}{x} \right)
  - \left( 1-z \right)^{a(\eta_0)} \log^2 \left( 1-z \right) \right]
\nonumber \\*&&
\stackrel{z \to 1}{\simeq}  -(1 - z)^{a(\eta_0)}
\Bigg\{\log(1 - z) \Big\{\log(1 - z) [\psi_0(a(\eta_0)+1) + \gE]
\nonumber \\*&&
\phantom{
  \stackrel{z \to 1}{\simeq}  -(1 - z)^{a(\eta_0)}
  \Bigg\{\log(1 - z) \Big\{
  }
+ 2 \psi_1(a(\eta_0)+1)\Big\} +  \psi_2(a(\eta_0)+1)\Bigg\}\,.
\eeqn
Upon using these results,
the two sides of eq.~(\ref{APexpl}) become linear combinations of terms
proportional to \mbox{$\log^p(1-z)$}, with $p=0,1,2,3$. By equating
the coefficients of such terms, one finds a system of differential
equations:
\begingroup \allowdisplaybreaks
\begin{align}
  \frac{d}{d\eta_0} a(\eta_0) &= 1 - \frac{5\aem}{9\pi} \NF\,, \\
  \frac{d}{d\eta_0} b(\eta_0) &= b(\eta_0) \left\{
  -[\psi_0(a(\eta_0)+1) + \gE]\left(1 - \frac{5\aem}{9\pi} \NF\right)
  + \left(\lambda_0 + \frac{\aem}{4\pi} \lambda_1\right) \right\}\,, \\
  \frac{d}{d\eta_0} e(\eta_0) &= e(\eta_0) \left\{
  -[\psi_0(a(\eta_0)+1) + \gE]\left(1 - \frac{5\aem}{9\pi} \NF\right)
  + \left(\lambda_0 + \frac{\aem}{4\pi} \lambda_1\right) \right\}\,, \\
  \frac{d}{d\eta_0} d(\eta_0) &= d(\eta_0) \left\{
  -[\psi_0(a(\eta_0)+1) + \gE]\left(1 - \frac{5\aem}{9\pi} \NF\right)
  + \left(\lambda_0 + \frac{\aem}{4\pi} \lambda_1\right) \right\}
  \nonumber \\
  & + \left(1 - \frac{5\aem}{9\pi} \NF\right)
  \Bigg\{ - 2\,e(\eta_0)\,\psi_1(a(\eta_0)+1) \Bigg\}\,, \\
  \frac{d}{d\eta_0} c(\eta_0) &= c(\eta_0) \left\{
  -[\psi_0(a(\eta_0)+1) + \gE]\left(1 - \frac{5\aem}{9\pi} \NF\right)
  + \left(\lambda_0 + \frac{\aem}{4\pi} \lambda_1\right) \right\}
  \nonumber \\
  & + \left(1 - \frac{5\aem}{9\pi} \NF\right) \Bigg\{
  - d(\eta_0)\,\psi_1(a(\eta_0)+1) - e(\eta_0)\,\psi_2(a(\eta_0)+1) \Bigg\}\,,
\end{align}
\endgroup
with solutions:
\begingroup \allowdisplaybreaks
\begin{align}
  a(\eta_0) &= \eta_0\left(1 - \frac{5\aem}{9\pi} \NF\right) + a_0
  \equiv \eta_1 + a_0 \,,
  \label{aeta0} \\
  b(\eta_0) &= b_0 \frac{e^{\hat{\eta}_1 - 
\gE \eta_1}}{\Gamma(\eta_1 + a_0 + 1)} \,,\\
  e(\eta_0) &= e_0 \frac{e^{\hat{\eta}_1 - 
\gE \eta_1}}{\Gamma(\eta_1 + a_0 + 1)} \,, \\
  d(\eta_0) &= e(\eta_0)\left[d_0 +
    \left(1 - \frac{5\aem}{9\pi} \NF\right)
    \int_{\eta_0}^1 dt\,2\,\psi_1(a(t)+1)\right]
  \nonumber \\*
  &=  e(\eta_0) \left[d_0 - 2\,\psi_0(\eta_1 + a_0 + 1)\right] \,, \\
  c(\eta_0) &= e(\eta_0)\left[c_0 +
    \left(1 - \frac{5\aem}{9\pi} \NF\right)
    \int_{\eta_0}^1 dt\,d(t)\,\psi_1(a(t)+1)
    + \psi_2(a(t)+1)\right]
  \label{ceta0}
  \nonumber \\
  &=  e(\eta_0) \left[c_0 - d_0\,\psi_0(\eta_1 + a_0 + 1)
    + \psi_0(\eta_1 + a_0 + 1)^2 - \psi_1(\eta_1 + a_0 + 1) \right] \,,
\end{align}
\endgroup
where $\eta_1$ and $\hat{\eta}_1$ have been defined in eqs.~(\ref{eta1def})
and~(\ref{heta1def}), respectively.

The arbitrary integration constants \mbox{$a_0\,,\ldots e_0$} can be found 
by matching with the initial condition. We observe that at $\mu=\mu_0$
the $\aem\to 0$ NLL result for the PDF must coincide with the LL one;
this implies that eq.~(\ref{initcondeqdiffLL}) must still hold true.
Because of this, one can expand eq.~(\ref{GammaxdepNLL}) by using the 
techniques employed in appendix~\ref{sec:lzexp} (see in particular
eq.~(\ref{lpomzexp1})), to obtain at $\ord(\aem)$ the same functional 
form as in eq.~(\ref{G1sol2}), which leads to the following results:
\beqn
c_0 &=& -\frac{7}{4} + \gE^2 + \frac{\pi^2}{6} + 
\left(\gE - \frac{3}{4}\right) \left(\log\frac{\mu_0^2}{m^2} - 1\right)\,,
\\
d_0 &=& 1 - 2\gE - \log\frac{\mu_0^2}{m^2}\,, 
\\
e_0 &=& -1\,.
\eeqn
By putting everything back together, one sees that eq.~(\ref{GammaxdepNLL})
coincides with eq.~(\ref{NLLsol3}).

\phantomsection
\addcontentsline{toc}{section}{References}
\bibliographystyle{JHEP}
\bibliography{eepdfmsb}

\end{document}